\documentclass[eqsecnum]{revtex4}

\usepackage{epsfig,amsmath,amsfonts}

\def\l{\langle}
\def\r{\rangle}
\def\be{\begin{eqnarray}}
\def\ee{\end{eqnarray}}
\def\a{\alpha}
\def\b{\beta}
\def\co{\hat{a}^{\dag}}
\def\ao{\hat{a}}
\def\bo{\begin{figure}}
\def\eo{\end{figure}}
\def\R{{\cal R}}

\def\nad #1.#2{\left({#1\atop #2}\right)}

\begin{document}

\title{Cold Trapped Ions as Quantum Information Processors}
\author{Marek \v Sa\v sura and Vladim{\'\i}r Bu\v zek}
\address{     
Research Center for Quantum Information, 
Slovak Academy of Sciences, 
D\'{u}bravsk\'{a} cesta 9, Bratislava 842 28, Slovakia
}
\date{September 13, 2001}

\begin{abstract}
In this tutorial we  review physical implementation of quantum computing 
using a system of cold trapped ions. We discuss systematically all the aspects for
making the implementation possible. 
Firstly, we go through the loading and confining 
of atomic ions in the linear Paul trap, then we describe 
the collective vibrational
motion of trapped ions. Further, we discuss interactions of the ions with 
a laser beam. We treat the interactions in the travelling-wave and
standing-wave configuration for dipole and quadrupole transitions. We 
review different types of laser cooling techniques associated with trapped
ions. We address Doppler cooling, sideband cooling in and beyond 
the Lamb-Dicke limit, sympathetic cooling and laser cooling using
electromagnetically induced transparency. After that we discuss the problem
of state detection using the electron shelving method. Then  
quantum gates are described. We introduce single-qubit rotations,
two-qubit controlled-NOT and multi-qubit controlled-NOT gates. We also
comment on more advanced multi-qubit logic gates. We describe how 
quantum logic networks may be
used for the synthesis of arbitrary pure quantum states. Finally, we discuss
the speed of quantum gates and we also give some numerical estimations for
them. A discussion of dynamics on off-resonant transitions associated with 
a qualitative estimation of the weak coupling regime and of the Lamb-Dicke
regime is included in Appendix.
\end{abstract}

\maketitle

\newpage

\tableofcontents

\newpage


\section{Introduction}
\label{impl}


Although trapped ions have found many applications in physics
\cite{NIST}, they caused a turning point in the evolution of quantum
computing when the paper entitled {\it Quantum computation with cold trapped
ions} was published by Cirac and Zoller in 1995 \cite{95-5}. This proposal
launched also an avalanche of other physical realizations of quantum
computing using different physical systems, from high finesse cavities
to widely manufactured
semiconductors \cite{sam}. Through the years we have learnt a lot, but also
revealed many peculiarities, about the physical realization of quantum
computing which has led to many discussions concerning the conditions under
which we could in principle implement quantum computing in 
certain quantum systems. 

Before we give the list of requirements for the physical implementation 
of quantum computing we will introduce the fundamental terminology 
to appear throughout this paper. 
We will follow the definitions in Ref. \cite{00-4}.
\begin{itemize}

\item A {\it qubit} is a quantum system in which the logical Boolean 
states 0 and 1 are represented by a prescribed pair of normalized and
mutually orthogonal quantum states labelled as $|0\r$ and $|1\r$. These two
states form a computational basis and any other (pure) state of the
qubit can be written as a superposition
\be
\label{q1}
|\psi\r=\a|0\r+\b|1\r
\ee
for some $\a$ and $\b$ such that $|\a|^2+|\b|^2=1$. 
It can be shown that we may choose $\a=\cos\vartheta$ and
$\b=e^{i\varphi}\sin\vartheta$. A~qubit is typically a
microscopic system, such as an atom, a nuclear spin or a polarized photon,
etc. In quantum optics a two-level atom with a selected ground $|g\r$
and excited $|e\r$ state represents a qubit. Hence the notation $|g\r$ and
$|e\r$ is used for the computational basis instead of $|0\r$ and $|1\r$.
For instance, 
some qubits can serve for logic operations or the storage of information. 
Then we refer to {\it logic qubits}. Some others can be used especially 
for sympathetic cooling of logic qubits and we may call them {\it cooling qubits}.
Some further qubits can be used as a quantum channel for
transferring the information between distinct 
logic qubits and then we refer to them as to a {\it quantum data bus}.

\item A {\it quantum register} of size $N$ refers to a collection 
of $N$ qubits.

\item A {\it quantum gate} is a device which performs a fixed unitary
operation on selected qubits in a fixed period of time.

\item A {\it quantum network} is a device consisting of quantum gates
whose computational steps are synchronized in time.

\item A {\it quantum computer (processor)} can be viewed 
as a quantum network or a~family of quantum networks.

\item A {\it quantum computation (computing)} 
is defined as a unitary evolution associated
with a set of networks which takes a initial quantum state (input)
into a final quantum state (output) and can be interpreted in terms of 
the theory of information processing.

\end{itemize}

For the moment we presume that the following five requirements 
(termed {\it DiVincenzo's checklist}) should be met in
order to realize quantum information processing on a quantum system
\cite{divin}. Actually, there are two more requirements for the case of the
transmission of qubits in space {\it (flying qubits)}. However, it appears
that all these requirements are necessary but not sufficient 
for successful experimental realization of a quantum processor
\cite{loss}.
\begin{itemize}

\item[(1)] The system must provide a well characterized qubit and 
the possibility to be scalable in order to create a quantum register.

\item[(2)] We must be able to initialize a simple initial state 
of the quantum register.

\item[(3)] Quantum gate operation times must be much shorter than 
decoherence times. The quantum gate operation time is the period 
required to perform a certain quantum gate on a single qubit or on a set of
qubits. The decoherence time approximately corresponds to the duration of
the transformation which turns a pure state of the qubit 
$|\psi\r=\a|0\r+\b|1\r$ into a mixture 
$\hat{\rho}=|\a|^2|0\r\l 0|+|\b|^2|1\r\l 1|$.

\item[(4)] We need a set of quantum gates, to perform any unitary
evolution operation that can be realized on the quantum system.
It has been shown that any unitary evolution can be
decomposed into a sequence of single qubit rotations and two-qubit 
controlled-NOT (CNOT) gates \cite{95-8}.

\item[(5)] The result of a quantum computational process must be efficiently
read out, i.e. the ability to measure distinct qubits is required.

\end{itemize}

Now we introduce briefly the physical system under consideration.
{\it Cold trapped ions} is a quantum system 
of $N$ atomic ions confined
in a linear trap. We assume an anisotropic and harmonic trapping
potential. The ions are laser cooled to a very low temperature, beyond 
the Doppler cooling limit, reaching the recoil cooling limit \cite{russia}. 
Hence the term cold trapped ions.
The ions form a linear crystal and oscillate in vibrational 
collective motional modes
around their equilibrium positions. In their internal structure, depending on
the choice of atomic species, we distinguish distinct atomic levels. The ions
are individually addressed with a laser or a set of lasers
in the travelling-wave or standing-wave configuration. We can detect 
the internal state of ions using optical detection devices.
Further, we address briefly the requirements for the physical implementation of
quantum computing (mentioned above) using cold trapped ions. 
\begin{itemize}

\item[(1)] The qubit is represented by a selected pair of internal atomic
states denoted as $|g\r$ and $|e\r$.
This selection is discussed in detail in Sec.\,\ref{lii}. 
The quantum register is realized by $N$ ions forming the ion
string in the linear trap, namely the linear Paul trap, which is reviewed in
Sec.\,\ref{ital}. A selected collective vibrational motional mode
(normal mode) is used as the quantum data bus. The vibrational motion of the
ions is treated in Sec.\,\ref{vmoti}.

\item[(2)] Different laser cooling techniques can be used 
for the proper initialization of the motional state of the ions.
They are described in Sec.\,\ref{lc}.
The initial internal state where all the ions are in the state $|g\r$ can be
reached by optical pumping to atomic states fast decaying to the ground
state $|g\r$ (Sec.\,\ref{lc} and \ref{es}).

\item[(3)] The influence of the decoherence on the motional state 
of the ions is suppressed by laser cooling to ground motional states 
of the normal modes. The internal levels of the ions representing the qubit
states $|0\r$ and $|1\r$ are selected such that they form slow transitions
with excited states of long lifetimes. A very detailed discussion 
of the decoherence bounds of trapped atomic ions can be 
found in Ref. \cite{98-5}.

\item[(4)] Single-qubit quantum rotations can be realized on any ion and
two-qubit controlled-NOT and multi-qubit
controlled-NOT quantum gates can be applied between chosen ions due to the possibility
of individual addressing with laser beams. The implementation of quantum
gates is discussed in Sec.\,\ref{qg}.

\item[(5)] The result of a computational process on cold trapped ions is
encoded into the final state of the internal atomic states. This information
can be very efficiently read out using the electron shelving method
addressed in Sec.\,\ref{es}.

\end{itemize}


\section{Ion trapping}
\label{ital}

\begin{figure}[htb]
\centerline{\epsfig{width=9cm,file=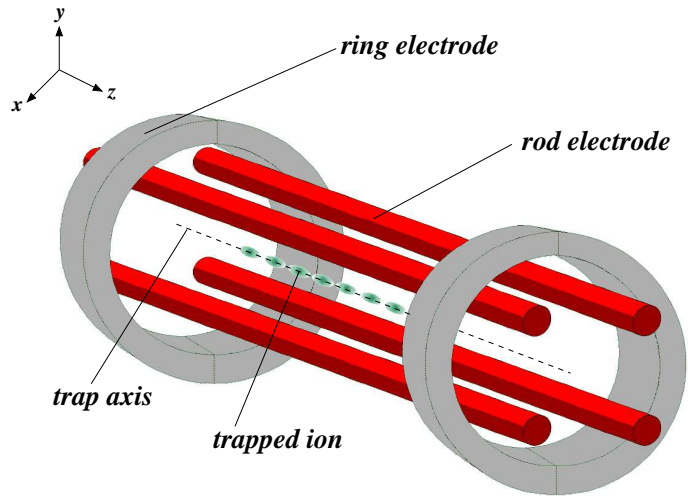}}
\caption{{\footnotesize
Linear Paul trap in the configuration with two ring electrodes spaced by
$2z_0$. The~diagonal distance between a pair of rod electrodes is $2r_0$.
Seven ions are confined at the~trap axis. The potential (\ref{trap1}) is
applied between two diagonally opposite rod electrodes. 
The other two are grounded. Ideally, equal static potential 
is applied on both ring electrodes. 
Used by kind permission of Rainer Blatt~\cite{innsbruck}.}}
\label{trap}
\end{figure}

Due to the charge of atomic ions, we can confine them  by particular
arrangements of electromagnetic fields. For studies of ions at low energy
two types of traps are used.
(i) {\it Penning trap} uses a combination of static electric and
magnetic fields and
(ii) {\it Paul trap} confines ions by oscillating electric fields.
Paul was awarded the Nobel Prize in 1990 for his work on trapping
particles in electromagnetic fields \cite{90-1}.
The operation of different ion traps is discussed in detail in Ref. \cite{ghosh}. 
For the~purpose considered in this paper we will
discuss only one trap configuration: the {\it linear Paul trap} 
(FIG.\,\ref{trap}).
We will follow Ref. \cite{ghosh} and \cite{LesH} for the mathematical treatment.
 
The linear Paul trap is basically a quadrupole mass filter, which is plugged 
at the ends with static electric potentials. An electric potential
\be
\label{trap1}
\phi_0=U_0+V_0\cos(\Omega t)
\ee
oscillating with the radiofrequency $\Omega$
is applied between two diagonally opposite rod electrodes. The electrodes
are coupled together with capacitors so that the
potential (\ref{trap1}) is constant as a function of the $z$ coordinate. 
The~other two rod electrodes are grounded. The resulting potential at the trap
axis (parallel with the $z$ direction) has the form
\be
\label{trap2}
\phi=\frac{\phi_0}{2r_0^2}(x^2-y^2)=
\frac{U_0+V_0\cos(\Omega t)}{2r_0^2}(x^2-y^2)\,,
\ee
where $r_0$ is the distance from the trap centre to the electrode surface.
In this field the (classical) equations of motion for an ion of the mass $m$
and charge $q$ are
\be
\label{trap3}
m\ddot{{\bf r}}=q{\bf E}=-q{\bf {\nabla}}\phi\,,\qquad
{\bf r}=(x,y,z)\,,
\ee
or rewritten in the components
\be
\label{trap4}
\ddot{x}+\frac{q}{mr^2_0}\bigg[U_0+V_0\cos(\Omega t)\bigg]x&=&0\,,\\
\label{trap5}
\ddot{y}-\frac{q}{mr^2_0}\bigg[U_0+V_0\cos(\Omega t)\bigg]y&=&0\,,\\
\label{trap6}
\ddot{z}&=&0\,.
\ee
After the substitution
\be
\label{trap6.1}
a=\frac{4qU_0}{mr^2_0\Omega^2},\qquad 
b=\frac{2qV_0}{mr^2_0\Omega^2},\qquad
\zeta=\frac{\Omega t}{2}\,,
\ee
Eq. (\ref{trap4}) and (\ref{trap5}) take the form of 
the~{\it Mathieu equation}
\be
\label{trap7}
\frac{d^2x}{d\zeta^2}+\bigg[a+2b\cos(2\zeta)\bigg]x&=&0\,,\\
\label{trap8}
\frac{d^2y}{d\zeta^2}-\bigg[a+2b\cos(2\zeta)\bigg]y&=&0\,.
\ee
The Mathieu equation can be solved, in general, using 
the~{\it Floquet solution}.
However, typically we have $a\ll b^2\ll 1$, then the approximate stable
solution of Eq. (\ref{trap7}) and (\ref{trap8}) are
\be
\label{trap9}
x(t)&\approx&x_0\left[1+\frac{b}{2}\cos(\Omega t)\right]
\cos(\omega_x t+\varphi_x)\,,\\
\label{trap10}
y(t)&\approx&y_0\left[1-\frac{b}{2}\cos(\Omega t)\right]
\cos(\omega_y t+\varphi_y)\,,
\ee
where
\be
\label{trap11}
\omega_x=\frac{\Omega}{2}\sqrt{\frac{b^2}{2}+a}\,,\qquad
\omega_y=\frac{\Omega}{2}\sqrt{\frac{b^2}{2}-a}
\ee
and $x_0$, $y_0$, $\varphi_x$, $\varphi_y$ are constants determined by initial
conditions. We see from Eq. (\ref{trap9}) and (\ref{trap10}) that the motion of
a single trapped ion in the radial direction is harmonic with the amplitude
modulated with the frequency $\Omega$. The harmonic oscillation
corresponding to the frequencies $\omega_x$ and $\omega_y$ 
is called the {\it secular motion}, whereas the small contribution oscillating 
at $\Omega$ is termed the {\it micromotion} \cite{nagerl, roos}.
We can eliminate the micromotion under certain conditions \cite{roos}. For
instance, well chosen voltages on additional compensation electrodes 
(not shown in FIG.\,\ref{trap}) null the micromotion.
Then the ion behaves as if it was confined 
in a harmonic pseudopotential $\psi_{2D}$ 
in the radial direction given by
\be
\label{trap12}
q\psi_{2D}=\frac{m}{2}\left(\omega_x^2x^2+\omega_y^2y^2\right)\,.
\ee
Typically, $U_0=0\,\mbox{V}$ and hence $a=0$, so the radial frequencies
$\omega_x$ and $\omega_y$
are degenerated. Then Eq. (\ref{trap12}) reduces to
\be
\label{trap13}
q\psi_{2D}=\frac{m\omega_r^2}{2}\left(x^2+y^2\right)\,,
\ee
where the {\it radial trapping frequency} $\omega_r$ is given by
\be
\label{trap14}
\omega_r=\frac{\Omega b}{2\sqrt{2}}=\frac{qV_0}{mr_0^2\Omega\sqrt{2}}\,.
\ee
In experiments \cite{nagerl, roos, 00-3, blatt1, blatt2}, 
typical operating parameters
are $V_0\simeq 300-800\,\mbox{V}$, 
$\Omega/2\pi\simeq 16-18\,\mbox{MHz}$, $r_0=1.2\,\mbox{mm}$, 
so we achieve the radial frequency 
$\omega_r/2\pi\simeq 1.4-2\,\mbox{MHz}$ for
Calcium ions $^{40}\mbox{Ca}^+$. In nature, $97\%$
of Calcium consists of this isotope.
To provide confinement along the $z$ direction, static potentials $U_1$ and
$U_2$ are applied on the ring electrodes. Ideally, $U_1=U_2=U_{12}$.
Numerical calculations show that the potential near the trap centre at the
trap axis is harmonic with the~approximate {\it axial trapping frequency} 
$\omega_z$ given by
\be
\label{trap15}
\frac{1}{2}m\omega_z^2z_0^2\approx{\xi}qU_{12}\,,
\ee 
where $z_0$ is the distance from the trap centre to the ring electrode and
$\xi$ is a geometric factor describing how much of
the static field from the ring electrodes is present along the trap axis
\cite{nagerl}.
Typical parameters are $\omega_z/2\pi\simeq 500-700\mbox{kHz}$ for
$U_{12}\simeq 2000\,\mbox{V}$ and $z_0=5\,\mbox{mm}$ \cite{blatt1,blatt2}.
The resulting pseudopotential for ions confined 
in the linear Paul trap in all three directions takes the form
\be
\label{trap16}
q\psi_{3D}=\frac{m\omega_r^2}{2}\left(x^2+y^2\right)+\frac{m\omega_z^2z^2}{2}\,,
\ee
where the radial trapping frequency $\omega_r$ is given by Eq. (\ref{trap14})
and the axial trapping frequency $\omega_z$ is defined by Eq. (\ref{trap15}).
For values of experimental parameters given above, we can calculate the depth
of the potential well in the axial direction 
($\omega_z/2\pi\simeq 700\,\mbox{kHz}$)
\be
\label{trap17}
V_z=\frac{m\omega_z^2z_0^2}{2}\simeq 100\,\mbox{eV}
\ee
and in the radial direction ($\omega_r/2\pi\simeq 2\,\mbox{MHz}$)
\be
\label{trap18}
V_r=\frac{m\omega_r^2r_0^2}{2}\simeq 820\,\mbox{eV}\,.
\ee
The potential well in the radial direction is almost several times deeper than
along the~trap axis, i.e. there is a strong binding in the radial direction.
Therefore we will not take into account radial oscillations of the ions 
in our further considerations.

Finally we briefly mention how ions are loaded into the trap. We will
follow the account of practical procedures in Ref. \cite{roos}.
Before starting the loading process, the trapping potentials are 
turned off for a while 
in order to get rid of any unwanted trapped residual ions.
The atomic oven producing Calcium atoms 
is switched on and heats up. This takes about a~minute. 
Then we turn on the electron gun ionizing neutral Calcium  atoms 
directly in the trapping volume. Cooling lasers are directed on the ion
cloud containing several hundreds of ions 
with a diameter of about 200\,$\mu$m. The~ion
cloud gradually relaxes into a steady state where the radiofrequency heating
(from the electrodes) is balanced by laser cooling. The number of trapped
ions is reduced by turning off the cooling. At low ion numbers, the ions
undergo a phase transition and form a linear crystal structure. Therefore,
we refer to the {\it ion crystal} or to the {\it ion string} or eventually
to the {\it ion chain}. The loading process itself takes normally about a minute.


\section{Collective vibrational motion}
\label{vmoti}


\subsection{Equilibrium positions}

We have learnt that the ions form a linear crystal structure in the linear
Paul trap after the loading process. We will assume a string of $N$
trapped ions. Due to the strong binding we can neglect the radial
oscillations. However, if a large number of ions is confined in the trap,
the radial vibrations become unstable and the ions undergo a phase
transition from a linear shape to an unstable zig-zag configuration. 
The relation
\be
\label{zig-zag}
\a_{crit}=cN^{\b}
\ee
determines a critical value for the ratio of the trapping frequencies
$\a=(\omega_z/\omega_r)^2$ for a given number of trapped ions
$N$. When $\a$ exceeds the critical value $\a_{crit}$, the ions are exposed
to a zig-zag motion. The experimental values of the constants in
Eq. (\ref{zig-zag}) are $c\simeq 3.23$ and $\b\simeq -1.83$. For experimental
details and the theoretical treatment we refer to Ref. \cite{zig-zag}. 

Further, we describe the collective vibrational motion of the ions. 
We will follow the treatment given by James in Ref. \cite{98-7}.
The ions are exposed to the harmonic potential (\ref{trap16}) due to the
trap electrodes and also to the repulsive Coulomb force from each other.
Taking into account all the assumptions given above, the potential energy of $N$
ions confined in the linear Paul trap is given by the expression
\be
\label{vibr1}
V=\sum_{i=1}^N\frac{m\omega_z^2z_i^2(t)}{2}+
\sum_{i,j=1\atop i<j}^N
\frac{q^2}{4\pi\varepsilon_0}\frac{1}{|z_i(t)-z_j(t)|}\,,
\ee
where $z_i(t)$ is the position of the $i$th ion numbering them from left to
right with the~origin in the trap centre, $m$ is the mass of the ion with 
the charge $q$, $\omega_z$ is the~axial trapping frequency (\ref{trap15}) and
$\varepsilon_0$ is the permitivity of the vacuum.

Assuming that the ions are cold enough, we can write for the position 
of the $i$th ion 
\be
\label{vibr2}
z_i(t)=\bar{z}_i+\Delta_i(t)\,,
\ee
where $\bar{z}_i$ is the equilibrium position and $\Delta_i(t)$ expresses
small vibrations around $\bar{z}_i$.
The ions placed in the equilibrium positions
minimize the potential energy. Hence these positions are determined by the
condition
\be
\label{vibr4}
\left[\ \frac{\partial V}{\partial z_i}\ \right]_{{\bf z}=\bar{{\bf
z}}}=0\,,\qquad i=1,\dots,N\,,
\ee
where ${\bf z}=(z_1,\dots,z_N)$ and 
$\bar{{\bf z}}=(\bar{z}_1,\dots,\bar{z}_N)$. 
We introduce a scaling factor $\gamma$ by the relation
\be
\label{vibr5}
\gamma^3={\frac{q^2}{4\pi\varepsilon_0m\omega_z^2}}
\ee
and the dimensionless equilibrium position as ${\cal Z}_i=\bar{z}_i/\gamma$. 
Then one can rewrite Eq. (\ref{vibr4}) to the form
\be
\label{vibr6}
{\cal Z}_i-
\sum_{j=1}^{i-1}
\frac{1}{({\cal Z}_i-{\cal Z}_j)^2}+
\sum_{j=i+1}^N
\frac{1}{({\cal Z}_i-{\cal Z}_j)^2}=0\,,\qquad i=1,\dots,N\,.
\ee
$N=1$ is a trivial case (${\cal Z}_1=0$).
We can find the analytical solution of Eq. (\ref{vibr6}) for two and three
ions:
\be
\label{vibr7}
&N=2\,,&\quad {\cal Z}_1=-\sqrt[3]{1/4}\,,\quad {\cal Z}_2=\sqrt[3]{1/4}\,,\\
&N=3\,,&\quad {\cal Z}_1=-\sqrt[3]{5/4}\,,\quad {\cal Z}_2=0\,,\quad
{\cal Z}_3=\sqrt[3]{5/4}\,.\nonumber
\ee
Numerical calculations are necessary for $N\geq 4$. For the~Calcium ions
$^{40}\mbox{Ca}^+$ and the~trap frequency $\omega_z/2\pi\simeq 700\,\mbox{kHz}$,
we may calculate the equilibrium positions as
\be
\label{vibr8}
&N=2\,,&\quad\Delta z_{min}\simeq 7.7\,\mu\mbox{m}\,,\\
&N=3\,,&\quad\Delta z_{min}\simeq 6.1\,\mu\mbox{m}\,.\nonumber
\ee
The minimum value $\Delta z_{min}$ of the distance between two neighbouring
ions in the trap occurs at the centre of the ion crystal, because 
the outer ions push the inner ions closer together. It has been
calculated from numerical data that this minimum distance is given 
approximately by the relation \cite{oxf, 98-7}
\be
\label{vibr9}
\Delta z_{min}(N)\approx\frac{2.018}{N^{0.559}}\gamma\,.
\ee
However, slightly different numerical results may be found in Ref. \cite{98-5}. 
The relation (\ref{vibr9}) happens to be important when one considers
individual ion addressing with a laser beam.
Quantum statistics of the ion ensemble is not considered here because 
the spatial spread of the zeropoint wavefunctions of the individual ions is of
the order of 10\,nm and the wavefunction overlap is then negligible 
\cite{oxf}.

\subsection{Normal modes}
\label{nm}

The (classical) Lagrangian of the ions in the trap is given by the formula
\be
\label{vibr10}
L\approx\frac{m}{2}\sum_{k=1}^N\dot{\Delta}^2_k-\frac{1}{2}\sum_{k,l=1}^N
\left[\ \frac{\partial^2V}{\partial z_k\partial z_l}\ \right]_
{{\bf z}=\bar{{\bf z}}}\Delta_k\Delta_l\,,
\ee
where we have expanded the potential energy (\ref{vibr1}) in a~Taylor
series about the equilibrium positions. 
In the expansion we have omitted the constant term and the~linear 
term which is zero [see Eq. (\ref{vibr4})]. 
Higher order terms ${\cal O}(\Delta^3_k)$ have been also neglected. 
However, they may cause a cross-coupling between different vibrational
modes which becomes a source of decoherence \cite{98-5}.
The~partial derivatives in Eq. (\ref{vibr10}) can be
calculated explicitly and we obtain the expression
\be
\label{vibr11}
L=\frac{m}{2}\left(\sum_{k=1}^N\dot{\Delta}_k^2-
\omega_z^2\sum_{k,l=1}^NV_{kl}\Delta_k\Delta_l\right)\,,
\ee
where
\be
\label{vibr12}
V_{kl}=\frac{1}{m\omega_z^2}
\left[\ \frac{\partial^2V}{\partial z_k\partial z_l}\ \right]_
{{\bf z}=\bar{{\bf z}}}=\left\{
\begin{array}{l}
1+\sum\limits_{j=1\atop j\neq k}^N
\frac{2}{|{\cal Z}_k-{\cal Z}_j|^3}\,,\ \hfill k=l\,,\\ \\
-\frac{2}{|{\cal Z}_k-{\cal Z}_l|^3}\,,\ \hfill k\neq l\,.
\end{array}
\right.
\ee
It follows from Eq. (\ref{vibr12}) that $V_{kl}=V_{lk}$.
The values of ${\cal Z}_j$ are given by Eq. (\ref{vibr7}) 
for $N=2$ and for $N=3$,
whereas they have to be calculated numerically for $N\geq 4$.

The dynamics of the trapped ions is governed by the Lagrange equations
\be
\label{vibr13}
\frac{d}{dt}\frac{\partial L}{\partial\dot{\Delta}_k}-
\frac{\partial L}{\partial\Delta_k}=0\,,\qquad k=1,\dots,N\,,
\ee
with the Lagrangian given by Eq. (\ref{vibr11}). 
We will search for a particular solution of Eq. (\ref{vibr13}) in the form
\be
\label{vibr14}
\Delta_k(t)=C_ke^{-i\nu t}\,,\qquad k=1,\dots,N\,,
\ee
where $C_k$ are constants. Substituting Eq. (\ref{vibr14}) into 
(\ref{vibr13}) we get the condition for $\nu$ in the form
\be
\label{vibr15}
\left\|\omega_z^2V_{kl}-\nu^2\delta_{kl}\right\|=0\,,
\ee
where $\delta_{kl}$ is the Kronecker symbol and $\|...\|$ denotes the
determinant. The equation (\ref{vibr15}) has in general up to $N$ real and
nonnegative solutions $\nu_{\a}$.
The frequencies $\nu_{\a}$ are
characteristic parameters of the system. They depend only on its physical
features (not on initial conditions). A general solution of Eq. (\ref{vibr13})
is a superposition of particular solutions (\ref{vibr14}) and we may write
\be
\label{vibr16}
\Delta_k(t)=\sum_{\a=1}^ND_k^{(\a)}\,Q_{\a}(t)\,,
\qquad k=1,\dots,N\,,
\ee
where
\be
\label{vibr17}
Q_{\a}(t)=C_{\a}e^{-i\nu_{\a}t}
\ee
By definition we will require the vectors
\be
\label{18.0}
{\bf D}^{(\a)}=\left(D_1^{(\a)},\dots,D_N^{(\a)}\right)\,,
\qquad \a=1,\dots,N\,, 
\ee
to be the eigenvectors of the matrix $V_{kl}$ defined in Eq. (\ref{vibr12}),
i.e.
\be
\label{vibr18}
\sum_{k=1}^NV_{kl}\,D_k^{(\a)}=\mu_{\a}D_l^{(\a)}\,,
\qquad l,\a=1,\dots,N\,,
\ee
and also to be orthogonal and properly normalized
\be
\label{vibr19}
\sum_{k=1}^ND_k^{(\a)}D_k^{(\b)}=\delta_{\a\b}\,,
\qquad \a,\b=1,\dots,N\,.
\ee
We will number the eigenvectors in order of the increasing
eigenvalues $\mu_{\a}$. It can be shown that the first two eigenvectors 
$(\a=1,2)$ always have the form
\be
\label{com}
{\bf D}^{(1)}&=&\frac{1}{\sqrt{N}}(1,1,\dots,1)\,,\qquad \mu_{1}=1\,,\\
\label{breath}
{\bf D}^{(2)}&=&\frac{1}{\sqrt{\sum_{k=1}^N{\cal Z}_k^2}}
({\cal Z}_1,{\cal Z}_2,\dots,{\cal Z}_N)\,,\qquad \mu_{2}=3\,.
\ee
We should emphasize that Eq. (\ref{com}) and (\ref{breath}) 
(they characterize two basic collective motional modes)
are not dependent on the number $N$ of the ions in the trap.
Next eigenvectors $(\a\geq 3)$ must be, in general, calculated numerically.
Substituting Eq. (\ref{com}) into (\ref{vibr19}) we get the
relation
\be
\label{vibr19.1}
\sum_{k=1}^ND^{(\a)}_k=0\,,\qquad \a=2,\dots,N\,.
\ee
We can determine analytically the eigensystem for two and three ions:
\be
\label{vibr19.2}
N=2\,,\qquad {\bf D}^{(1)}&=&\frac{1}{\sqrt{2}}(1,1)\,,\qquad \mu_{1}=1\,,\\
{\bf D}^{(2)}&=&\frac{1}{\sqrt{2}}(-1,1)\,,\qquad \mu_{2}=3\,,\nonumber\\
N=3\,,\qquad {\bf D}^{(1)}&=&\frac{1}{\sqrt{3}}(1,1,1)\,,\qquad \mu_{1}=1\,,\\
{\bf D}^{(2)}&=&\frac{1}{\sqrt{2}}(-1,0,1)\,,\qquad \mu_{2}=3\,,\nonumber\\
{\bf D}^{(3)}&=&\frac{1}{\sqrt{6}}(1,-2,1)\,,\qquad \mu_{3}=29/5\,.
\ee
For larger $N$, the eigenvectors and eigenvalues must be
computed numerically. The numerical values for up to ten ions can
be found in Ref. \cite{98-7}.

Substituting Eq. (\ref{vibr16}) into (\ref{vibr11}) 
we get a new expression for the Lagrangian
\be
\label{vibr20}
L=\frac{m}{2}\sum_{\a=1}^N\left(\dot{Q}^2_{\a}-\nu^2_{\a}Q^2_{\a}\right)\,,
\ee
where
\be
\label{vibr21}
\nu_{\a}=\omega_z\sqrt{\mu_{\a}}\,.
\ee
The Lagrangian (\ref{vibr20}) has split into $N$ uncoupled terms, where $Q_{\a}$
[Eq. (\ref{vibr17})] refer to the {\it normal modes} and
$\nu_{\a}$ defined in Eq. (\ref{vibr21}) are termed 
the~{\it normal frequencies}.
Finally, the position of the $i$th ion in the trap can be 
rewritten in terms of Eq. (\ref{vibr16}) using (\ref{vibr2}) to the~form
\be
\label{vibr22}
z_i(t)=\bar{z}_i+
\Re\left\{
\sum_{\a=1}^NC_{\a}D^{(\a)}_i\,e^{-i\,\nu_{\a}t}
\right\}\,,\qquad i=1,\dots,N\,,
\ee
where $\Re\{...\}$ denotes the real part and
$C_{\a}$ are constants given by initial conditions.
The collective vibrational motion of trapped ions determined by the
eigenvector ${\bf D}^{(1)}$ [Eq. (\ref{com})] refers to 
the normal mode called the {\it center-of-mass (COM) mode} 
\be
\label{vibr23}
z^{(1)}_i(t)=\bar{z}_i+
\Re\left\{
\frac{1}{\sqrt{N}}C_1\,e^{-i\omega_zt}
\right\}\,,\qquad i=1,\dots,N\,,
\ee
and corresponds to all of the~ions oscillating back and forth 
as if they were a rigid body.
The motion determined by the next eigenvector ${\bf D}^{(2)}$, 
[Eq. (\ref{breath}] refers to the {\it breathing mode} 
\be
\label{vibr24}
z_i^{(2)}(t)=\bar{z}_i+
\Re\left\{
\frac{\bar{z}_i}{\sqrt{\sum_{k=1}^N\bar{z}^2_k}}C_2\,
e^{-i(\omega_z\sqrt{3})t}
\right\}\,,\qquad i=1,\dots,N\,.
\ee
It corresponds to each ion oscillating with the amplitude proportional 
to its equilibrium distance from the trap center.
The COM motional mode can be excited in experiments by applying 
an additional AC voltage on one of the ring electrodes. For exciting 
the breathing motional mode, a 300-times higher voltage must be applied
\cite{00-3}. Higher motional modes require gradient field excitation due
to the nontrivial configuration of the ions in the ion string.
However, in the limit of large ion trap dimension in comparison 
with the ion crystal dimension, the electrode electric fields are almost uniform
across the ion crystal and the COM mode is very susceptible to heating due
to these fields. Therefore, it seems to be more advantageous 
to use rather the breathing mode, 
which is much less influenced by uniform fields, as the quantum data bus. 
This will be discussed in more detail later on in the section
on sympathetic cooling (Sec.\,\ref{sympcool}).
On the other hand, the ions can be easily addressed
with a laser beam in the COM mode, while higher modes require accurate
bookkeeping when addressing distinct ions in the ion crystal
\cite{98-7,symp1}.

\subsection{Quantized vibrational motion}

The normal modes $Q_{\a}$ are uncoupled in Eq. (\ref{vibr20}), so the
corresponding canonical momentum conjugated to $Q_{\a}$ is 
$P_{\a}=m\dot{Q_{\a}}$ and one may write the (classical) Hamiltonian
\be
\label{vibr25}
H=\frac{1}{2m}\sum_{\a=1}^NP_{\a}^2+
\frac{m}{2}\sum_{\a=1}^N\nu_{\a}^2Q_{\a}^2\,.
\ee
The quantum motion of the ions can be considered by introducing the operators
\be
\label{vibr26}
Q_{\a}\ \rightarrow\ \hat{Q}_{\a}=\sqrt{\frac{\hbar}{2m\nu_{\a}}}
\left(\co_{\a}+\ao_{\a}\right)\,,\\
\label{vibr27}
P_{\a}\ \rightarrow\ \hat{P}_{\a}=i\sqrt{\frac{\hbar m\nu_{\a}}{2}}
\left(\co_{\a}-\ao_{\a}\right)
\ee
with the corresponding commutation relations 
\be
\label{vibr27.1}
[\hat{Q}_{\a},\hat{P}_{\b}]=i\hbar\delta_{\a\b}\,,\qquad
[\ao_{a},\co_{\b}]=\delta_{\a\b}\,.
\ee
The Hamiltonian operator associated with the external (vibrational) 
degrees of freedom of the trapped ions is then expressed as follows
$(H\rightarrow\hat{H}_{ext})$
\be
\label{vibr28}
\hat{H}_{ext}=\sum_{\a=1}^N\hbar\nu_{\a}\left(\co_{\a}\ao_{\a}+1/2\right)\,,
\ee
where $\ao_{\a}$ and $\co_{\a}$ are the usual annihilation and creation
operators referring to the $\a$th normal mode. We use the standard notation
for the number states associated with the collective vibrational 
motion of the ions
\be
\label{vibr-pom1}
\co_{\a}\ao_{\a}|n_{\a}\r=n_{\a}|n_{\a}\r\,,
\ee
where $|n_{\a}\r$ refers to the state of the $\a$th normal mode and $n_{\a}$
denotes the number of vibrational phonons in this mode. The states
$\{|n_{\a}\r\}$ form the complete and orthonormal basis 
\be
\label{vibr-pom2}
\l m_{\a}|n_{\b}\r&=&\delta_{\a\b}\, \delta_{mn}\,.
\ee 
We can quantize the motion of the ions by applying Eq. (\ref{vibr26}) 
to the relation (\ref{vibr16}) and expressing the displacement operator
of the $i$th ion in the time-independent picture 
\be
\label{vibr29}
\hat{\Delta}_i&=&\sum_{\a=1}^ND^{(\a)}_i\sqrt{\frac{\hbar}{2m\nu_{\a}}}
(\co_{\a}+\ao_{\a})=\sum_{\a=1}^N{\cal K}_i^{(\a)}z_0(\co_{\a}+\ao_{\a})\,,
\qquad i=1,\dots,N\,,
\ee
where [see Eq. (\ref{vibr21})]
\be
\label{vibr30}
{\cal K}_i^{(\a)}=\frac{D^{(\a)}_i}{\sqrt[4]{\mu_{\a}}}\,,\qquad
z_0=\sqrt{\frac{\hbar}{2m\omega_z}}\,.
\ee 
We can easily calculate from Eq. (\ref{com}) that for the COM mode applies
\be
\label{vibr31}
{\cal K}_i^{(1)}=\frac{1}{\sqrt{N}}
\ee
and for the breathing mode [Eq. (\ref{breath})]
\be
\label{vibr31}
{\cal K}_i^{(2)}=
\frac{\bar{{\cal Z}}_i}{\sqrt{\sum_{l=1}^N\bar{{\cal Z}}_l^2}}
\frac{1}{\sqrt[4]{3}}=
\frac{\bar{z}_i}{\sqrt{\sum_{l=1}^N\bar{z}_l^2}}
\frac{1}{\sqrt[4]{3}}\,.
\ee 

Although we have not considered the radial vibrations due 
to the strong binding of the ions in the radial direction, a detailed treatment
of the ion motion in the trap would require the extension to all three
dimensions. Then Eq. (\ref{vibr2}) has to be replaced with
\be
\label{3D-1}
{\bf q}_i=\bar{{\bf q}}_i+{\Delta}{\bf q}_i\,,\quad i=1,\dots,N\,,
\ee
where $\bar{{\bf q}}_i$ denotes the equilibrium position of the $i$th ion 
in the 3D space and
${\Delta}{\bf q}_i$ is its displacement from the equilibrium
position. We can write 
\be
\label{3D-2}
\bar{{\bf q}}_i&=&
\bar{x}_i\,{\bf x}+\bar{y}_i\,{\bf y}+\bar{z}_i\,{\bf z}\,,\nonumber\\
\label{3D-3}
{\Delta}{\bf q}_i&=&
\Delta_i\,{\bf x}+\Delta_{N+i}\,{\bf y}+\Delta_{2N+i}\,{\bf z},
\quad i=1,2,\dots,N\,,
\ee
where $\bar{x}_i$, $\bar{y}_i$, $\bar{z}_i$ are the equilibrium positions 
of the $i$th ion and {\bf x}, {\bf y}, {\bf z} are unit vectors
in the 3D space. The free Hamiltonian associated with 
the vibrational motion in the 3D space reads
\be
\label{3D-4}
\hat{H}_{ext}^{(3\!D)}=
\sum_{\a=1}^{3N}\hbar\nu_{\a}\left(\co_{\a}\ao_{\a}+1/2\right)
\ee
and the displacement operators in Eq. (\ref{3D-3}) are given as follows
\be
\label{3D-5}
\hat{\Delta}_i=\sum_{\a=1}^{3N}{\cal K}_i^{(\a)}z_0(\co_{\a}+\ao_{\a})\,,
\qquad i=1,\dots,3N\,,
\ee
where the numerical factors ${\cal K}_i^{(\a)}$ in general 
have to be determined numerically.

   
\section{Laser-ion interactions}
\label{lii}

\begin{figure}[h!]
\centerline{\epsfig{width=14cm,file=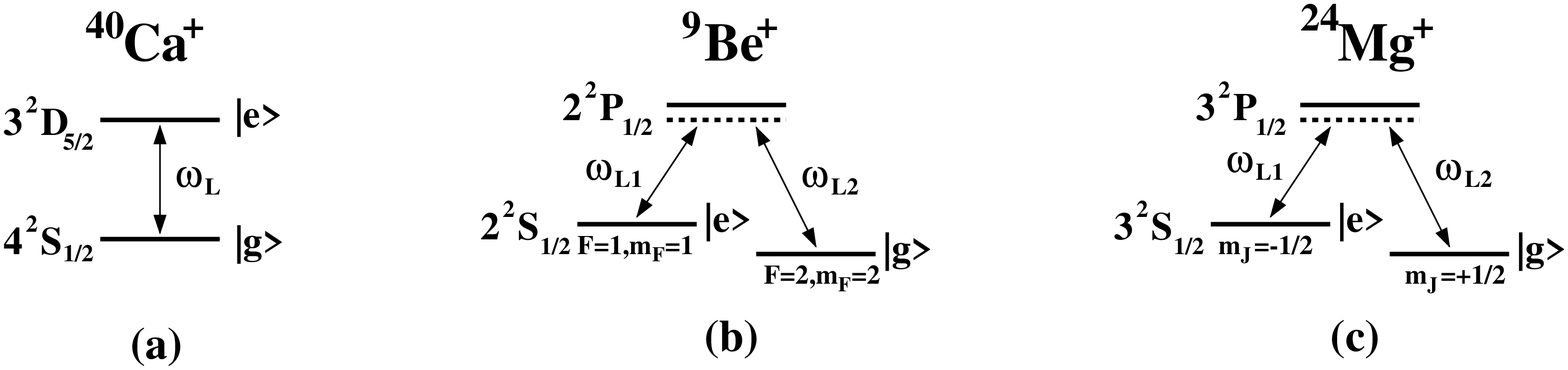}}
\caption{{\footnotesize Three possible choices for two internal atomic states
representing the qubit:
(a) a ground and a metastable excited state, (b) sublevels of a ground state and
(c) Zeeman sublevels of a ground state, where $\omega_L$, $\omega_{L1}$ and
$\omega_{L2}$ refer to the laser frequencies \cite{zeil}.}} 
\label{3types}
\end{figure}

Information is encoded in internal (atomic) states, while
it is transferred via external (motional) states of the ions.
We can manipulate these states due to laser-ion interactions.
It can be accomplished in the travelling-wave and standing-wave
configurations. We will address in detail both approaches in what follows. 
However, we should first comment on the selection of the two internal 
atomic levels to form the qubit. 
There are three possibilities \cite{98-7}:
\begin{itemize}

\item We can employ a ground and metastable fine structure excited state.
This applies for ions with zero nuclear angular momentum 
[FIG.\,\ref{3types}(a)]. In this
case we refer to the {\it single beam scheme} and we can drive transitions
on optical frequencies. This configuration is used, for example, by the
group in Innsbruck using Calcium ions $^{40}\mbox{Ca}^{+}$ 
\cite{innsbruck, 00-11}.

\item We can also choose two sublevels of a ground state within the
hyperfine structure (ions with nonzero nuclear angular momentum)
[FIG.\,\ref{3types}(b)].
The spacing of such two sublevels is in the range of GHz. Thus, a two-beam
{\it Raman scheme} via a~third virtual level is required in
order to resolve the individual sublevels. Experiments in this configuration with
Beryllium ions $^9\mbox{Be}^+$ were performed in Boulder 
\cite{NIST, 98-5, exp}.

\item It also possible to apply a magnetic field and consider two
Zeeman sublevels of the ground state [FIG.\,\ref{3types}(c)].
This scheme also requires Raman excitation.
In this class, we can mention, for example, Magnesium ions
$^{24}\mbox{Mg}^+$ used by the group in Garching \cite{garching}.

\end{itemize}
We have to mention also other active groups running experiments towards quantum
logic with trapped ions.  
For instance (in alphabetical order) 
IBM Almaden using $^{138}\mbox{Ba}^+$ \cite{IBM}, 
Imperial College ($^{40}\mbox{Ca}^+$, $^{199}\mbox{Hg}^+$) \cite{imperial},
JPL in Los~Angeles ($^{199}\mbox{Hg}^+$) \cite{JPL},
Los Alamos National Laboratory ($^{40}\mbox{Ca}^+$) \cite{LANL, lanl}, 
Oxford University ($^{40}\mbox{Ca}^+$) \cite{oxf, oxford},
University of Aarhus ($^{40}\mbox{Ca}^+$) \cite{aarhus},
University of Hamburg ($^{138}\mbox{Ba}^+$, $^{171}\mbox{Yb}^+$) \cite{hamburg}
and University of Mainz ($^{40}\mbox{Ca}^+$) \cite{mainz}.

\begin{figure}[htb]
\centerline{\epsfig{width=10cm,file=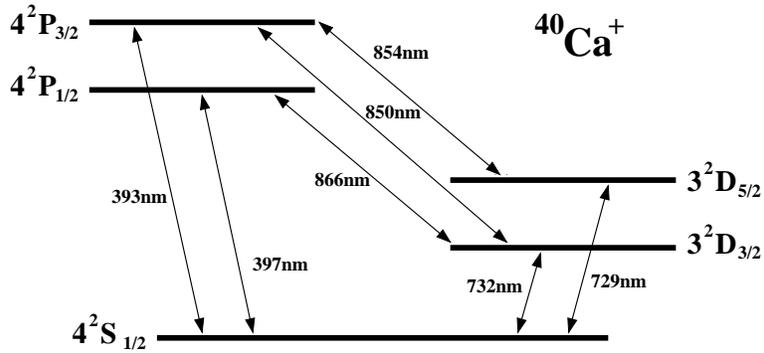}}
\caption{{\footnotesize
Five lowest available atomic levels of the Calcium ion $^{40}\mbox{Ca}^+$
\cite{roos}. All transitions are accessible with solid state diode lasers. 
The nuclear spin of this isotope is zero, i.e. there is no hyperfine 
structure. For the spectroscopic notation of the levels see the text.}}
\label{Ca}
\end{figure}

We can use {\it dipole} and {\it quadrupole} transitions.
Theoretically, the difference is only in the interaction constants 
as we will see later on in this section. 
On the other hand, in experiments
quadrupole transitions have much longer lifetimes (one second for Calcium ions)
comparing to fast decaying dipole transitions ($10^{-8}\,\mbox{s}$).
Experiments on an {\it octupole}
transition in an Ytterbium ion has also been realized.
The predicted theoretical lifetime in this system is
of the order of $10^8\,\mbox{s}$ \cite{oct}. However, in this case one deals 
with very weak transitions with very stringent demands on the laser sources 
used in the experiment (although they are of major interest as potential ion
trap {\it clocks}). Moreover, weak transitions have to be driven with 
a very intense laser which enhances the possibility for off-resonant
excitations.
From now on we will describe in this paper all experimental procedures 
for Calcium ions $^{40}\mbox{Ca}^+$ (FIG.\,\ref{Ca}).

In the following we will deal with the single beam scheme, i.e. transitions
being driven by a single laser beam. We will not treat here the Raman
scheme. The derivation of the Hamiltonian in this scheme can be found in
Ref. \cite{raman}. 
We just mention that the final Hamiltonian in the Raman scheme
has the same form as the one in the single beam scheme, except for differences
in coupling constants and for atomic frequencies which are Stark light shifted. 
In the Raman scheme the resulting effective light field has the direction
(frequency) determined by the difference of the wavevectors (frequencies) of
the two participating laser beams, where each beam is represented 
(in a semiclassical approach) with a monochromatic travelling wave. Finally, 
the single beam scheme requires a very high laser frequency stability, 
while in the Raman scheme we only need to control the relative
frequency stability between the two laser beams which is technically less
demanding. With the Raman scheme we can also ensure the relative wavevector
of the two beams to be parallel to the trap axis which suppresses the coupling to
radial motional modes. On the other hand, the Raman scheme can introduce 
significant Stark light shifts \cite{98-5}.

In the rest of the paper
we will use the standard atomic level notation $n\,^{2S+1}L_J$, where $n$ is 
the principal quantum number, $S$ is the spin angular momentum, $L$ is the
orbital angular momentum and $J$ is the total angular momentum of electrons.
For the fine structure case the notation is $n\,^{2S+1}L_J(m_J)$ where $m_J$ is
the projection of $J$ onto the quantization axis. In the case of 
the hyperfine structure we denote $n\,^{2S+1}L_J(F,m_F)$ where $F$ is the
total angular momentum of the atom (electrons + nucleus) and $m_F$ is the
projection of $F$ onto the quantization axis.

Let us consider that the ion has two internal levels, denoted $|g\r$ (lower) 
and $|e\r$ (upper) with corresponding energies $E_g$ and $E_e$, where the
transition frequency is $\omega_0=(E_e-E_g)/\hbar$. Then the free
Hamiltonian associated with the internal degrees of freedom is given by
\be
\label{int0}
\hat{H}_{int}=E_e|e\r\l e|+E_g|g\r\l g|=
\frac{\hbar\omega_0}{2}\sigma_z+\frac{E_e+E_g}{2}\openone_{int}\,,
\ee
where $\sigma_z=|e\r\l e|-|g\r\l g|$ and 
$\openone_{int}=|e\r\l e|+|g\r\l g|$.
Finally, we can write the total free Hamiltonian for the $j$th ion of $N$
ions confined in the trap communicating via one of the collective
vibrational modes [see Eq. (\ref{vibr28})]
\be
\label{int1}
\hat{H}_{0j}=\hat{H}_{int}+\hat{H}_{ext}=
\frac{\hbar\omega_0}{2}\sigma_{zj}+\hbar\nu\co\ao\,,
\ee
where we have omitted constant terms $(E_e+E_g)/2$, $\hbar\nu/2$ and
dropped down the index $\a$ denoting a vibrational mode.
The motional mode used for manipulations (especially quantum logic
operations) with the ions is called the {\it quantum data bus} because, as
we will see later, it serves to transfer the information between
distinct ions within the ion crystal (representing a quantum register).
We will consider for this purpose only the COM mode $(\nu=\omega_z)$ or the
breathing mode $(\nu=\omega_z\sqrt{3})$.

Further, we assume a powerful laser, i.e. the interaction with the ions has 
no influence on the laser photon statistics. Therefore, we will employ
a semiclassical description of the laser beam. We will consider the laser beam
in the~(i)~{\it travelling-wave} and (ii)~{\it standing-wave configuration}.

\subsection{Travelling-wave configuration}

There are two different ways for addressing the ions. 
We can set the laser beam at a fixed position and shift the ion
string by a very slight variation of the DC voltage on the ring electrode. 
On the other hand, we can fix the ion string and scan the laser
across the string. In this case an acousto-optical modulator is used for 
laser beam deflection \cite{00-3}.

\begin{figure}[htb]
\centerline{\epsfig{width=10cm,file=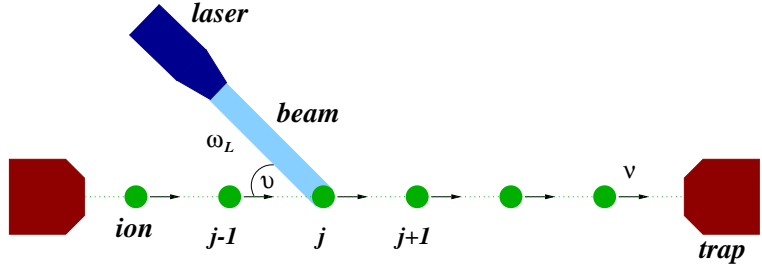}}
\caption{{\footnotesize
The travelling-wave configuration corresponds to illuminating
the $j$th ion in the ion string with the laser beam of the frequency
$\omega_L$ at the angle $\vartheta$ to the trap axis.}}
\label{trav}
\end{figure}

Let us approximate the laser beam as a monochromatic travelling
wave (FIG.\,\ref{trav}). We can write
\be
\label{int2}
{\bf E}&=&E_0\boldsymbol{\epsilon}
\cos\big(\omega_Lt-\boldsymbol{\kappa}\cdot{\bf q}+\phi\big)\,,\\
&=&\frac{E_0\boldsymbol{\epsilon}}{2}\left[
e^{-i(\omega_Lt-\boldsymbol{\kappa}\cdot{\bf q}+\phi)}+
e^{i(\omega_Lt-\boldsymbol{\kappa}\cdot{\bf q}+\phi)}
\right]\,,\nonumber
\ee 
where $E_0$ is the real amplitude, $\boldsymbol{\epsilon}$ 
is the polarization vector with $|\boldsymbol{\epsilon}|=1$,
$\omega_L$ is the laser frequency, 
$\boldsymbol{\kappa}=\kappa{\bf n}=(\omega_L/c){\bf n}$ is
the wavevector with $|{\bf n}|=1$, 
$c$ is the speed of light, ${\bf q}$ is the position 
vector and $\phi$ is the phase factor.
The full Hamiltonian for the $j$th ion is given by
\be
\label{int3}
\hat{H}_j=\hat{H}_{0j}+\hat{V}_j\,,
\ee
where the interaction Hamiltonian 
(assuming a hydrogen-like atomic configuration) 
expanded to second order 
(neglecting magnetic dipole interaction) has only two terms
\be
\label{int3.1}
\hat{V}_j=\hat{V}_j^{D\!P}+\hat{V}_j^{Q\!D}\,.
\ee
The electric dipole (DP) term is defined as follows
\be
\label{int4}
\hat{V}^{D\!P}_j=
-q_e\sum_{a}(\hat{{\bf r}}_j)_{a}\,E_{a}(t,\hat{{\bf R}}_j)=
-q_e\hat{{\bf r}}_j\cdot {\bf E}(t,\hat{{\bf R}}_j)\,,
\ee
summing over $a=x,y,z$. We refer to Eq. (\ref{int4}) 
as the {\it dipole approximation}.
The electric quadrupole (QD) term reads 
\be
\label{int4.1}
\hat{V}^{Q\!D}_j=
-\frac{q_e}{2}\sum_{a,b}
(\hat{{\bf r}}_j)_{a}(\hat{{\bf r}}_j)_{b}
\frac{\partial E_{b}(t,\hat{{\bf R}}_j)}{\partial q_{a}}\,,
\ee
where the sum is applied over $a,b=x,y,z$ and we refer 
to Eq. (\ref{int4.1}) as the {\it quadrupole approximation}. We denote
$q_e$ to be the electron charge, 
$\hat{{\bf r}}_j$ is the internal position operator associated with 
the position of the valence electron in the $j$th ion and 
$\hat{{\bf R}}_j=(0,0,\hat{z}_j)$ is the external position operator
corresponding to the position of the $j$th ion in the trap.

For the present we will
consider only the dipole term (\ref{int4}), regarding to the situation 
when the dipole interaction is present and the quadrupole contribution
(\ref{int4.1}) is then negligible. 
Later we will also comment on the quadrupole interaction.
If we consider consider only a single motional mode, we get from 
Eq.~(\ref{vibr29}) for the external position operator of the $j$ ion
\be
\label{int5.0}
\hat{z}_j=\bar{z}_j+{\cal K}_jz_0(\co+\ao)\,.
\ee
Then we can sandwich the internal position operator 
$\hat{{\bf r}}_j$ with the unity
operator $\openone_j=|e_j\r\l e_j|+|g_j\r\l g_j|$ and rewrite Eq. (\ref{int4}) 
to the form
\be
\label{int5}
\hat{V}_j=
-q_e\bigg[
({\bf r}_{eg})_j\hat{\sigma}_{+j}+({\bf r}_{eg})_j^*\hat{\sigma}_{-j}
\bigg]
\cdot\frac{E_0\boldsymbol{\epsilon}}{2}
\left\{e^{-i\left[\omega_Lt-\eta_j(\co+\ao)+\phi_j\right]}
+\mbox{H.c.}\right\}\,,
\ee
where  
$({\bf r}_{eg})_j=\l e_j|\hat{{\bf r}}_j|g_j\r$, 
$\hat{\sigma}_{+j}=|e_j\r\l g_j|$,
$\hat{\sigma}_{-j}=|g_j\r\l e_j|$,
$\kappa=\omega_L/c$,
$\eta_j={\cal K}_j\bar{\eta}$,
$\bar{\eta}=\kappa_{\vartheta}z_0$,
$\kappa_{\vartheta}=\kappa\cos\vartheta$,
$\phi_j=\phi-\kappa_{\vartheta}\bar{z}_j$
with ${\cal K}_j$ and $z_0$ defined by Eq. (\ref{vibr30}).
In Eq. (\ref{int5}) we consider that
$\l e_j|\hat{{\bf r}}_j|e_j\r=\l g_j|\hat{{\bf r}}_j|g_j\r=0$, because 
we assume spatial symmetry of the wavefunctions associated 
with the internal atomic
states $|g_j\r$ and $|e_j\r$.
The schematic configuration is depicted in FIG.\,\ref{trav}.
It is useful to transform to the interaction picture defined by the
prescription
\be
\label{int6}
i\hbar\frac{\partial}{\partial t}|\Psi\r=\hat{H}|\Psi\r
\quad &\longrightarrow&\quad
i\hbar\frac{\partial}{\partial t}|\psi\r=\hat{{\cal H}}|\psi\r\,,\qquad
|\psi\r=\hat{U}_0^{\dag}|\Psi\r\,,
\nonumber\\ \nonumber\\
\hat{H}=\hat{H}_0+\hat{V}
\quad &\longrightarrow&\quad
\hat{{\cal H}}=\hat{U}_0^{\dag}\,\hat{V}\,\hat{U}_0\,,\qquad
\hat{U}_0=\exp\left(-\frac{i\hat{H}_0t}{\hbar}\right)\,.
\ee
The Hamiltonian (\ref{int5}) after the transformation to the interaction
picture (\ref{int6}) reads
\be
\label{int7}
\hat{{\cal H}}_j=
\frac{\hbar\lambda_j}{2}\hat{\sigma}_{+j}
\exp\left[
i\eta_j\bigg(\co e^{i\nu t}+\ao e^{-i\nu t}\bigg)
\right]e^{-i\delta t}
+\mbox{H.c.}\,,
\ee
where $\delta=\omega_L-\omega_0$ and we have neglected rapidly oscillating
terms at the frequency $\omega_L+\omega_0$ compared with low-frequency
terms at $\omega_L-\omega_0$. In practice $\omega_L\simeq\omega_0$,
therefore to a good degree of approximation for times of interest,
high-frequency terms average to zero \cite{louisell}. 
This approximation is called 
the {\it rotating wave approximation (RWA)}. 
In Eq. (\ref{int7}) we substitute $\eta_j=\bar{\eta}/\sqrt{N}$ for the COM
mode or $\eta_j=\bar{\eta}\bar{z}_j/(\sqrt[4]{3}\sum_{l=1}^N\bar{z}_l^2)$ 
for the breathing mode.
The laser coupling constant $\lambda_j$ introduced 
in Eq. (\ref{int7}) is defined by the~relation
\be
\label{int8}
\lambda^{D\!P}_j=
-\frac{q_eE_{0}}{\hbar}
\bigg[\sum_{a}\l e_j|(\hat{{\bf r}}_j)_{a}|g_j\r\epsilon_{a}\bigg]\,
e^{-i\phi_j}=
-\frac{q_eE_{0}}{\hbar}
\bigg[({\bf r}_{eg})_j\cdot\boldsymbol{\epsilon}\bigg]\,e^{-i\phi_j}\,.
\ee
However, for a dipole forbidden transition when 
$\l e_j|\hat{{\bf r}}_j|g_j\r=0$, 
the dipole term (\ref{int4}) 
does not contribute ($\lambda^{D\!P}_j=0$) and the key role 
is played by the weaker quadrupole interaction. 
In that case the laser coupling constant in the Hamiltonian (\ref{int7}) reads
\be
\label{int8.1}
\lambda^{Q\!D}_j=
-\frac{iq_eE_0\omega_L}{2\hbar c}
\bigg[\sum_{a,b}
\l e_j|(\hat{{\bf r}}_j)_{a}(\hat{{\bf r}}_j)_{b}|g_j\r n_{a}\epsilon_{b}
\bigg]e^{-i\phi_j}\,,
\ee
where all parameters are defined in Eq. (\ref{int2}).

\begin{figure}[bht]
\centerline{\epsfig{width=13cm,file=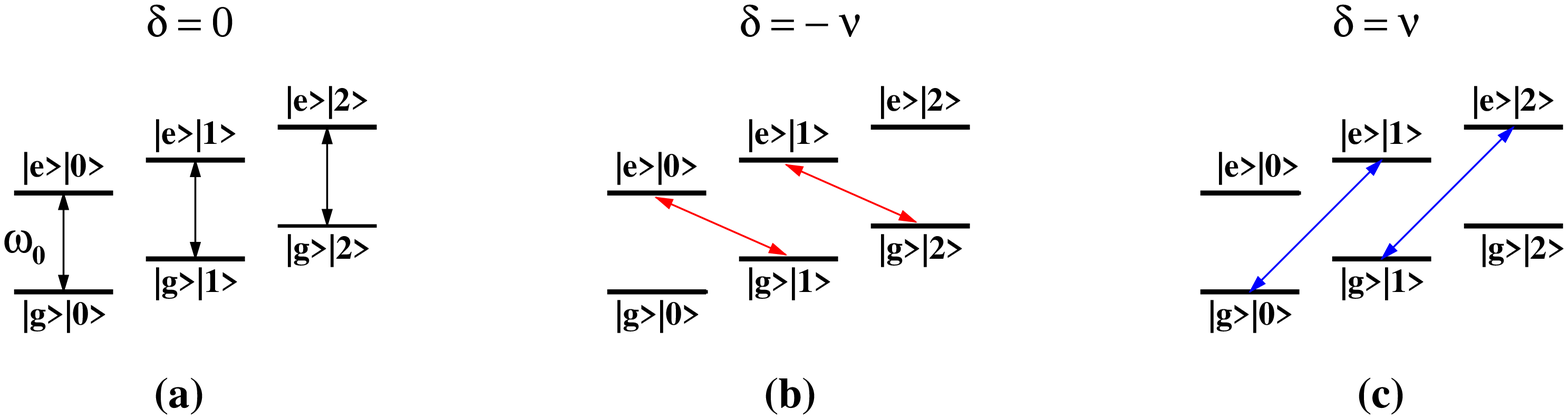}}
\caption{{\footnotesize
Scheme for the transition driven on the carrier (a), on the first red
sideband (b) and on the first blue sideband (c), where $\omega_L$ denotes 
the laser frequency, $\omega_0$ is the atomic frequency and $\nu$ is
the vibrational frequency of the respective motional mode. 
The parameter $\delta$ is the detuning defined as $\delta=\omega_L-\omega_0$.}}
\label{3freq}
\end{figure}

Next, let us assume the detuning $\delta$ of the laser frequency $\omega_L$
from the atomic frequency $\omega_0$ for the vibrational frequency $\nu$ in
the form
\be
\label{int9.0}
\delta=\omega_L-\omega_0=k\nu\,,\qquad  k=0,\pm 1,\pm 2,\dots\,,
\ee
and apply the Baker-Campbell-Hausdorff theorem \cite{louisell} 
$e^{\hat{A}+\hat{B}}=e^{\hat{A}}e^{\hat{B}}e^{-\frac{1}{2}[\hat{A},\hat{B}]}$
Eq. (\ref{int7}). Then we can write
\be
\label{int9}
\hat{{\cal H}}_j=
\frac{\hbar\lambda_j}{2}\hat{\sigma}_{+j}\,e^{-(\eta_j^2/2)}
\sum_{\a,\b=0}^{\infty}\left(i\eta_j\right)^{\a+\b}
\frac{(\co)^{\a}}{\a!}\frac{\ao^{\b}}{\b!}\,
e^{i\nu t(\a-\b-k)}+\mbox{H.c.}\,.
\ee
If the laser is tuned at the frequency $\omega_L$ such that $k>0$, 
the spectral line is
termed the~$k$th {\it blue sideband}. For $k=0$ the line is called 
the {\it carrier} and for $k<0$ refers the $k$th {\it red sideband}
because the laser is red (blue) detuned from the atomic 
frequency $\omega_0$ (FIG.\,\ref{3freq}).

When the constant $\lambda_j$ is sufficiently small 
we can assume that there are no excitations on
off-resonant transitions ({\it weak coupling regime}).
Then the level structure of the ion can be 
considered as a series of isolated two-level systems \cite{97-2}.
Precisely what is meant by sufficiently small is detailed in 
Appendix \ref{Off}. Assuming the weak coupling regime,
we can neglect off-resonant terms ($\a-\b-k\neq 0$) and 
rewrite Eq. (\ref{int9}) for $k\geq 0$ to the~form 
\be
\label{int10}
\hat{{\cal H}}_j^{(+)}=
\frac{\hbar\lambda_j}{2}\,\hat{\sigma}_{+j}\,(\co)^{|k|}\,
{\cal F}_{k}(\co\ao)+
\frac{\hbar\lambda_j^*}{2}\,\hat{\sigma}_{-j}\,
{\cal F}_{k}^{\dag}(\co\ao)\,\ao^{|k|}
\ee
and for $k<0$
\be
\label{int11}
\hat{{\cal H}}_j^{(-)}=
\frac{\hbar\lambda_j}{2}\,\hat{\sigma}_{+j}\,
{\cal F}_{k}(\co\ao)\,\ao^{|k|}+
\frac{\hbar\lambda_j^*}{2}\,\hat{\sigma}_{-j}\,(\co)^{|k|}\,
{\cal F}_{k}^{\dag}(\co\ao)
\,.
\ee
In the last relation we introduce the operator function
\be
\label{int12}
{\cal F}_{k}(\co\ao)=e^{-(\eta_j^2/2)}
\left(i\eta_j\right)^{|k|}
\sum_{\a=0}^{\infty}(-\eta_j^2)^{\a}
\frac{(\co\ao)^{\a}}{\a!(\a+|k|)!}\,.
\ee
Although we allow the parameter $k$ to be positive or negative, we
keep writing its absolute value $|k|$ in both cases in order to avoid the tricky
notation of form $(\co)^{-k}$ and $\ao^{-k}$ in the relation
(\ref{int11}) and also in what follows next. 
The final form of the Hamiltonian is given by
\be
\label{int13}
\hat{{\cal H}}_j^{(+)}=\hbar
\sum_{n=0}^{\infty}\left[
\frac{\Omega_{j}^{n,k}}{2}\bigg(|e_j\r\l g_j|\otimes |n+|k|\r\l n|\bigg)+
\frac{(\Omega_{j}^{n,k})^*}{2}\bigg(|g_j\r\l e_j|\otimes |n\r\l n+|k||
\bigg)
\right]\nonumber\\
\
\ee
and
\be
\label{int14}
\hat{{\cal H}}_j^{(-)}=\hbar 
\sum_{n=0}^{\infty}\left[
\frac{\Omega_{j}^{n,k}}{2}\bigg(|e_j\r\l g_j|\otimes |n\r\l n+|k||\bigg)+
\frac{(\Omega_{j}^{n,k})^*}{2}\bigg(|g_j\r\l e_j|\otimes |n+|k|\r\l n|
\bigg)\right]\,.\nonumber\\
\
\ee
We have defined a new coupling constant
\be
\label{int15}
\Omega_j^{n,k}=
\lambda_j\,e^{-(\eta_j^2/2)}
\left(i\eta_j\right)^{|k|}
\sqrt{\frac{n!}{(n+|k|)!}}\,L_n^{|k|}\left(\eta_j^2\right)\,,
\ee
where
\be 
\label{int16}
L_n^{a}(x)=\sum_{m=0}^n(-1)^m\,\frac{x^m}{m!}
\left({n+a\atop n-m}\right)
\ee 
is the generalized Laguerre polynomial and 
$\nad {n+a}.{n-m}=\frac{(n+a)!}{(n-m)!(a+m)!}$.
Finally, we may write the unitary evolution operator for time-independent
Hamiltonians (\ref{int13}) and (\ref{int14}) 
\be
\label{int17}
\hat{{\cal U}}_j^{(\pm)}=
\exp\left(-\frac{i\hat{{\cal H}}_j^{(\pm)}t}{\hbar}\right)\,,
\ee
which is given for $k\geq 0$ by the formula
\be
\label{int18}
\hat{{\cal U}}_j^{(+)}&=&
\sum_{n=0}^{\infty}\cos\left(\frac{|\Omega_{j}^{n,k}|t}{2}\right)
\Bigg[\bigg(
|e_j\r\l e_j|\otimes |n+|k|\r\l n+|k||\bigg)+
\bigg(|g_j\r\l g_j|\otimes |n\r\l n|\bigg)\Bigg]
\nonumber\\
&-&i\sum_{n=0}^{\infty}\sin\left(\frac{|\Omega_{j}^{n,k}|t}{2}\right)
\Bigg[
\bigg(|e_j\r\l g_j|\otimes |n+|k|\r\l n|\bigg)e^{-i\tilde{\phi}_j}+
\bigg(|g_j\r\l e_j|\otimes |n\r\l n+|k||\bigg)e^{i\tilde{\phi}_j}
\Bigg]
\nonumber\\
&+&\sum_{n=0}^{|k|-1}|e_j\r\l e_j|\otimes |n\r\l n|
\ee
and for ${k<0}$
\be
\label{int19}
\hat{{\cal U}}_j^{(-)}&=&
\sum_{n=0}^{\infty}\cos\left(\frac{|\Omega_{j}^{n,k}|t}{2}\right)
\Bigg[\bigg(
|e_j\r\l e_j|\otimes |n\r\l n|\bigg)+
\bigg(|g_j\r\l g_j|\otimes |n+|k|\r\l n+|k||
\bigg)\Bigg]
\nonumber\\
&-&i\sum_{n=0}^{\infty}\sin\left(\frac{|\Omega_{j}^{n,k}|t}{2}\right)
\Bigg[
\bigg(|e_j\r\l g_j|\otimes |n\r\l n+|k||\bigg)e^{-i\tilde{\phi}_j}+ 
\bigg(|g_j\r\l e_j|\otimes |n+|k|\r\l n|\bigg)e^{i\tilde{\phi}_j}
\Bigg]
\nonumber\\
&+&\sum_{n=0}^{|k|-1}|g_j\r\l g_j|\otimes |n\r\l n|\,.
\ee
We have denoted $\tilde{\phi}_j=\phi_j-\frac{\pi}{2}|k|$. 
\label{phase} 
For each value of $k$ the phase factor $\tilde{\phi}_j$ can be chosen
arbitrarily for the first application of $\hat{{\cal U}}_j^{(\pm)}$.
However, once chosen, it must be kept track of if subsequent applications of
$\hat{{\cal U}}^{(\pm)}_{j}$ are performed on the $j$th ion \cite{98-5}.
The real parameter $|\Omega_{j}^{n,k}|$ is called 
the {\it Rabi frequency} of the transition 
$|e_j\r|n\r\leftrightarrow|g_j\r|n+|k|\r$ or 
$|e_j\r|n+|k|\r\leftrightarrow|g_j\r|n\r$, respectively. This term comes
originally from the field of nuclear magnetic resonance (NMR), 
where it refers to the
periodic flipping of a nuclear spin in the magnetic field. 
It follows that:
\begin{itemize}

\item A $4\pi$-pulse ($|\Omega_{j}^{n,k}|t=4\pi$)
returns the system back to its initial state. For example 
\be
|e_j\r|n\r\stackrel{4\pi}{\longrightarrow}|e_j\r|n\r\,.
\ee

\item A $2\pi$-pulse ($|\Omega_{j}^{n,k}|t=2\pi$) changes the sign of the
state. For instance 
\be
|g_j\r|n+|k|\r\stackrel{2\pi}{\longrightarrow}-|g_j\r|n+|k|\r\,.
\ee

\item A $\pi$-pulse ($|\Omega_{j}^{n,k}|t=\pi$) implies that
\be
|e_j\r|n+|k|\r\stackrel{\pi}{\longrightarrow}|g_j\r|n\r\,,
\ee
where we set the phase factor $\tilde{\phi}_j$ to be zero.
Other cases may be easily calculated from the formulas 
(\ref{int18}) and (\ref{int19}) given above.
\end{itemize}

In what follows we will assume that all motional modes are 
in the {\it Lamb-Dicke regime} characterized by the {\it Lamb-Dicke limit} 
(Appendix \ref{ldl}). Hence $\eta_j$ introduced in Eq. (\ref{int5}) is called 
the {\it Lamb-Dicke parameter}. The Lamb-Dicke regime facilitates the ground
state cooling (Sec.\,\ref{sideband}) and enables to maintain the contrast of
Rabi oscillations on a longer time scale [see Eq.\,(\ref{eit2})].
Then the coupling constant (\ref{int15}) simplifies to the form
\be
\label{int20}
\Omega_j^{n,k}\approx\lambda_j
\left(i\eta_j\right)^{|k|}
\sqrt{\frac{(n+|k|)!}{n!}}\,\frac{1}{|k|!}\,.
\ee
For the purpose of coherent manipulations with internal states of cold
trapped ions we will be primarily interested in the interaction on the carrier
($k=0$) and on the first red sideband ($k=-1$) which will be used 
for the~construction of a wide class of quantum logic
gates. The corresponding unitary evolution operators in the Lamb-Dicke
regime for the transition on the carrier $(\hat{{\cal A}})$ 
and on the first red sideband $(\hat{{\cal B}})$ 
may be determined from Eq. (\ref{int18}) and (\ref{int19}) as follows
\be
\label{int21}
\hat{{\cal A}}_j&=&
\sum_{n=0}^{\infty}\cos\left(\frac{|A_j^n|t}{2}\right)
\Bigg[
\bigg(|e_j\r\l e_j|\otimes |n\r\l n|\bigg)+
\bigg(|g_j\r\l g_j|\otimes |n\r\l n|\bigg)
\Bigg]\\
&-&i\sum_{n=0}^{\infty}\sin\left(\frac{|A_j^n|t}{2}\right)
\Bigg[
\bigg(|e_j\r\l g_j|\otimes |n\r\l n|\bigg)e^{-i\phi_j}+
\bigg(|g_j\r\l e_j|\otimes |n\r\l n|\bigg)e^{i\phi_j}
\Bigg]
\nonumber 
\ee
and
\be
\label{int22}
\hat{{\cal B}}_j&=&
\sum_{n=0}^{\infty}\cos\left(\frac{|B_j^n|t}{2}\right)
\Bigg[
\bigg(|e_j\r\l e_j|\otimes |n\r\l n|\bigg)+
\bigg(|g_j\r\l g_j|\otimes |n+1\r\l n+1|\bigg)
\Bigg]\\
&-&i\sum_{n=0}^{\infty}\sin\left(\frac{|B_j^n|t}{2}\right)
\Bigg[
\bigg(|e_j\r\l g_j|\otimes |n\r\l n+1|\bigg)e^{-i\tilde{\phi}_j}+
\bigg(|g_j\r\l e_j|\otimes |n+1\r\l n|\bigg)e^{i\tilde{\phi}_j}
\Bigg]
\nonumber\\
&+&|g_j\r\l g_j|\otimes |0\r\l 0|\,.\nonumber
\ee
The respective Rabi frequencies in the Lamb-Dicke limit [Eq. (\ref{int20})] 
for $k=0$ and $k=-1$ are given by
\be
\label{conA}
|A_j^n|&=&|\lambda_j|\,,\\
\label{int24}
\label{conB}
|B_j^n|&=&|\lambda_j|\eta_j\sqrt{n+1}\,.
\ee
We could by analogy obtain evolution operators for other 
sideband transitions.

\subsection{Standing-wave configuration}

\begin{figure}[htb]
\centerline{\epsfig{width=10cm,file=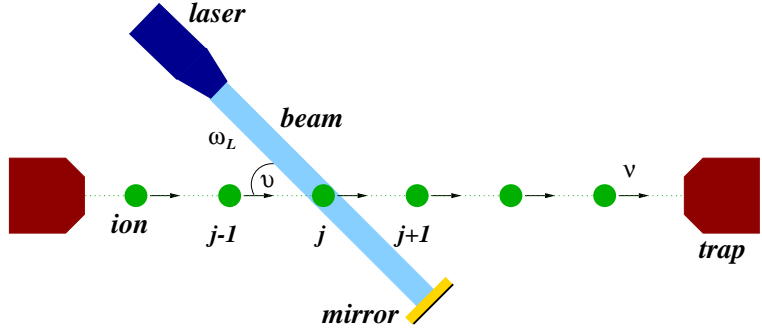}}
\caption{{\footnotesize
The standing-wave configuration corresponds to illuminating 
the $j$th ion in the ion crystal with the laser beam of the frequency
$\omega_L$ at the angle $\vartheta$ to the trap axis.}}
\label{stan}
\end{figure}

As an alternative approach to the laser-ion interactions we could choose 
a standing light field (FIG.\,\ref{stan}). 
One can place a mirror in the setup and let the
laser beam reflect from it. The counter propagating waves interfere and create
a standing-wave configuration with nodes and antinodes. However, it is
experimentally very demanding to place an ion precisely 
to a node or an antinode. Let us approximate
the incident laser beam as a monochromatic travelling wave
\be
\label{int25}
{\bf E}_i=E_0\boldsymbol{\epsilon}
\cos\big(\omega_Lt-\boldsymbol{\kappa}\cdot{\bf q}+\phi\big)
\ee
and the reflected beam as a counter propagating travelling wave
\be
\label{int26}
{\bf E}_r=E_0\boldsymbol{\epsilon}
\cos\big(\omega_Lt+\boldsymbol{\kappa}\cdot{\bf q}+\phi-\pi\big)\,,
\ee
where the reflected wave acquires an additional phase $\pi$ on the reflection
at the perfect lossless mirror. Then we can write for the resulting 
standing wave
\be
\label{int27}
{\bf E}={\bf E}_i+{\bf E}_r=2E_0\boldsymbol{\epsilon}
\sin\big(\omega_Lt+\phi\big)\sin\big(\boldsymbol{\kappa}\cdot{\bf q}\big)\,,
\ee
where the notation is adopted from Eq. (\ref{int2}).             
Following Eq. (\ref{int4}) and (\ref{int4.1})
we can write the corresponding relations for the standing wave 
in the semiclassical representation 
\be
\label{int28}
{\bf E}(t,\hat{{\bf R}}_j)=-iE_0\boldsymbol{\epsilon}
\left[e^{i(\omega_Lt+\phi)}-e^{-i(\omega_Lt+\phi)}\right]
\sin\left[\chi_j+\eta_j(\co+\ao)\right]\,,
\ee
and
\be
\label{int29}
\frac{\partial E_{b}(t,\hat{{\bf R}}_j)}{\partial q_{a}}=
-iE_0\kappa_{a}\epsilon_{b}
\left[e^{i(\omega_Lt+\phi)}-e^{-i(\omega_Lt+\phi)}\right]
\cos\left[\chi_j+\eta_j(\co+\ao)\right]\,,
\ee
where the new parameter $\chi_j=\kappa_{\vartheta}\bar{z}_j$ 
determines the position of the $j$th ion in the standing wave 
and $\kappa_{a}=(\omega_L/c)n_{a}$. The notation is adopted from 
Eq. (\ref{int2}) and (\ref{int5}).
The condition $\chi_j=0$ refers to the $j$th ion placed in the node, whereas 
$\chi_j=\pi/2$ refers to the ion positioned in the antinode of the
standing wave. 

Following the derivation for the travelling-wave configuration,
one can easily derive the Hamiltonian in
the interaction picture for the standing-wave configuration. It takes the
form of the expressions (\ref{int10}) and (\ref{int11}), except that 
the laser coupling constant $\lambda_j$ and 
the operator function ${\cal F}_k$ are
replaced with $\tilde{\lambda}_j$ and $\tilde{\cal F}_k$. They are given 
in the dipole approximation (assuming a dipole allowed transition) by
\be
\label{int32}
\tilde{\lambda}^{D\!P}_j&=&
-\frac{i2q_eE_{0}}{\hbar}
\bigg[({\bf r}_{eg})_j\cdot\boldsymbol{\epsilon}\bigg]\,e^{-i\phi}\,,\\
\label{int33}
\tilde{{\cal F}}_{k}^{D\!P}(\co\ao)&=&e^{-(\eta_j^2/2)}
\sin\bigg(\chi_j+\frac{\pi}{2}|k|\bigg)
\eta_j^{|k|}
\sum_{\a=0}^{\infty}
\left(-\eta_j^2\right)^{\a}
\frac{(\co\ao)^{\a}}{\a!(\a+|k|)!}
\ee
and in the quadrupole approximation 
(assuming a dipole forbidden transition) by
\be
\label{int34}
\tilde{\lambda}^{Q\!D}_j&=&
-\frac{iq_eE_0\omega_L}{\hbar c}
\bigg[\sum_{a,b}
\l e_j|(\hat{{\bf r}}_j)_{a}(\hat{{\bf r}}_j)_{b}|g_j\r n_{a}\epsilon_{b}
\bigg]e^{-i\phi}\,,\\
\label{int35}
\tilde{{\cal F}}_{k}^{Q\!D}(\co\ao)&=&e^{-(\eta_j^2/2)}
\cos\bigg(\chi_j+\frac{\pi}{2}|k|\bigg)
\eta_j^{|k|}
\sum_{\a=0}^{\infty}
\left(-\eta_j^2\right)^{\a}
\frac{(\co\ao)^{\a}}{\a!(\a+|k|)!}\,.
\ee
Comparing the expressions for the coupling constant in the travelling-wave
configuration [Eq. (\ref{int8}) and (\ref{int8.1})] with those ones for the
standing-wave configuration [Eq. (\ref{int32}) and (\ref{int34})], 
we find out that $|\tilde{\lambda}_j^{D\!P}|=2|{\lambda}_j^{D\!P}|$ and
$|\tilde{\lambda}_j^{Q\!D}|=2|{\lambda}_j^{Q\!D}|$. The factor 2 arises
from the expression of the standing wave (\ref{int27}) where we have 
superposed two travelling waves with equal amplitudes.
Finally, the Hamiltonian can be written in the form given by Eq. (\ref{int13})
and (\ref{int14}) with the coupling constant 
in the dipole approximation
\be
\label{int36}
(\tilde{\Omega}_j^{n,k})^{D\!P}=
\lambda_j^{D\!P}\,e^{-(\eta_j^2/2)}
\sin\bigg(\chi_j+\frac{\pi}{2}|k|\bigg)
\eta_j^{|k|}
\sqrt{\frac{n!}{(n+|k|)!}}\,L_n^{|k|}(\eta_j^2)  
\ee
and in the quadrupole approximation
\be
\label{int37}
(\tilde{\Omega}_j^{n,k})^{Q\!D}=
\lambda_j^{Q\!D}\,e^{-(\eta_j^2/2)}
\cos\bigg(\chi_j+\frac{\pi}{2}|k|\bigg)
\eta_j^{|k|}
\sqrt{\frac{n!}{(n+|k|)!}}\,L_n^{|k|}(\eta_j^2)\,.
\ee
It follows from Eq. (\ref{int36}) that for the $j$th ion 
in the dipole approximation 
placed in the node of the standing wave ($\chi_j=0$)
only transitions on odd sidebands ($|k|=2p+1$) are present. 
For the same ion in the antinode
($\chi_j=\pi/2$) only even sidebands ($|k|=2p$) are present, 
where $p$ is an integer or the zero. 
In the quadrupole approximation
the statements above are valid in the opposite order 
[compare Eq. (\ref{int36}) with (\ref{int37})]. 
The reason for missing transitions in the standing-wave
configuration comes from the destructive interference between the two
counter propagating travelling waves in the standing-wave field.

We could easily write the coupling constant $\tilde{\Omega}^{n,k}_j$ in the
Lamb-Dicke limit [see Eq. (\ref{int20})]. We could also write 
the unitary evolution operator for the standing-wave configuration. However, 
it differs from the evolution operator in the travelling-wave configuration 
[Eq. (\ref{int18}) and (\ref{int19})] 
only in the coupling constant and in the phase
factor, but it produces no fundamental problem for further applications. 
Therefore, in what follows we will
consider the expressions and formulas for the travelling-wave configuration
keeping in mind the way how to convert to a standing-wave configuration.


\section{Laser cooling}
\label{lc}


{\it Laser cooling} is the process in which the kinetic energy of atoms is
reduced through the action of one or more laser beams. The last decade brought 
rapid progress in this research field and this effort culminated in 1997
with the award of the Nobel Prize in physics for laser cooling and trapping
of atoms \cite{98-8,98-9,98-10}. A recent review of different experimental
techniques for laser cooling can be found in Ref. \cite{metcalf}.

One of the requirements for the practical implementation of quantum
computing is the ability to prepare well defined initial states 
of the qubits \cite{divin}. In our case the qubits are represented by
trapped ions with vibrational (external) and atomic (internal) degrees of
freedom. {\it Laser cooling} enables the preparation of well defined
initial states of motion and {\it electron shelving} serves for 
the proper initialization of the ion register. We will describe
this method later on in Sec.\,\ref{es}.
Laser cooling of trapped ions with the axial trapping frequency $\omega_z$ 
has two stages depending on the linewidth $\Gamma$ 
of the cooling transition \cite{nagerl,roos}:
\begin{itemize}

\item {\it Doppler cooling} is applied when the vibrational frequency of the
ions is smaller than the linewidth of a transition used for cooling
($\omega_z\leq\Gamma$). In other words, this means that the velocity 
of the ion due to the trapping potential changes on a longer time scale 
than the time it takes the ion to absorb or emit a photon (strong laser
driving is assumed). Therefore, we can assume that these processes 
change the momentum of the ion instantaneously. For $\omega_z\leq\Gamma$ we
refer to the weak confinement regime (in the sense of weak binding of 
the~ions to the ion trap).

\item {\it Sideband cooling} is used for further cooling below the Doppler
cooling limit and requires the vibrational frequency to be much bigger
than the linewidth ($\omega_z\gg\Gamma$). Under this condition the
ion develops well resolved sidebands and cooling to a lowest vibrational
state is realized through driving a lower sideband. For $\omega_z\gg\Gamma$
we refer to the strong confinement regime. One can use instead a novel
technique called {\it laser cooling using electromagnetically induced
transparency}.

\end{itemize}

\subsection{Doppler cooling}

This stage of laser cooling is based on the Doppler effect. 
The technique is
based on the fact that moving atoms absorb photons from a counter propagating 
red detuned laser beam (tuned slightly below the atomic
frequency) and emit spontaneously in a random direction. After several
such cooling cycles (absorption followed by spontaneous emission) we
can write for the total momentum {\bf p} of atoms
\be
\label{dopp1}
{\bf p}={\bf p}_0+
\sum_j\hbar{\bf k}_j^{(abs)}+\sum_j\hbar{\bf k}_j^{(em)}\,,
\ee
where ${\bf p}_0$ is the initial momentum of atoms, ${\bf k}_j^{(abs)}$ and
${\bf k}_j^{(em)}$ denotes the wavevectors of the absorbed and emitted
photons in the $j$th cooling cycle.
We usually use a fast decaying dipole transition for the Doppler
cooling, therefore the spontaneous emission is much faster than 
stimulated emission. The average total momentum of atoms after many cooling
cycles takes the form
\be
\label{dopp2}
\l {\bf p}\r={\bf p}_0+\hbar{\bf k}_L\l n\r\,,
\ee
where $\l n\r$ is the average number of absorption and emission events 
(typically $\l n\r\simeq 10^3-10^4$) and by the definition 
${\bf k}_L={\bf k}_j^{(abs)}$ (${\bf k}_L$ is the wavevector associated with 
the laser light). The spontaneous contribution averages to zero because it
is randomly distributed over the solid angle $4\pi$. If the laser is red
detuned and counter propagating to the motion of atoms 
(${\bf p}_0\uparrow\downarrow{\bf k}_L$), then the velocity of the atoms is
significantly decreased [see Eq. (\ref{dopp2})]. For more details see
ref. \cite{LesH, metcalf}. 

The discussion above is valid for free atoms but it also applies for trapped
ions \cite{86-1}, where the motion towards the laser is provided by the
periodic vibrations. The Doppler cooling limit corresponds to the final
temperature $T_{dopp}=\hbar\Gamma/2k$, where $\Gamma$ is 
the linewidth of the cooling transition and $k$ is the Boltzmann constant. This
temperature is typically of the order of mK \cite{zeil}.
However, the Doppler cooling limit can be also translated into 
the minimum average phonon number in the axial
direction \cite{LesH}
\be
\label{dopp3}
\l n_z\r_{min}=\frac{\Gamma}{\omega_z}
\left(\frac{1+\a}{4}\right)
\left(\frac{\Gamma}{\delta}+\frac{\delta}{\Gamma}\right)-\frac{1}{2}\,,
\ee
where $\Gamma$ is the natural linewidth of the cooling transition,
$\a$ is determined from the angular distribution of the emitted radiation
and $\delta$ is the laser detuning from the atomic frequency.
For a dipole radiation pattern we get $\a=2/5$. 
The cooling is optimal for 
the detuning $\delta=\Gamma\gg\omega_z$. Concerning this condition 
we can rewrite
Eq. (\ref{dopp3}) for a dipole transition to the form
\be
\label{dopp4}
\l n_z\r_{min}\simeq\frac{7}{10}\frac{\Gamma}{\omega_z}\,.
\ee
We have omitted the factor 1/2 corresponding to
the zero point energy because it has a negligible contribution. 
The~Doppler cooling limit is associated with the recoil of the atoms
at the spontaneous emission. 

\begin{figure}[htb]
\centerline{\epsfig{width=8cm,file=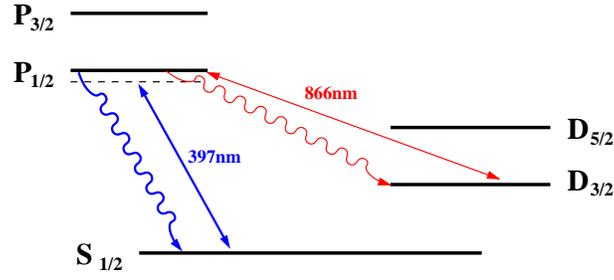}}
\caption{{\footnotesize
Doppler cooling. The $S_{1/2}\leftrightarrow P_{1/2}$ is
the~cooling transition. The~cooling laser on 397\,nm is red detuned by
10\,MHz. The ion spontaneously decays from the $P_{1/2}$
level. There is a $6\%$ probability of the decay to the metastable 
state $D_{3/2}$, therefore the pumping laser on 866\,nm is switched on.}} 
\label{dopp}
\end{figure}

The relation (\ref{dopp4}) 
leads us to a discussion of how to choose the axial trapping frequency
$\omega_z$. In order to be able to address individually each ion with 
a single laser beam, the minimum spacing between the ions 
[see Eq. (\ref{vibr9})] has to be large enough, which requires small
$\omega_z$. On the other hand, 
we do not want the frequency $\omega_z$ to be too small
in order to make the result of Doppler cooling as efficient as possible 
[Eq. (\ref{dopp4})]. Thus, the design of ion traps is also determined
by the trade-off between these two options.

For Calcium ions $^{40}\mbox{Ca}^+$ the $S_{1/2}\leftrightarrow P_{1/2}$
transition with the natural linewidth 
$\Gamma=20\,\mbox{MHz}$ is used for the Doppler cooling. 
The lifetime of the $P_{1/2}$ level is about 7\,ns and 
this level decays with $6\%$ probability to the metastable $D_{3/2}$
level (FIG.\,\ref{dopp}).
Therefore, optical pumping between the $P_{1/2}$ and $D_{3/2}$
levels is present on 866\,nm. The laser on the cooling transition 
$S_{1/2}\leftrightarrow P_{1/2}$ is red detuned by 
$\Gamma/2\simeq 10\mbox{MHz}$.
On the other hand, the pumping laser at 866nm is kept on the resonance in
order to prevent population trapping in the superposition of the
$S_{1/2}$ and $D_{3/2}$ levels \cite{roos}. For $\a=2.5$,
$\omega_z/2\pi\simeq 700\,\mbox{kHz}$ and $\delta=\Gamma/2$ we can
calculate from Eq. (\ref{dopp3}) the minimum average phonon number to be 
$\l n_z\r_{min}\simeq 3.5$. This number differs from 
experimentally measured values (which are bigger) 
because Eq. (\ref{dopp3}) has been
derived for a two-level system while the experimental realization of Doppler
cooling involves a three-level system. Nevertheless,
$\l n_z\r_{min}\simeq 3.5$ is still not $\l n_z\r_{min}=0$
required for proper operation of the quantum processor with cold trapped
ions. Therefore a second cooling stage must be launched.

\subsection{Sideband cooling}
\label{sideband}

Doppler cooling represents the precooling stage 
in experiments with trapped
ions. The final stage can be realized by the sideband cooling technique which
may prepare the ions to the ground motional state, i.e. a well defined
initial quantum state. Firstly, we address the basic idea of sideband
cooling. Then we illustrate this cooling technique on two trapped ions in
and outside the Lamb-Dicke regime.
Finally, we describe how sideband cooling is realized experimentally.

In the strong confinement 
regime ($\omega_z\gg\Gamma$) a single trapped ion exhibits in
its absorption spectrum well resolved sidebands at 
$\omega_0\pm k\omega_z$ ($k$ is an integer) spaced on both sides of
the carrier on the atomic frequency $\omega_0$. Sideband cooling occurs
when the cooling laser is tuned to a lower sideband at
$\omega_L=\omega_0-k\omega_z$. In the Lamb-Dicke limit cooling works
efficiently with the laser tuned on the first red sideband at
$\omega_L=\omega_0-\omega_z$. Then the ion absorbs photons of the energy
$\hbar(\omega_0-\omega_z)$ and spontaneously emitted photons of the average
energy $\hbar\omega_0-E_r$ bring the ion back to its initial internal state
(see Appendix \ref{ldl}). In every cooling cycle (absorption + emission) the
motional energy of the ion is damped by one vibrational quantum if
$\hbar\omega_z\gg E_r$. This condition implies that in this form 
the sideband cooling requires the ion to be in the Lamb-Dicke limit. The whole  
process consists of cooling cycles in which the absorption is followed by
the spontaneous emission until the ion reaches the ground motional state
$|n=0\r$ and decouples from the cooling laser. 
For the sideband cooling of the single trapped ion initially 
in the internal state $|g\r$ and in the
motional state $|n\r$, where $a$ denotes the absorption and $e$ stays for
the spontaneous emission, we may schematically write
\be
|g\r|n\r\ \stackrel{a}{\rightarrow}\ |e\r|n-1\r\ \stackrel{e}{\rightarrow}
|g\r|n-1\r\ \stackrel{a}{\rightarrow}\ \dots\ \stackrel{e}{\rightarrow}
|g\r|1\r\ \stackrel{a}{\rightarrow}|e\r|0\r\ \stackrel{e}{\rightarrow}
|g\r|0\r\,.\nonumber
\ee
The minimum average phonon number in the axial direction that can be
reached by sideband cooling is then given by \cite{LesH}
\be
\label{side1}
\l n_z\r_{min}=\left(\frac{\Gamma}{\omega_z}\right)^2
\left(\a+\frac{1}{2}\right)\,,
\ee
where the parameter $\a$ has been defined in Eq. (\ref{dopp1}). 
It is evident that now one can 
achieve efficient cooling to the ground motional state, i.e.
$\l n_z\r_{min}\simeq 0$, assuming the strong confinement regime 
($\omega_z\gg\Gamma$). The limit for the sideband cooling 
[Eq. (\ref{side1})]
is constrained by the recoil of the ion and is determined by the equilibrium 
between cooling and heating processes. Heating is caused mainly by 
off-resonant excitations on the carrier ($|g\r|n\r\leftrightarrow|e\r|n\r$) and
on the first blue sideband ($|g\r|n\r\leftrightarrow|e\r|n+1\r$).
The sideband cooling of a single ion beyond the Lamb-Dicke limit also exists and is
based on the creation of a dark state in the energy level structure
\cite{morigi0}.
A single trapped Mercury $^{198}\mbox{Hg}^+$
ion was firstly cooled to the ground motional state in 1989 in Boulder
\cite{89}, while sideband cooling of a single Beryllium ion $^9\mbox{Be}^+$
in all three dimensions was firstly reported in 1995 also 
by the group in Boulder \cite{95-11}. 

\subsubsection{Sideband cooling of two ions}

The key difference between one and more ions lies in the energy spectrum
\cite{oxf, 98-7}.
While single ions have discrete energy levels (three motional degrees of
freedom), a chain of $N$ oscillating ions ($3N$ motional degrees of freedom)
exhibits a quasicontinuous energy spectrum due to the incommensurate
frequencies of the motional modes.
For instance, in the axial direction we have the frequencies
$\omega_z,\,\omega_z\sqrt{3},\,\omega_z\sqrt{5.8},\,\omega_z\sqrt{9.3}$,
etc. 

\begin{figure}[htb]
\centerline{\epsfig{width=10cm,file=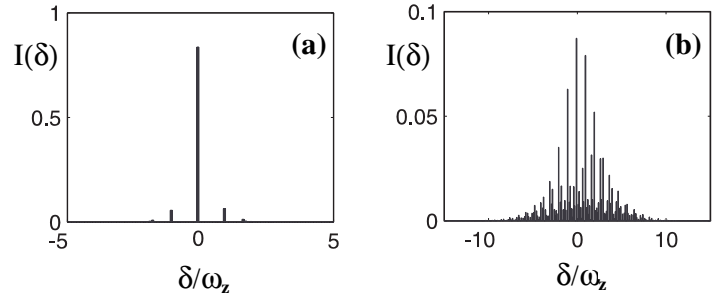}}
\caption{{\footnotesize
Absorption spectrum $I(\delta)$ of two ions for a thermal distribution 
$P({\bf n})$ in the Lamb-Dicke regime (a) and outside
the Lamb-Dicke regime (b) as a function of detuning $\delta$ (in units of the
axial vibrational frequency $\omega_z$) where $I(\delta)$ is defined 
in Eq.~(\ref{abs}) and $\delta$ 
in Eq.~(\ref{int9.0}). 
Used by kind permission of Giovanna Morigi and
J\"{u}rgen Eschner \cite{morigi}.}}
\label{LDL}
\end{figure}

We will discuss in detail the case of two trapped ions to illustrate 
the situation of sideband cooling of more than a single ion \cite{morigi}.
The state of each ion ($j=1,2$) will be expressed in the basis 
$\{|g_j\r|{\bf n}\r, |e_j\r|{\bf n}\r\}$, where 
${\bf n}=(n_1,n_2)$, $n_1$ is the vibrational number associated with the COM
mode ($\nu_1=\omega_z$) and $n_2$ with the breathing mode 
($\nu_2=\omega_z\sqrt{3}$). The absorption spectrum of the $j$th ion 
will be considered in the form \cite{comm1} 
\be
\label{abs}
I_j(\delta)=\sum_{E_{\bf n}-E_{\bf m}=\delta}
|\l {\bf n}|\exp{(i\Delta_j\kappa_{\vartheta})|{\bf m}\r|^2P({\bf n})}\,,
\ee
where $E_{\bf n}=\hbar\omega_zn_1+\hbar\omega_z\sqrt{3}n_2$, $\delta$ is 
the detuning [Eq. (\ref{int9.0})], $\kappa_{\vartheta}$ is defined 
by Eq. (\ref{int5})
and $\Delta_j$ is the displacement operator of the $j$th ion 
[Eq. (\ref{vibr29})].
$P({\bf n})=P(n_1,n_2)$ is a probability distribution associated with 
the vibrational motion of the ions.
The Lamb-Dicke parameter distinguishes between two very different 
regimes of sideband cooling of more ions: 
\begin{itemize}

\item
In the Lamb-Dicke regime (Appendix \ref{ldl}) and 
in the strong confinement regime ($\omega_z\gg\Gamma$)
only the first sidebands of the motional modes 
at $\omega_0\pm\omega_z$ and $\omega_0\pm\omega_z\sqrt{3}$ appear around the
significant carrier peak at $\omega_0$ in the absorption spectrum 
[FIG.\,\ref{LDL}(a)].
The higher sidebands are suppressed due to their strength being proportional to
higher powers in the Lamb-Dicke parameter than $\eta$ denoted as 
${\cal O}(\eta^2)$. Tuning the laser on the first red sideband of the COM
mode ($\delta=-\omega_z$) we can reach its ground state $|n_1=0\r$ 
at the same cooling rate as for a single ion \cite{morigi}. However, 
in the case of two ions the breathing mode is decoupled from the COM mode
and its cooling is almost frozen. 
Simultaneous cooling of more modes requires the modes to be coupled to
the cooling laser and then the requirement of the
strong coupling regime ($\omega_z\gg\Gamma$) has to be reconsidered or one
has to use alternative techniques.

\item
Outside the Lamb-Dicke regime higher sidebands 
with the strength proportional to ${\cal O}(\eta^2)$ also
contribute and the absorption
spectrum exhibits the structure with many overlapping sidebands 
at $\omega_0\pm k\omega_z\pm l\omega_z\sqrt{3}$ where $k, l$
are integers [FIG.\,\ref{LDL}(b)].
In this situation the laser tuned on a lower sideband 
(it does not have to be strictly the first red sideband of the COM mode) 
excites simultaneously all sideband transitions around this lower sideband
in the interval of the linewidth $\Gamma$. Then the COM and the breathing mode
are coupled and cooled at once. 
However, the cooling process is much slower
in comparison to cooling of a single ion beyond the Lamb-Dicke limit. It is
partly caused by (i) the increasement of the number of the motional modes 
but also by (ii) the appearance of dark states \cite{morigi}. The dark
states are almost decoupled from a resonantly excited state because their
motional wave function after the absorption overlaps with the motional
wavefunction of the excited state only a little. Thus, the ions may be
trapped in these dark states and it slows the cooling process down. This
problem can be solved by escaping from the strong confinement regime
($\omega_z\gg\Gamma$), i.e. by increasing the linewidth $\Gamma$. 
It will cause that a single level will be coupled to more levels (more
sidebands are in the resonance) and the ions will be cooled more efficiently
due to more cooling channels. As a result the dark states will
disappear because more channels provide more ways for the ion to escape from 
dark (population trapping) states. Moreover, the rate of the cooling
cycles (absorption + emission) is proportional to the linewidth $\Gamma$.
Summarizing both effects we can conclude that the total cooling time beyond
the Lamb-Dicke limit can be shortened significantly for $\Gamma\simeq\omega_z$. 

\end{itemize}

\subsubsection{Experimental sideband cooling}

Two ions were cooled for the first time to the ground motional state in 1998
in Boulder.
It was achieved on Beryllium ions $^9\mbox{Be}^+$ illuminating both
ions at once \cite{king}. However, it is sufficient to illuminate only one
ion from the entire ion string because other ions are cooled sympathetically
due to the strong Coulomb coupling. Although we need only one motional mode
(COM or breathing axial mode) as the quantum data bus, which has to be in
the ground motional state, we require also other
modes to be cooled close to the ground state. Uncooled motional modes with
thermal phonon distributions significantly affect the Rabi frequency in the
data mode and spoil the fidelity of the coherent state manipulation
[see Eq. (\ref{eit2})].

\begin{figure}[htb]
\centerline{\epsfig{width=6.5cm,file=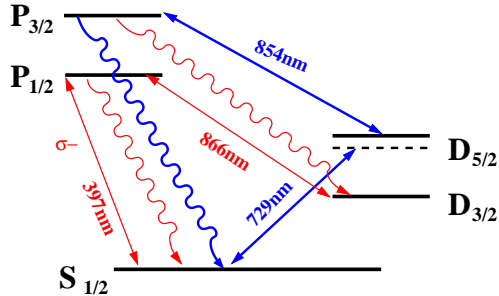}}
\caption{{\footnotesize
Sideband cooling with cooling and pumping transitions.
The blue transitions correspond to the cooling cycle.
The~laser at 729\,nm is tuned on the first red sideband. The laser at 854\,nm
couples the metastable $D_{5/2}$ level with the fast decaying $P_{3/2}$
level. The cooling cycle is closed by the spontaneous emission  
on the~fast $S_{1/2}\leftrightarrow P_{3/2}$ dipole transition. 
The $P_{3/2}$ level may
decay to the state $D_{3/2}$. The~ion is recycled to the state
$P_{1/2}$ by the laser at 866\,nm followed by the spontaneous emission back
to the state $S_{1/2}$. Decay to the sublevels $S_{1/2}$($m$=+1/2) is
counteracted by driving the~$S_{1/2}\leftrightarrow P_{1/2}$ transition with
the $\sigma^-$ polarized laser at 397\,nm.}}
\label{side}
\end{figure}

The group in Innsbruck has realized different approaches in sideband cooling 
of two Calcium ions \cite{blatt1, blatt2}. If they cool only one motional
mode, while the other modes are left in the thermal states, they achieve the
ground state population greater than 95\% ($\l n\r\simeq 0.05$) in the
respective mode. However, they can cool sequentially all motional modes 
close to the ground state. For this purpose they use a small modification 
in the sideband cooling scheme. The laser frequency and laser power has 
to be set
sequentially for the respective first red sideband of the given motional
mode. After sequential cooling all modes the corresponding average
phonon numbers are from $\l n\r\simeq 0.05$ to $\l n\r\simeq 2.3$ because 
the recoil energy from the spontaneous emission in the cooling process of
one motional mode reheat other modes. 

For Calcium ions $^{40}\mbox{Ca}^+$ the quadrupole transition between 
the two Zeeman sublevels $S_{1/2} (m_J=-1/2)$ and $D_{5/2} (m_J=-5/2)$ is used
for the sideband cooling (FIG.\,\ref{side}).
The laser on 729\,nm is tuned on the first red sideband of the respective
motional mode and a weak magnetic field is applied for Zeeman splitting
of energy levels. The lifetime of the metastable $D_{5/2}$ 
level is about a second,
therefore there is the pumping on 854\,nm to the fast decaying 
$P_{3/2} (m_J=-3/2)$ level in order to decrease the duration of one cooling
cycle, i.e. to increase the cooling rate. The $P_{3/2} (m_J=-3/2)$ level
decays spontaneously to the initial state $S_{1/2} (m_J=-1/2)$ and closes
the cooling cycle. However, the $P_{3/2}$ level may decay with a small
probability to the $D_{3/2}$ level, therefore the pumping laser on 866\,nm
recycles the population to the $P_{1/2}$ state which decays to the $S_{1/2}$
state. The $S_{1/2}\leftrightarrow P_{1/2}$ transition is driven with 
the $\sigma^-$ polarized laser on 397\,nm to counteract the population of 
the $S_{1/2} (m_J=+1/2)$ level \cite{roos}.

\subsection{Sympathetic cooling}
\label{sympcool}

In the previous section we have mentioned that it is sufficient to
illuminate with cooling lasers only one ion from the ion string because
the other ions are cooled sympathetically due to the Coulomb interaction
between them. Hence the term {\it sympathetic cooling}. However, instead of
identical ions one can consider different atomic species (eventually
isotopes) in the ion crystal \cite{symp2}. 
Then the addressing of {\it cooling} ions avoids the disturbance 
of internal states of {\it logic} ions which store the information
\cite{dfs}.

\begin{figure}[h!]
\centerline{\epsfig{width=10.5cm,file=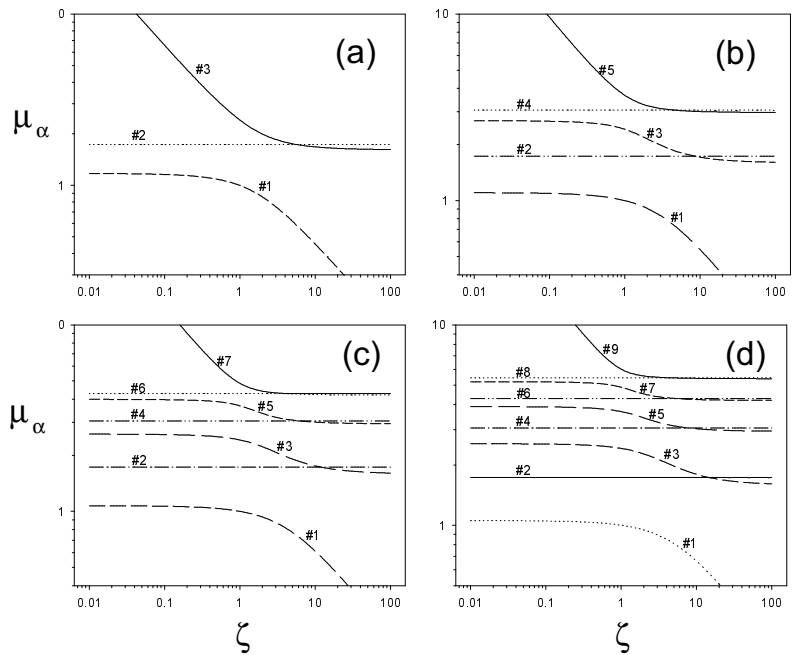}}
\caption{{\footnotesize
Normalized axial frequencies (\ref{vibr21}) as a function of $\zeta$ 
for (a) three, (b) five, (c) seven and (d) nine ions. 
Used by kind permission of David Wineland and David Kielpinski \cite{symp1}.}}
\label{symp1}
\end{figure}

Electric fields from the trap electrodes are one of the sources of 
the motional decoherence of the ion crystal due to heating of collective
vibrational motional modes (normal modes). 
If one assumes the dimension of the ion trap to be much larger
than the dimension of the ion crystal, then we can expect the electrode
electric fields to be nearly uniform across the ion crystal. Such uniform
fields influence and heat only collective motional modes  
involving the centre-of-mass (COM) motion of the ion crystal.
Uniform electric fields can directly heat up the normal mode used for quantum
logic as the quantum data bus. We can overcome this constrain by selecting 
a specific normal mode for quantum logic which is decoupled from heating.
However, not all motional modes are prevented from heating. In what follows
we will discuss this point following Ref. \cite{symp1}.

Let us consider
the ion crystal with an odd number $N$ of ions which consists of $N-1$ ions 
of mass $m$ and of a central ion of the mass $M$ defining the ratio
$\zeta=M/m$. 
Now we can follow the lines in Sec.\,\ref{vmoti} 
and find the normal modes and frequencies of the ion string with unequal ions. 
We find out that (i) there are $(N-1)/2$ axial normal modes 
for which the central ion 
does not move and corresponding eigenvectors ${\bf D}^{(\a)}$ and eigenfrequencies
$\nu_{\a}=\omega_z\sqrt{\mu_{\a}}$ do not depend on the parameter
$\zeta$. Moreover, these modes do not have a component associated with the
axial COM motion. 
(ii) There are also
other $(N+1)/2$ modes having a component of the COM motion and coupling
to any uniform electric field which causes their heating.

\begin{figure}[h!]
\centerline{\epsfig{width=10.5cm,file=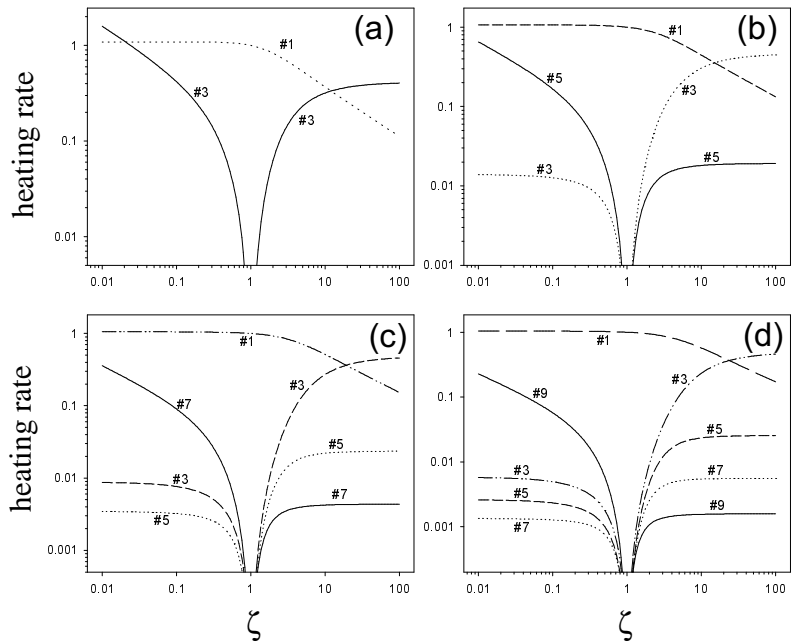}}
\caption{{\footnotesize
Normalized heating rates for hot axial modes as a function of $\zeta$ 
for (a) three, (b) five, (c) seven and (d) nine ions. 
Used by kind permission of David Wineland and David Kielpinski \cite{symp1}.}}
\label{symp2}
\end{figure}

For very small or very large values of $\zeta$ the motional modes 
become degenerate and pair up (FIG.\,\ref{symp1}).
From this point of view the value $\zeta\simeq 1$ seems to be
suitable. It has also been calculated that those modes having the central
ion at rest (neglecting gradient electric fields) do not heat at all
(we refer to these as {\it cold modes}), while all other motional 
modes heat to some extent depending on the value of $\zeta$. 
Their heating rate (average number of phonons gained per second) drops 
rapidly for $\zeta\to 1$ (FIG.\,\ref{symp2}). 
We refer to these modes as {\it hot modes}.

Thus, it seems that the optimal choice 
is $\zeta\simeq 1$. That means the central (cooling) ion should be chosen 
such that it is identical to $N-1$ other (logic) ions or is an isotope 
of logic ions. For $\zeta\simeq 1$ we can choose the lowest cold mode 
(the second lowest motional mode called breathing mode) 
to be used for quantum logic as the quantum data bus because in the case of
$\zeta\simeq 1$ only the lowest motional mode
(corresponding to the COM mode for equal ions) will heat significantly and
can be cooled via the central cooling ion. If the value of $\zeta$ differs 
very much from 1, we have to cool all $(N+1)/2$ hot modes via the central ion.

The group in Garching runs experiments where the ion string
consists of Indium ($^{115}\mbox{In}^+$) and Magnesium
($^{25}\mbox{Mg}^+$) ions. The numerical analysis for the ion
crystal containing these two atomic species ordered in different 
configuration can be found in Ref. \cite{symp2}. 
The mass ratio is $\zeta=4.6$ which is not within the optimal range
discussed above ($\zeta\simeq 1$).
On the other hand, it can quite advantageous because the heavy
ion fulfills the Lamb-Dicke limit (Appendix \ref{ldl}) easier 
than the light ion. Distinct atomic species can also have 
a very different atomic spectrum which may be found convenient when
laser addressing closely spaced ions. However,
for the heavy central cooling ion we pay the price in the form of
heating rates of higher motional modes.
By all means, Indium ions can be efficiently cooled to
the ground motional state \cite{In} and Magnesium ions can serve for quantum
logic operations and storing the information.

Finally, we have to mention that the demonstration of sympathetic
cooling using two different atomic species is very demanding on current
experimental technology due to problems of
loading the ion trap with distinct atoms in a~desired configuration.

\subsection{Laser cooling using electromagnetically induced transparency}

Quantum computing with cold trapped ions requires one of the motional modes
(the one used as the quantum data bus) to be cooled to the motional ground
state and other modes to be inside the Lamb-Dicke regime. For this purpose
one could eventually use Doppler cooling assuming the axial trapping
frequency $\omega_z$ comparable with the linewidth of the cooling transition
$\Gamma$ [Eq.~(\ref{dopp4})]. However, it would cause very close spacing of
the ions in the trap [Eq.~(\ref{vibr9})] with difficulties at individual
addressing with the laser beam and optical resolving. On the other hand, we
can use sequential sideband cooling of the motional modes described in 
Sec.\,\ref{sideband}. However, cooling of one motional mode causes heating of
the other modes. Moreover, sideband cooling requires a very narrow bandwidth
to excite the first red sideband of the respective motional mode only. 
Otherwise, off-resonant transitions (especially carriers) are also 
driven what causes heating as well \cite{blatt1}.

\begin{figure}[htb]
\centerline{\epsfig{width=8cm,file=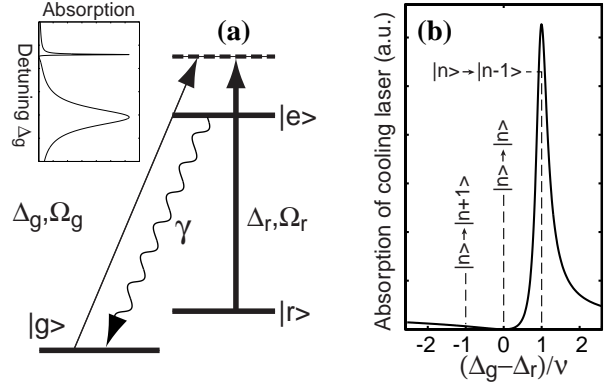}}
\caption{{\footnotesize
(a) Levels and transitions of the cooling technique using electromagnetically 
induced transparency (EIT). The inset shows the absorption on 
$|g\r\leftrightarrow|e\r$ while strongly driving the transition 
$|r\r\leftrightarrow|e\r$. (b) Absorption on the carrier
($|n\r\rightarrow|n\r$) and on the first sidebands 
($|n\r\rightarrow|n\pm1\r$). Used by kind permission of Giovanna Morigi and 
J\"{u}rgen Eschner \cite{EIT1}.}}
\label{EIT}
\end{figure}

A novel cooling technique was developed in 2000 with a lower cooling limit
than Doppler cooling and with a wider cooling bandwidth than sideband
cooling. It was named {\it laser cooling using electromagnetically induced
transparency (EIT)} \cite{EIT1, EIT2}. It is based on a quantum interference
effect called EIT or {\it coherent population trapping} or also {\it dark
resonance} \cite{EIT3}. It employes a three-level system with a ground state
$|g\r$, a stable or metastable state $|r\r$ and an excited state $|e\r$ 
[FIG.\,\ref{EIT}(a)]. The transition $|r\r\leftrightarrow|e\r$ is driven
with a laser beam of the intensity $I_r$ blue detuned by $\Delta_r$. The
transition $|g\r\leftrightarrow|e\r$ is coupled by a weak laser with the
intensity $I_g$ (where $I_r/I_g\simeq 100$) also blue detuned by $\Delta_g$.
The intense laser with $|\Omega_r|^2\propto I_r$ introduces a significant 
Stark light shift $\Delta\omega$ where \cite{comm2}
\be
\label{eit1}
\Delta\omega=(\sqrt{\Delta_r^2+|\Omega_r|^2}-|\Delta_r|)/2\,.
\ee
Thus, the laser on the transition $|r\r\leftrightarrow|e\r$ designs the
absorption spectrum seen by the weak laser on $|r\r\leftrightarrow|e\r$ via
the level $|e\r$. Then there is a broad resonance at $\Delta_g\simeq 0$, 
a dark resonance (EIT) at $\Delta_g=\Delta_r$ and a bright narrow resonance
at $\Delta_g=\Delta_r+\Delta\omega$ [see the inset in FIG.\,\ref{EIT}(a)].
Therefore, (i) taking into account also the motional degrees of freedom,
(ii) setting the detunings such that $\Delta_g=\Delta_r$ and (iii) setting 
the Stark light shift equal to the vibrational frequency 
($\Delta\omega\simeq\nu$) we obtain the absorption spectrum depicted in 
FIG.\,\ref{EIT}(b). We see that the absorption on the first red sideband (cooling
transition) is enhanced while the absorption on the carrier (heating
transition) is eliminated.

The bright resonance width can be wide enough to cover several motional
modes which can be consecutively cooled at once. It was experimentally
demonstrated in Innsbruck on two motional modes separated in the frequency
by 1.73\,MHz. The modes were cooled to their ground motional states with 74\%
($\l n\r\simeq 0.35$) and 58\% ($\l n\r\simeq 0.72$) occupation \cite{EIT2}.
A great improvement of these results should possible for a rebuilt apparatus
allowing optimal access of laser beams to the ions \cite{danny}.

Following the advantages of laser cooling using EIT it has been estimated
that all $3N$ motional modes can be cooled to a mean phonon number 
$\l n\r<1$ for $\omega_z/2\pi=700\,\mbox{kHz}$ and $N=10$
\cite{EIT3}. It is very important to cool $3N-1$ spectator motional modes to
the Lamb-Dicke regime. Otherwise, thermally excited spectator modes cause
the fractional fluctuations (blurring) in the Rabi frequency of the mode
used as the quantum data bus for quantum logic operations \cite{98-5}. These
fluctuations in the Rabi frequency of the $\alpha$th mode can be estimated as
\cite{comm3}
\be
\label{eit2}
\left(\frac{\Delta\Omega}{\Omega}\right)_{\alpha}\approx
\sqrt{\sum_{\beta\neq\alpha}
\eta_{\beta}^4\,\l n_{\beta}\r\,(\,\l n_{\beta}\r+1)}\,,
\ee
where $\eta_{\beta}$ is the Lamb-Dicke parameter of the $\beta$th motional
mode and $\l n_{\beta}\r$ is the respective average phonon number. The ratio
$(\Omega/\Delta\Omega)$ determines the maximal number of Rabi cycles
\cite{blatt2}. A detailed description of experimental laser cooling using 
EIT on Calcium ions can be found in Ref. \cite{EIT3}.


\section{Electron shelving}
\label{es}

\begin{figure}[htb]
\centerline{\epsfig{width=8cm,file=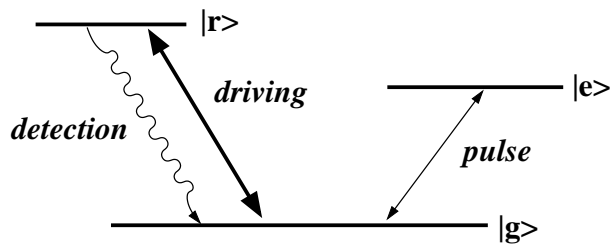}}
\caption{{\footnotesize 
Electron shelving. The $|g\r\leftrightarrow |e\r$
transition is coupled with a laser pulse forming the superposition 
$\a|g\r+\b|e\r$. The $|g\r\leftrightarrow |r\r$ transition is driven with a
strong laser. The fluorescence detection signal is collected
on the $|g\r\leftrightarrow |r\r$ transition if the ion collapses 
to the state $|g\r$. If the ion is shelved in the dark state $|e\r$, no
fluorescence is observed because the $|e\r$ state is a metastable state.}}
\label{shel}
\end{figure}

{\it Electron shelving} is the experimental method for the discrimination
between two electronic levels with an efficiency approaching $100\%$. It was
firstly demonstrated in 1986 \cite{86-2}. 
Let us assume a three-level atom consisting of a ground level $|g\r$,
a metastable excited state $|e\r$ and an auxiliary excited fast
decaying state $|r\r$ (FIG.\,\ref{shel}).
The $|g\r\leftrightarrow|e\r$ transition is coupled by a weak laser
forming a superposition $\a|g\r+\beta|e\r$, while 
the $|g\r\leftrightarrow|r\r$ transition is driven with a strong 
laser. If the atom collapses to the $|g\r$ state during the measurement, 
a strong fluorescence signal is collected on the fast transition 
$|g\r\leftrightarrow|r\r$, i.e. the atom is excited from $|g\r$ to $|r\r$
and spontaneously decays back to the $|g\r$ state what is observed as the
fluorescence. However, if the atom stays shelved in the metastable excited
state $|e\r$, no fluorescence can be observed on the driven
$|g\r\leftrightarrow|r\r$ transition. Hence the name electron shelving.
Even though the detection efficiency is
low, we can keep exciting the measuring transition $|g\r\leftrightarrow|r\r$
and detect some spontaneously emitted photons. Thus, we are able
to discriminate the $|g\r$ and $|e\r$ states with almost $100\%$ efficiency.
We can also obtain the occupation probability $|\a|^2$ for 
the $|g\r$ state and $|\beta|^2$ for the $|e\r$ state averaging 
over many repetitions of the same experiment \cite{00-3}.

In the case of Calcium ions, the ground state $|g=S_{1/2}\r$ and
and the metastable excited state $|e=D_{5/2}\r$ form the qubit \cite{roos}. 
The auxiliary state corresponds to the $P_{1/2}$ level. 
The $S_{1/2}\leftrightarrow D_{5/2}$ transition is illuminated with a weak laser
pulse on 729\,nm and the $S_{1/2}\leftrightarrow P_{1/2}$ transition 
is driven with a strong laser at 397\,nm. However, the ion can decay from the
$P_{1/2}$ level with a small probability to the $D_{3/2}$ level. Therefore,
there is pumping on the $P_{1/2}\leftrightarrow D_{3/2}$ transition at
866\,nm (see FIG.\,\ref{dopp}). 

To clarify the efficiency of the electron shelving method 
we report briefly some results measured in
Innsbruck \cite{00-3}. When the ion is found in the $S_{1/2}$
state it scatters about 2000 photons in 100\,ms to the detector. 
However, for the ion 
in the dark $D_{5/2}$ state the number of events drops to only about 150 
photons in 100\,ms. These 150 photons appear due to dark counts 
of the photomultiplier and some scattered light from the laser at 397\,nm.
The ion string in the linear ion trap may represent a quantum register,
where the internal state of each ion, i.e. the state of the qubit, can be
detected using a CCD camera. Then the ion in the $|g=S_{1/2}\r$ state 
appears as a bright spot or a dark spot if the ion is found in the
$|e=D_{5/2}\r$ state \cite{nagerl, roos, 00-3}.


\section{Quantum gates}
\label{qg}


One of the requirements for the physical implementation of quantum computing
in
a certain quantum system is a set of quantum gates that can be realized in 
the quantum system under consideration. It has been shown that any unitary
operation can be composed of single-qubit rotations and two-qubit
controlled-NOT gates \cite{95-8}. In what follows we will describe how 
these and some more complex quantum gates can be implemented on cold trapped
ions. We will use the notation $|g\r$ and $|e\r$ of the
logical states for the qubit rather than $|0\r$ and $|1\r$ due 
to the representation of the qubit by the internal states of the ion.

\subsection{Single-qubit rotations}

\begin{figure}[htb]
\centerline{\epsfig{width=2.5cm,file=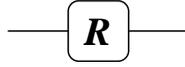}}
\caption{{\footnotesize
Schematical representation
of a single-qubit rotation. $R$ is defined by
Eq. (\ref{qg2}) in the basis $\{|g\r,|e\r\}$.}}
\label{rot}
\end{figure}

A general {\it single-qubit gate} corresponds to a unitary evolution 
operator that acts on a single qubit and is represented 
in the basis $\{|g\r,|e\r\}$ by the matrix
\be
\label{qg1}
W=\left(
\begin{array}{ll}
W_{gg} & W_{ge}\\
W_{eg} & W_{ee}
\end{array}
\right)
\ee 
A special case of the single-qubit gates 
is a {\it single-qubit rotation} (FIG.\,\ref{rot}).
Its parameterization depends on the choice of the coordinates on the Bloch
sphere. We will define it in the matrix form in the basis 
$\{|g\r,|e\r\}$ as follows
\be
\label{qg2}
\R(\theta,\phi)=
\left(
\begin{array}{cc}
\R_{gg} & \R_{ge}\\
\R_{eg} & \R_{ee}
\end{array}
\right)=
\left(
\begin{array}{cc}
\cos(\theta/2) & e^{i\phi}\sin(\theta/2)\\
-e^{-i\phi}\sin(\theta/2) & \cos(\theta/2)
\end{array}
\right)\,,
\ee
where $\theta$ refers to the rotation and $\phi$ to the relative phase shift
of the states $|g\r$ and $|e\r$ in the corresponding Hilbert space.

The single-qubit rotation can be performed on a selected ion 
from the ion string in the Lamb-Dicke regime by applying the unitary evolution
operator (\ref{int21}). We may rewrite this operator to the form
\be
\label{qg3}
\hat{{\cal A}}_j^{\ell}(\phi_j)&=&
\sum_{n=0}^{\infty}\cos (\ell\pi/2)
\Bigg[
\bigg(|e_j\r\l e_j|\otimes |n\r\l n|\bigg)+
\bigg(|g_j\r\l g_j|\otimes |n\r\l n|\bigg)
\Bigg]\\
&+&\sum_{n=0}^{\infty}\sin (\ell\pi/2)
\Bigg[
-\bigg(|e_j\r\l g_j|\otimes |n\r\l n|\bigg)e^{-i\phi_j}+
\bigg(|g_j\r\l e_j|\otimes |n\r\l n|\bigg)e^{i\phi_j}
\Bigg]\,,
\nonumber 
\ee
where we have applied the arbitrary choice of the phase factor 
($\phi_j\rightarrow\phi_j+\pi/2$) with respect to the remark below 
Eq. (\ref{int19}). The operator (\ref{qg3}) corresponds to the
$j$th ion illuminated with the laser beam 
on the carrier ($\omega_L=\omega_0$) with the laser pulse duration
$t=\ell\pi/|\lambda_j|$, where the laser coupling constant $\lambda_j$ 
depends on the type of (i) the driven transition and (ii) the laser
configuration (see Sec.\,\ref{lii}). We will refer to the operation expressed
by Eq. (\ref{qg3}) as the $\ell\pi$-pulse on the carrier.

\subsection{Two-qubit controlled-NOT gates}
\label{CNOT}

\begin{figure}[htb]
\centerline{\epsfig{width=2.5cm,file=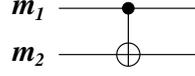}}
\caption{{\footnotesize
Schematical representation
of a two-qubit controlled-NOT (CNOT) quantum gate. 
The $m_1$ ($m_2$) qubit is control (target).
The gate is defined by the truth table (\ref{qg4}).}}
\label{2-CNOT}
\end{figure}

A two-qubit controlled-NOT (CNOT or XOR) gate acts on two qubits denoted 
as a control and a target qubit (FIG.\,\ref{2-CNOT}).
If the control qubit ($m_1$) is in the state $|e\r$, then the state 
of the target qubit ($m_2$) is flipped.
Otherwise, the gate acts trivially, i.e. as the unity operator $\openone$.
We may characterize this gate with the help of the following truth table
\be
\label{qg4}
\begin{array}{lll}
|g_{m_1}\r|g_{m_2}\r & \longrightarrow & |g_{m_1}\r|g_{m_2}\r\,,\\
|g_{m_1}\r|e_{m_2}\r & \longrightarrow & |g_{m_1}\r|e_{m_2}\r\,,\\
|e_{m_1}\r|g_{m_2}\r & \longrightarrow & |e_{m_1}\r|e_{m_2}\r\,,\\
|e_{m_1}\r|e_{m_2}\r & \longrightarrow & |e_{m_1}\r|g_{m_2}\r\,.
\end{array}
\ee

The implementation of the two-qubit CNOT gate on two selected ions in the
ion string requires the introduction of a third auxiliary internal level
$|r\r$. In the original proposal \cite{95-5} the selective
excitation of two sublevels of the $|e\r$ level is used instead. 
The selection depends on the laser polarization, where the $|e,p=0\r$ and
$|e,p=1\r$ sublevels are considered. There have also appeared proposals how
to avoid the establishment of the auxiliary internal level to the scheme
\cite{chuang, monroe, jon}. Two of them will be discussed later on in his
section.

Now we are ready to write two unitary evolution operators
corresponding to laser pulses driven on the first red sideband in the
Lamb-Dicke regime on the $j$th ion [Eq. (\ref{int22})] between the internal
levels $|g\r\leftrightarrow|e\r$ with the atomic frequency
$\omega_0^{eg}=(E_e-E_g)/\hbar$
and for $|g\r\leftrightarrow|r\r$ with $\omega_0^{rg}=(E_r-E_g)/\hbar$.
They are given by
\be
\label{qg5}
\hat{{\cal B}}_j^{\ell,I}(\phi_j)&=&
\cos(\ell\pi/2)
\Bigg[
\bigg(|e_j\r\l e_j|\otimes |0\r\l 0|\bigg)+
\bigg(|g_j\r\l g_j|\otimes |1\r\l 1|\bigg)
\Bigg]\nonumber\\
&-&i\sin(\ell\pi/2)
\Bigg[
\bigg(|e_j\r\l g_j|\otimes |0\r\l 1|\bigg)e^{-i{\phi}_j}+
\bigg(|g_j\r\l e_j|\otimes |1\r\l 0|\bigg)e^{i{\phi}_j}
\Bigg]\nonumber\\
&+&|g_j\r\l g_j|\otimes |0\r\l 0|+{\cal O}
\ee
and
\be
\label{qg6}
\hat{{\cal B}}_j^{\ell,II}(\phi_j)&=&
\cos(\ell\pi/2)
\Bigg[
\bigg(|r_j\r\l r_j|\otimes |0\r\l 0|\bigg)+
\bigg(|g_j\r\l g_j|\otimes |1\r\l 1|\bigg)
\Bigg]\nonumber\\
&-&i\sin(\ell\pi/2)
\Bigg[
\bigg(|r_j\r\l g_j|\otimes |0\r\l 1|\bigg)e^{-i{\phi}_j}+
\bigg(|g_j\r\l r_j|\otimes |1\r\l 0|\bigg)e^{i{\phi}_j}
\Bigg]\nonumber\\
&+&|g_j\r\l g_j|\otimes |0\r\l 0|+{\cal O}\,,
\ee
where we have applied again the arbitrary choice of the phase factor
($\tilde{\phi}_j=\phi_j-\pi/2\rightarrow\phi_j$). The symbol ${\cal O}$
in Eq.~(\ref{qg5}) and (\ref{qg6})
correspond to the terms in Eq. (\ref{int22}) associated with the dynamics on
higher vibrational levels for $n\geq 2$. We do not have to consider them because 
the ions are assumed to be cooled to the ground motional state $|n=0\r$. We use
the Hilbert space spanned only by the motional states $|n=0\r$ and $|n=1\r$  
forming an auxiliary qubit used as the quantum data bus.

The operators (\ref{qg5}) and (\ref{qg6}) correspond
to $k\pi$-pulses on the first red sideband
($\omega_L=\omega_0^{eg}-\nu$ and $\omega_L=\omega_0^{rg}-\nu$)
for $n=0$ with the laser pulse duration $t=\ell\pi/|\lambda_j|\eta_j$. 
Finally, the two-qubit CNOT gate on two ions corresponds to the evolution
operator sequence (acting from right to left) \cite{95-5}
\vspace{2mm}
\be
\label{qg7}
\hat{{\cal A}}_{m_2}^{1/2}(\pi)\,
\hat{{\cal B}}_{m_1}^{1,I}\,
\hat{{\cal B}}_{m_2}^{2,II}\,
\hat{{\cal B}}_{m_1}^{1,I}\,
\hat{{\cal A}}_{m_2}^{1/2}(0)\,, 
\ee
where $\hat{{\cal A}}_{m_2}^{1/2}(0)$ and $\hat{{\cal A}}_{m_2}^{1/2}(\pi)$
are given by Eq. (\ref{qg3}) and stand for the $\pi/2$-pulses on
the carrier ($\omega_L=\omega_0^{eg}$)
on the $m_2$th ion with the phase $\phi_j=0$ and $\phi_j=\pi$,
respectively. The operator $\hat{{\cal B}}_{m_1}^{1,I}$ is defined by 
Eq. (\ref{qg5}) and represents the $\pi$-pulse on the first red
sideband ($\omega_L=\omega_0^{eg}-\omega_z$) on the $m_1$th ion with the
phase factor $\phi_j=0$. The operator $\hat{{\cal B}}_{m_2}^{2,II}$ defined
in Eq. (\ref{qg6}) stands for the $2\pi$-pulse on the first red sideband 
($\omega_L=\omega_0^{rg}-\omega_z$) on the $m_2$th ion with $\phi_j=0$.
The middle sequence
$\hat{{\cal B}}_{m_1}^{1,I}\,
\hat{{\cal B}}_{m_2}^{2,II}\,
\hat{{\cal B}}_{m_1}^{1,I}$
in the evolution operator (\ref{qg7}) can be schematically represented as
follows
\be
\label{qg8}
\begin{array}{rcrcrcr}

& \hat{{\cal B}}_{m_1}^{1,I} & & \hat{{\cal B}}_{m_2}^{2,II} & & 
\hat{{\cal B}}_{m_1}^{1,I}\\

|g_{m_1}\r|g_{m_2}\r|0\r & \longrightarrow & 
|g_{m_1}\r|g_{m_2}\r|0\r & \longrightarrow & 
|g_{m_1}\r|g_{m_2}\r|0\r & \longrightarrow & 
|g_{m_1}\r|g_{m_2}\r|0\r\,,\\

|g_{m_1}\r|e_{m_2}\r|0\r & \longrightarrow & 
|g_{m_1}\r|e_{m_2}\r|0\r & \longrightarrow &
|g_{m_1}\r|e_{m_2}\r|0\r & \longrightarrow & 
|g_{m_1}\r|e_{m_2}\r|0\r\,,\\

|e_{m_1}\r|g_{m_2}\r|0\r & \longrightarrow & 
-i|g_{m_1}\r|g_{m_2}\r|1\r & \longrightarrow &
i|g_{m_1}\r|g_{m_2}\r|1\r & \longrightarrow & 
|e_{m_1}\r|g_{m_2}\r|0\r\,,\\

|e_{m_1}\r|e_{m_2}\r|0\r & \longrightarrow & 
-i|g_{m_1}\r|e_{m_2}\r|1\r & \longrightarrow &
-i|g_{m_1}\r|e_{m_2}\r|1\r & \longrightarrow & 
-|e_{m_1}\r|e_{m_2}\r|0\r\,.

\end{array}
\ee
Finally, the evolution operator (\ref{qg7}) refers 
to the transformation
\be
\label{qg9}
\begin{array}{rcr}
|g_{m_1}\r|g_{m_2}\r|0\r & \longrightarrow & |g_{m_1}\r|g_{m_2}\r|0\r\,,\\
|g_{m_1}\r|e_{m_2}\r|0\r & \longrightarrow & |g_{m_1}\r|e_{m_2}\r|0\r\,,\\
|e_{m_1}\r|g_{m_2}\r|0\r & \longrightarrow & |e_{m_1}\r|e_{m_2}\r|0\r\,,\\
|e_{m_1}\r|e_{m_2}\r|0\r & \longrightarrow & |e_{m_1}\r|g_{m_2}\r|0\r\,,
\end{array}
\ee
on two selected ions labelled as $m_1$ and $m_2$ in the string of $N$ ions.

It is evident from the discussion above that this realization of
the CNOT logic gate on the ion system requires the ions to be cooled 
to the ground motional state $|n=0\r$ in order to maintain the fidelity of
the computational process. Otherwise, as the ions heat up,
higher terms ${\cal O}$ in Eq. (\ref{qg5}) and (\ref{qg6}) 
also contribute and introduce 
significant imperfections into the implementation of the quantum gate.
The two-qubit CNOT was firstly demonstrated in Boulder in 1995
\cite{95-11}. A single Beryllium ion was used, where the control qubit was
stored into two lowest vibrational states $|n=0\r$ and $|n=1\r$ and 
the target qubit was represented by two hyperfine levels 
$|g=S_{1/2}(F=2,m_F=2)\r$ and $|e=S_{1/2}(F=1,m_F=1)\r$.

\subsection{Alternative implementation of two-qubit controlled-NOT gates}

\subsubsection{Simplified quantum logic}
\label{monroe}

Monroe et al. have proposed the realization of the two-qubit quantum logic
gate based on the precise setting of the Lamb-Dicke parameter \cite{monroe}.
The control qubit is assumed to be encoded into two lowest 
vibrational states $|n=0\r$ and $|n=1\r$ of the considered collective
vibrational mode, while the target qubit is represented by two internal
levels $|g_j\r$ and $|e_j\r$ of the $j$th ion from the string of $N$ ions
in the linear Paul trap. The CNOT gate under consideration 
is then described by the truth table
\be
\label{mon1}
\begin{array}{rcr}
|0\r|g_{j}\r & \longrightarrow & |0\r|g_{j}\r\,,\\
|0\r|e_{j}\r & \longrightarrow & |0\r|e_{j}\r\,,\\
|1\r|g_{j}\r & \longrightarrow & |1\r|e_{j}\r\,,\\
|1\r|e_{j}\r & \longrightarrow & |1\r|g_{j}\r\,.
\end{array}
\ee
Further, we adopt the main idea of the original proposal \cite{monroe}.
Driving the $j$th ion with the laser on the carrier ($\omega_L=\omega_0$)
is described by the evolution operator (\ref{int18}).
The coupling constant $\Omega_j^{n,k}$ is
introduced by the expression (\ref{int15}). 
Then for $n=0$ and $n=1$ with $k=0$ we get
\be
\label{mon2}
\Omega_j^{0,0}&=&\lambda_je^{-{\eta_j}^2/2}\,,\\
\label{mon3}
\Omega_j^{1,0}&=&\lambda_je^{-{\eta_j}^2/2}(1-{\eta_j}^2)\,.
\ee
Let us set the Lamb-Dicke parameter such that
\be
\label{mon4}
{\eta_j}^2=\frac{1}{2p}\,,
\ee
where $p$ is an integer. The realization of the transformation (\ref{mon1})
requires driving the carrier transition on the $j$th ion with the duration
$t$ such that 
\be
\label{mon4.1}
\Omega_j^{0,0}t=2p\pi\,.
\ee
We can calculate using Eq. (\ref{mon4}) that 
\be
\label{mon4.2}
\Omega_j^{1,0}t=(2p-1)\pi\,.
\ee
Substituting Eq. (\ref{mon4.1}) and (\ref{mon4.2})
into Eq. (\ref{int18}) we find out that the internal
state of the $j$th ion is flipped only if the collective vibrational state
is $|n=1\r$. We can write
\be
\label{mon7}
\begin{array}{rcl}
|0\r|g_j\r & \longrightarrow & |0\r|g_j\r\,,\\
|0\r|e_j\r & \longrightarrow & |0\r|e_j\r\,,\\
|1\r|g_j\r & \longrightarrow & ie^{-i\phi_j}|1\r|e_j\r\,,\\
|1\r|e_j\r & \longrightarrow & ie^{+i\phi_j}|1\r|g_j\r\,.
\end{array}
\ee
This transformation corresponds to the CNOT gate (\ref{mon1}) apart from
the phase factor $\phi_j$ which can be eliminated by the appropriate phase
settings of subsequent operations. 

The CNOT gate between distinct ions representing two logic qubits can be
implemented (using the proposal being discussed) by two additional laser
pulses on the first red sideband ($\omega_L=\omega_0-\nu$). We have on
mind the CNOT gate given by the truth table (\ref{qg9}). Firstly, we apply
a $\pi$-pulse on the first red sideband on the $m_1$th ion (corresponding
to the evolution operator (\ref{int19}) with $\Omega_j^{0,1}t=\pi$)
mapping the internal state of this ion onto the collective 
vibrational state. We can write
\be
\label{mon8}
\begin{array}{rcl}
|g_{m_1}\r|0\r & \longrightarrow & |g_{m_1}\r|0\r\,,\\
|e_{m_1}\r|0\r & \longrightarrow & -ie^{+i\tilde{\phi}_{m_1}}|g_{m_1}\r|1\r\,,\\
|g_{m_1}\r|1\r & \longrightarrow & -ie^{-i\tilde{\phi}_{m_1}}|e_{m_1}\r|0\r\,.
\end{array}
\ee   
Secondly, we apply a laser pulse on the carrier on the $m_2$th ion
representing the {\it reduced} CNOT gate (\ref{mon1}) and finally, we map
back the collective vibrational state onto the internal state of the $m_1$th
ion by reapplying a $\pi$-pulse on the first red sideband on this ion.
This sequence of three laser pulses corresponds to the {\it complete} CNOT
gate (\ref{qg9}) between two distinct ions with the appropriate choice of
the phase factors.

Comparing this scheme to the original proposal of Cirac and Zoller
(\ref{qg7}), we need fewer laser pulses to realize a two-qubit CNOT gates on
trapped ions and there is no need for a third internal auxiliary level. 
However, more important is the overall time needed to complete the gate and
the sensitivity to imprecisions. The main limitation of this Monroe scheme
is that it is slow compared with other methods at the same level of
infidelity (caused by off-resonant transitions) and it is 
rather sensitive to imprecision in the laser intensity 
($|\Omega|^2\propto I$) \cite{stn}.   

\subsubsection{Fast quantum gates}

\begin{figure}[h!]
\centerline{\epsfig{width=12cm,file=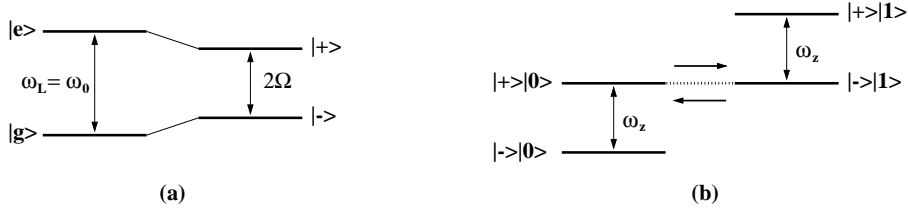}}
\caption{{\footnotesize
Fast quantum gates on cold trapped ions are based: (a) on the splitting of
the dressed states $|\pm\r=1/\sqrt{2}(|g\r\pm|e\r)$ 
by amount proportional to the laser intensity [$\Omega=\Omega(I)$]
when the laser is tuned on the carrier transition ($\omega_L=\omega_0$)
and (b) on setting the splitting such that it is proportional 
to one motional quantum ($\nu=2\Omega$).}}
\label{jon}
\end{figure}

Jonathan et al. have proposed another alternative realization of two-qubit
quantum gates on cold trapped ions scalable on $N$ ions \cite{jon}. It is
based on (i) using carrier transitions and (ii) taking into
account Stark light shifts of atomic levels.
Following Ref. \cite{jon} the basic idea of fast quantum gates is that
the resonant driving of a carrier transition 
($|g\r|n\r\leftrightarrow |e\r|n\r$) with an intense laser causes 
the splitting of dressed states $|\pm\r=1/\sqrt{2}(|g\r\pm|e\r)$ in the
interaction picture by amount $2\hbar\Omega$, where the coupling constant
$\Omega$ is proportional to the laser intensity $I$ [FIG.\,\ref{jon}(a)]. 
When we set the laser intensity such that the splitting 
of the dressed states $|-\r$
and $|+\r$ is equal to one motional quantum $\hbar\nu$.
Then Rabi oscillations appear between the state $|+\r|0\r$
and $|-\r|1\r$ with $|0\r$ and $|1\r$ referring to the lowest collective
vibrational states of the ions [FIG. \ref{jon}(b)]. Using this swapping
between the $|+\r|0\r$ and $|-\r|1\r$ states one can construct a CNOT gate
between two distinct ions following the truth table (\ref{qg9}). 
Quantum gates using this idea are faster than standard quantum gates 
on trapped ions (discussed in Sec.\,\ref{CNOT}) approximately 
by the factor of $1/{\eta}$ assuming the Lamb-Dicke regime. 
The speed of quantum gates will be discussed in Sec.\,\ref{speed}.

\subsection{Multi-qubit controlled-NOT gates}
\label{mqcnot}

\begin{figure}[htb]
\centerline{\epsfig{width=2.75cm,file=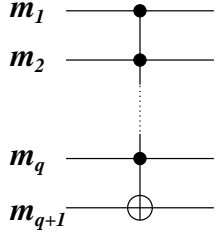}}
\caption{{\footnotesize
Schematical representation
of a multi-bit $(\mbox{controlled})^q$-NOT gate acting on $q+1$ qubits
with $q$ control qubits and $m_{q+1}$th qubit to be target. 
The gate is defined by the truth table (\ref{qg20}).}} 
\label{multi-CNOT}
\end{figure}

A multi-qubit controlled-NOT gate is defined by analogy to the two-qubit CNOT
gate. The only difference is the number of control 
qubits (FIG.\,\ref{multi-CNOT}).
The multi-qubit (controlled)$^q$-NOT gate acts on $q+1$ qubits with $q$ control
qubits $(m_1,\dots,m_q$) and the $m_{q+1}$th qubit is target. If all control
qubits are in the state $|e\r$, then the state of the target qubits is
flipped. Otherwise, the gate acts as the unity operator $\openone$. The
truth table of the multi-qubit (controlled)$^q$-NOT gate acting on
$m_1,\dots,m_{q+1}$ qubits is
\be
\label{qg20}
\begin{array}{llll}
|\Psi_{no}\r|g_{m_{q+1}}\r & \longrightarrow & 
\quad |\Psi_{no}\r|g_{m_{q+1}}\r\,, &
\quad |\Psi_{no}\r\neq\prod\limits_{j=1}^q\otimes|e_{m_j}\r\,,\\
|\Psi_{no}\r|e_{m_{q+1}}\r & \longrightarrow & 
\quad |\Psi_{no}\r|e_{m_{q+1}}\r\,, & \\
|\Psi_{yes}\r|g_{m_{q+1}}\r & \longrightarrow & 
\quad |\Psi_{yes}\r|e_{m_{q+1}}\r\,, &
\quad |\Psi_{yes}\r=\prod\limits_{j=1}^q\otimes|e_{m_j}\r\,,\\
|\Psi_{yes}\r|e_{m_{q+1}}\r & \longrightarrow & 
\quad |\Psi_{yes}\r|g_{m_{q+1}}\r\,. &
\end{array}
\ee
The multi-qubit (controlled)$^q$-NOT gate acting on $q+1$ ions ($m_1,\dots,m_q$
ions represent the control qubits, while the $m_{q+1}$th ion stands for the
target qubit) can be realized by applying the evolution operator (acting
from right to left)
\be
\label{qg21}
\hat{{\cal A}}_{m_{q+1}}^{1/2}(\pi)\,
\hat{{\cal B}}_{m_1}^{1,I}\,
\left[\prod_{j=2}^q\hat{{\cal B}}_{m_j}^{1,II}\right]\,
\hat{{\cal B}}_{m_{q+1}}^{2,II}\,
\left[\prod_{j=q}^2\hat{{\cal B}}_{m_j}^{1,II}\right]\,
\hat{{\cal B}}_{m_1}^{1,I}\,
\hat{{\cal A}}_{m_{q+1}}^{1/2}(0)\,,
\ee
where the $\hat{{\cal B}}$ operators are taken for the value $\phi=0$.
However, this choice of the phase factor has no fundamental importance.
Eq. (\ref{qg21}) applies for three and more ions and
the scheme requires again the auxiliary qubit encoded into two lowest
levels $|n=0\r$ and $|n=1\r$ of the collective vibrational mode used as the
quantum data bus.

Now we verify whether the evolution operator (\ref{qg21})
corresponds to the truth table of the multi-qubit CNOT gate given 
by Eq.~(\ref{qg20}). At first we consider only the $\hat{{\cal B}}$
operators and then we comment on the action of the $\hat{{\cal A}}$ operators.
It holds for $q+1$ ions involved in the multi-qubit CNOT gate that:
\begin{itemize}

\item If the $m_1$th ion is in the ground state $|g_{m_1}\r$, then the
action of the $\hat{{\cal B}}$ operators in Eq. (\ref{qg21}) corresponds to
the unity operator.

\item If the $m_1$th ion is excited with all other ions 
in the ground state $|e_{m_1}\r|g\r^q$, we get
\be
\label{*1}
|e_{m_1}\r|g\r^q|0\r
\stackrel{\hat{{\cal B}}^{1,I}_{m_1}}{\longrightarrow}
-i|g\r^{q+1}|1\r
\stackrel{\hat{{\cal B}}^{1,II}_{m_2}}{\longrightarrow}
-|r_{m_2}\r|g\r^q|0\r
\stackrel{\hat{{\cal B}}^{1,II}_{m_2}}{\longrightarrow}
i|g\r^{q+1}|1\r
\stackrel{\hat{{\cal B}}^{1,I}_{m_1}}{\longrightarrow}
|e_{m_1}\r|g\r^q|0\r\,.
\ee
Thus, the transformation is performed on the $m_1$th ion and then 
on the first next ion in the ground state. The state of 
all other ions in the ground state is not transformed. 
If more ions (besides the $m_1$th one) are excited 
(except if they all are excited), 
their state does not change because the $\hat{{\cal B}}^{\ell, II}_{m_j}$ 
operator acts only  in the Hilbert space spanned by 
$\{|g_{m_j}\r, |r_{m_j}\r\}$ [see Eq. (\ref{qg6})].

\item If all the ions are excited, i.e. $|e\r^{q+1}$, it follows that
\be
\label{*2}
|e\r^{q+1}|0\r
\stackrel{\hat{{\cal B}}^{1,I}_{m_1}}{\longrightarrow}
-i|g_{m_1}\r|e\r^q|1\r
\stackrel{\hat{{\cal B}}^{1,I}_{m_1}}{\longrightarrow}
-|e\r^{q+1}|0\r\,.
\ee
\end{itemize}
Finally, the $\hat{{\cal A}}$ operators complete 
the operation (\ref{qg21}) such that it corresponds 
to the transformation (\ref{qg20}) 
by analogy to Eq. (\ref{qg8}) and (\ref{qg9}).

\subsection{Multi-qubit controlled-$R$ gates}
\label{mqcrot}

\begin{figure}[htb]
\centerline{\epsfig{width=10cm,file=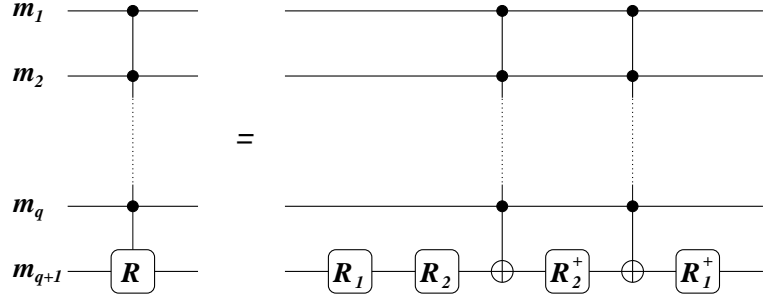}}
\caption{{\footnotesize
Scheme of a multi-qubit $(\mbox{controlled})^q$-$R$ quantum gate. The gate acts
on $q+1$ qubits with $q$ control qubits and the $m_{q+1}$th qubit is target.  
$R$ is defined by Eq. (\ref{qg22}) in the~basis 
$\{|g\r_{m_{q+1}},|e\r_{m_{q+1}}\}$ of the target qubit. $R_1$, 
$R_1^{\dag}$, $R_2$ and $R_2^{\dag}$ 
are also defined by Eq. (\ref{qg22}) in the same basis.}}
\label{CROT}
\end{figure}

A multi-qubit (controlled)$^q$-$R$ gate acts again on $q+1$ qubits. However, it
performs a single-qubit operation (\ref{qg2}) on the $m_{q+1}$th (target)
qubit if all $m_1,\dots,m_q$ control qubits 
are in the state $|e\r$. Otherwise, it acts trivially (FIG.\,\ref{CROT}).
Speaking precisely, if all control qubits are in the state
$|e\r$, then the rotation 
$R=R_1^{\dag}\,\sigma\,R_2^{\dag}\,\sigma\,R_2\,R_1$ is applied (from right
to left) on the target qubit. In the basis of the target qubit
$\{|g\r_{m_{q+1}},|e\r_{m_{q+1}}\}$ we introduce the matrices
\be
\label{qg22}
\begin{array}{cc}
R=
\left(
\begin{array}{cc}
\cos\theta & e^{i2\phi}\sin\theta\\
-e^{-i2\phi}\sin\theta & \cos\theta
\end{array}
\right)\,,
&
\sigma=
\left(
\begin{array}{cc}
0 & 1 \\
1 & 0
\end{array}
\right)\,,
\\ \\
R_1= \left(
\begin{array}{cc}
0 & e^{i\phi}\\
-e^{-i\phi} & 0
\end{array}
\right)\,,
&
R_1^{\dag}=
\left(
\begin{array}{cc}
0 & -e^{i\phi} \\
e^{-i\phi} & 0
\end{array}
\right)\,,
\\ \\
R_2=
\left(
\begin{array}{rc}
\cos(\theta/2) & \sin(\theta/2) \\
-\sin(\theta/2) & \cos(\theta/2)
\end{array}
\right)\,,
&
R_2^{\dag}=
\left(
\begin{array}{cr}
\cos(\theta/2) & -\sin(\theta/2) \\
\sin(\theta/2) & \cos(\theta/2)
\end{array}
\right)\,,
\end{array}
\ee
where 
$R_1=\R(\pi,\phi)$, $R_1^{\dag}=\R^{\dag}(\pi,\phi)$,$R_2=\R(\theta,0)$ and
$R_2^{\dag}=\R(\theta,0)$. The rotation $\R(\theta,\phi)$ is defined by 
Eq. (\ref{qg2}). The matrix $\sigma$ denotes the NOT operation.
If not all control qubits are in the state $|e\r$,
then the gate performs on the target qubit the unity operator 
$\openone=R_1^{\dag}\,\openone\,R_2^{\dag}\,\openone\,R_2\,R_1$.
Finally, we may write the truth table of the multi-qubit 
$(\mbox{controlled})^q$-$R$ gate as follows 
\be
\label{qg23}
\begin{array}{lll}
|\Psi_{no}\r|g_{m_{q+1}}\r & \longrightarrow & 
\quad |\Psi_{no}\r|g_{m_{q+1}}\r\,,\\ \\
|\Psi_{no}\r|e_{m_{q+1}}\r & \longrightarrow & 
\quad |\Psi_{no}\r|e_{m_{q+1}}\r\,,\\ \\
|\Psi_{yes}\r|g_{m_{q+1}}\r & \longrightarrow & 
\quad |\Psi_{yes}\r
\big(\cos\theta\,|g_{m_{q+1}}\r-e^{-i2\phi}\sin\theta\,|e_{m_{q+1}}\r\big)\,,
\\ \\
|\Psi_{yes}\r|e_{m_{q+1}}\r & \longrightarrow & 
\quad |\Psi_{yes}\r
\big(e^{i2\phi}\sin\theta\,|g_{m_{q+1}}\r+\cos\theta\,|e_{m_{q+1}}\r\big)\,,
\end{array}
\ee
where $|\Psi_{no}\r$ and $|\Psi_{yes}\r$ are defined in Eq. (\ref{qg20}).
The multi-qubit controlled-R (CROT) gate (FIG.\,\ref{CROT}) is performed 
on cold trapped ions by applying the evolution operator (\ref{qg21}) for 
the multi-qubit CNOT gates and the corresponding operator for the
single-qubit rotations [Eq. (\ref{qg3})].

\begin{figure}[htb]
\centerline{\epsfig{width=8cm,file=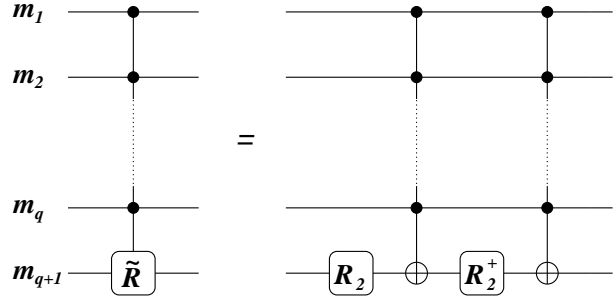}}
\caption{{\footnotesize
Scheme of a reduced multi-qubit $(\mbox{controlled})^q$-$\tilde{R}$ quantum gate.
The gate acts
on $q+1$ qubits with $q$ control qubits and the $m_{q+1}$th qubit is target.  
$\tilde{R}$ is defined by Eq. (\ref{qg24}). $R_2$ and
$R_2^{\dag}$ are defined by Eq. (\ref{qg22}) in the~basis 
$\{|g\r_{m_{q+1}},|e\r_{m_{q+1}}\}$ of the target qubit.}}
\label{CROTr}
\end{figure}

If the preparation of a particular class of quantum states does not
require the introduction of a relative phase shift $\phi$ between the basis
states $|g\r$ and $|e\r$, then a reduced quantum logic network is 
sufficient (FIG.\,\ref{CROTr}).
In particular, the rotation $\tilde{R}=\sigma\,R_2^{\dag}\,\sigma\,R_2$ on
the target qubit conditioned by the state of control qubits 
can be realized according to the following truth table
\be
\label{qg24}
\begin{array}{lll}
|\Psi_{no}\r|g_{m_{q+1}}\r & \longrightarrow & 
\quad |\Psi_{no}\r|g_{m_{q+1}}\r\,,\\ \\
|\Psi_{no}\r|e_{m_{q+1}}\r & \longrightarrow & 
\quad |\Psi_{no}\r|e_{m_{q+1}}\r\,,\\ \\
|\Psi_{yes}\r|g_{m_{q+1}}\r & \longrightarrow & 
\quad |\Psi_{yes}\r
\big(\cos\theta\,|g_{m_{q+1}}\r-\sin\theta\,|e_{m_{q+1}}\r\big)\,,\\ \\
|\Psi_{yes}\r|e_{m_{q+1}}\r & \longrightarrow & 
\quad |\Psi_{yes}\r
\big(\sin\theta\,|g_{m_{q+1}}\r+\cos\theta\,|e_{m_{q+1}}\r\big)\,.
\end{array}
\ee

The results for the multi-qubit controlled-$R$ gates are 
compatible with the scheme proposed in Ref.~\cite{95-8}, where 
the decomposition of multi-qubit CNOT gates into
the network of two-qubit CNOT gates has been presented as well.
However, this decomposition may require many elementary operations in a
particular realization of quantum logic gates.
It seems to be more appropriate for some practical implementations 
of quantum computing to implement directly multi-qubit CNOT gates 
(see Sec.\,\ref{speed}).

\section{Quantum logic networks}
\label{networks}


In this section we present quantum logic networks as effective
tools for the synthesis of quantum coherent superpositions of internal atomic
states. We provide two particular networks, where both of them 
apply to an arbitrary register of qubits. The networks
consist of single-qubit rotations, multi-qubit controlled-NOT and
multi-qubit controlled-$R$ gates. Their implementation on cold trapped
ions is described in detail in Sec.\,\ref{mqcnot} and \ref{mqcrot}.  
The generation 
of nonclassical motional states of a trapped ion experimentally 
is described in Ref. \cite{gen}.

\begin{figure}[htb]
\centerline{\epsfig{width=7.5cm, file=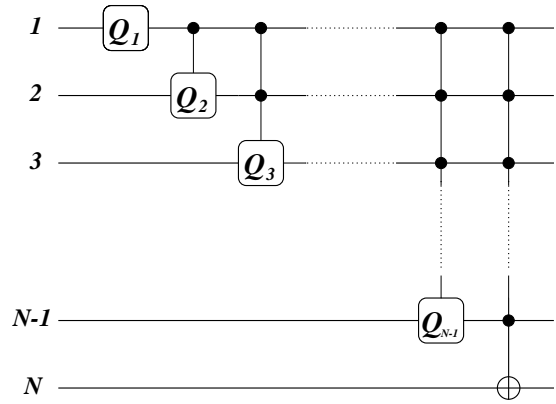}}
\caption{{\footnotesize
The network for the synthesis of the symmetric entangled
state (\ref{net1}). The rotations $Q_j$ are given 
by Eq.~(\ref{net2}). $N$ qubits are assumed to be initially prepared 
in the state  $|eee...e\r$.}}
\label{fnet1}
\end{figure}

We keep the notation $|g\r$ and $|e\r$ for the logical states of the qubit
also in this section.
Firstly, let us introduce the network for the preparation of a totally
symmetric state (with respect to the permutations) of $N$ qubits, such that all
qubits except one are in the excited state  
\be
\label{net1}
|\Psi\r
=\frac{1}{\sqrt{N}}\bigg(
|gee\dots e\r+|ege\dots e\r+|eeg\dots e\r+\dots+|eee\dots g\r
\bigg)\,.
\ee
It has been shown that the maximal degree of bipartite entanglement
measured in the concurrence \cite{conc} 
is equal to $2/N$ and is achieved when a system
of $N$ qubits is prepared just in the state (\ref{net1}). The synthesis of
this state realizes the network in FIG.\,\ref{fnet1} assuming all qubits to
be initially prepared in the state $|e\r$. The rotations $Q_j$ are given as
follows
\be
\label{net2}
Q_j=\left(
\begin{array}{cc}
\sqrt{\frac{N-j}{N-j+1}} & \frac{1}{\sqrt{N-j+1}}\\
-\frac{1}{\sqrt{N-j+1}} & \sqrt{\frac{N-j}{N-j+1}}
\end{array}
\right)\,,\quad
j=1,\dots,N-1\,.
\ee
For more details we refer to our original paper \cite{ms}.

\begin{figure}[htb]
\centerline{\epsfig{width=11cm, file=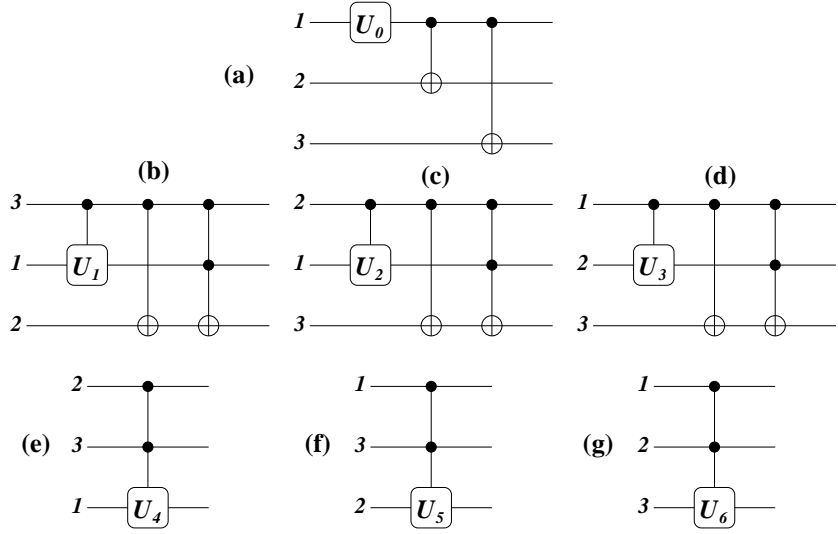}}
\caption{{\footnotesize
An array of networks for the synthesis of an arbitrary pure
quantum state (\ref{net3}) on three qubits. The initial state is $|ggg\r$ and
the rotations $U_j$ are given by Eq.~(\ref{net4})--(\ref{net6}). 
The networks (a)--(g) generate gradually the respective terms 
in the superposition (\ref{net3}).}}
\label{fnet2}
\end{figure}

\begin{figure}[htb]
\centerline{\epsfig{width=11cm, file=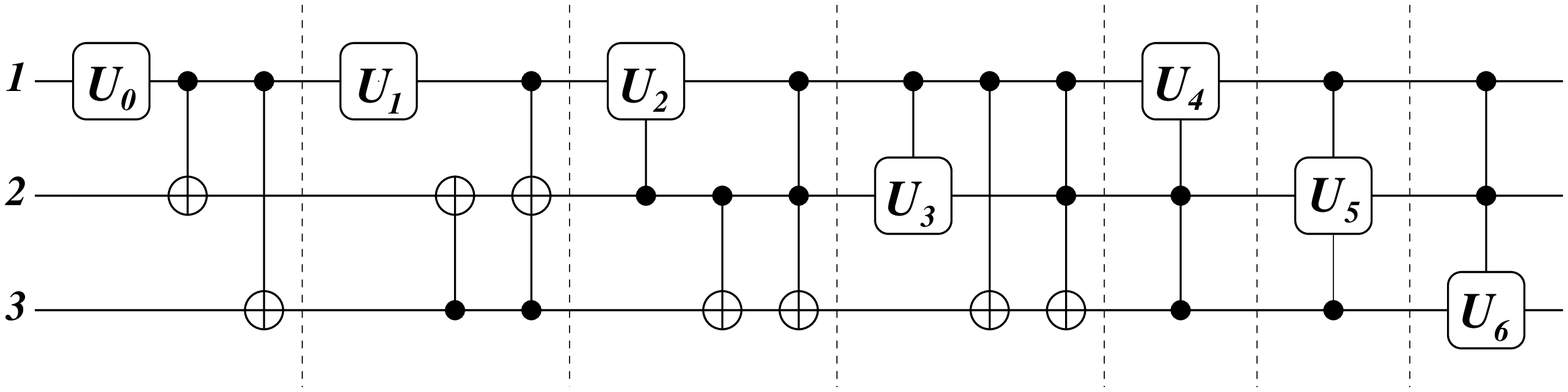}}
\caption{{\footnotesize
A compact form of the array of the networks 
shown in FIG.\,\ref{fnet2}.}}
\label{fnet3}
\end{figure}

Secondly, we propose an array of quantum logic networks 
for the synthesis of an arbitrary pure quantum state for illustration 
depicted on three qubits. However, the scheme is quite easily scalable on $N$ 
qubits \cite{ms}. Let us assume a general state of three qubits 
in the form
\be
\label{net3}
|\psi\r&=&
\a_0|ggg\r+e^{i\varphi_1}\a_1|gge\r+e^{i\varphi_2}\a_2|geg\r+
e^{i\varphi_3}\a_3|egg\r\nonumber\\
&+&e^{i\varphi_4}\a_4|gee\r+e^{i\varphi_5}\a_5|ege\r+
e^{i\varphi_6}\a_6|eeg\r+e^{i\varphi_7}\a_7|eee\r\,,
\ee
The state (\ref{net3}) can be realized by applying
the array of the networks in FIG.\,\ref{fnet2} (shown in a more compact form in
FIG.\,\ref{fnet3}) on the initial state $|ggg\r$, i.e. all three 
qubits in the state $|g\r$. We have denoted the rotations $U_j$ as
follows
\be
\label{net4}
U_j=\left(
\begin{array}{cc}
a_j & e^{i2\phi_j}b_j\\
-e^{-i2\phi_j}b_j & a_j
\end{array}
\right)\,,\qquad j=0,\dots,6\,,
\ee
where $a_j=\cos\theta_j$ and $b_j=\sin\theta_j$. The state (\ref{net3}) is
given by 14 real parameters and the network in FIG.\,\ref{fnet3} preparing 
this state is also determined by 14 parameters (seven rotations), where
\be
\label{net5}
\phi_0=\frac{1}{2}(\pi-\varphi_7)\,, \qquad
\phi_j=\frac{1}{2}(\varphi_j-\varphi_7)\,,\qquad j=1,\dots,6
\ee
and
\be
\label{net6}
b_0=\sqrt{1-\alpha_0^2}\,,\qquad
b_j=\frac{\alpha_j}{\sqrt{1-\sum\limits_{k=0}^{j-1}\alpha_k^2}}\,,\qquad
j=1,\dots,6\,.
\ee
Thus, the mapping between the state under preparation (\ref{net3}) 
and the network (FIG. \ref{fnet3}) is clearly defined.


\section{Speed of quantum gates}
\label{speed}


One of the most important requirements for the implementation of quantum
logic (Sec.\,\ref{impl}) on a particular candidate quantum system 
is the physical realization of quantum gates on time scales which are 
much shorter compared to time
scales associated with decoherence effects. 
We have represented quantum gates on cold trapped ions with unitary
evolution operators (\ref{qg3}), (\ref{qg5}) and (\ref{qg6}) associated with
laser pulses on the carrier and on the first red sideband. However, these
operators are valid only in the Lamb-Dicke and weak coupling regime. Taking
into account the complete Hamiltonian (\ref{int9})
we have to deal with resonant and off-resonant transitions
accompanied with Stark light shifts of the energy levels (Appendix \ref{Off}).
A detailed treatment of this problem was presented by Steane et al. 
in Ref. \cite{speed} and we adopt some of their results in this section. 
The analysis of the speed of gate operations in ion traps was firstly
discussed by Plenio and Knight in Ref. \cite{plen1, plen2}.
At first we discuss the speed of single-qubit rotations and two-qubit CNOT
gates. Then we include some estimations for the speed of multi-qubit CNOT 
gates with cold trapped ions.

\begin{itemize}

\item
The single-qubit rotations are associated with the transition on the carrier
(\ref{qg3}) with the duration $T_{{\cal A}}=\ell\pi/|\lambda|$, 
i.e. a $\ell\pi$-pulse on the carrier applied on a given ion. However, 
we have to consider rather the evolution operator corresponding to the
Hamiltonian (\ref{int9})
when we want to discuss unwanted off-resonant transitions. Directing 
the laser beam such that it is perpendicular to the $z$ axis, 
the Lamb-Dicke parameter $\eta$ becomes equal to zero and off-resonant
transitions $|e\r|n\r\leftrightarrow |g\r|n\pm |k|\r$ for $k\neq 0$ do
not appear in the dynamics. Therefore, one can make the laser coupling
constant $|\lambda|$ large without the restriction on the weak coupling
regime characterized by the condition $|\lambda|\ll\nu$ 
(Appendix \ref{Off}).
We can assume $|\lambda|/2\pi\simeq 300\,\mbox{kHz}$, then we get typically 
for a $\pi/2$-pulse on the carrier $T_{{\cal A}}\simeq 1\,\mu\mbox{s}$.

\item
The two-qubit CNOT gate (\ref{qg7}) is realized by two $\pi/2$-pulses on the
carrier [$\hat{{\cal A}}_{m_2}^{1/2}(0)$, $\hat{{\cal A}}_{m_2}^{1/2}(\pi)$], 
two $\pi$-pulses ($\hat{{\cal B}}_{m_1}^{1,I}$) and a single $2\pi$-pulse 
($\hat{{\cal B}}_{m_2}^{2,II}$) on the first red sideband. 
A pulse on the first red sideband is represented with 
the unitary evolution operator (\ref{int19}) for $k=-1$. However, it was derived
for an ideal case when off-resonant transitions and Stark light shifts were
not considered. We can correct for the light shifts 
by tuning the laser on the frequency 
$\omega_L=\omega_0-\omega_z+\Delta\omega$, where $\Delta\omega$ 
corresponds to the light shifts caused by the presence of the carrier
transitions. The imprecision caused by the excitation of off-resonant
transitions can be corrected by applying a correction laser pulse 
with a correspondingly adjusted phase. This was accomplished in Ref. \cite{speed}
and the limit for the duration of the operation corresponding 
to the $\pi$-pulse on the first red sideband is given as
\be
\label{sp1}
\frac{1}{T_{{\cal B}}}\leq
2\sqrt{2}\epsilon\,\sqrt{\frac{E_r}{Nh}\frac{\omega_z}{2\pi}}\,,
\ee
where $\epsilon=\sqrt{1-F}$ is the imprecision defined via the fidelity $F$,
$E_r=\hbar^2\kappa_{\vartheta}^2/2m$ is the recoil energy of a single ion of
the mass $m$, $\kappa=(2\pi/\Lambda)\cos\vartheta$, 
$\Lambda$ is the laser wavelength, 
$h=2\pi\hbar$ and $\omega_z$ is the axial trapping frequency.
The limit for the duration of the operation
corresponding to the $2\pi$-pulse on the first red sideband 
is the double of the expression given by Eq. (\ref{sp1}).  

\end{itemize}
In TABLE \ref{tab} we give the estimations for Calcium ions $^{40}\mbox{Ca}^+$.
We assume the angle between the laser beam and the $z$ axis to be 
$\vartheta=60^{\circ}$,
the laser wavelength is $\Lambda=729\,\mbox{nm}$ and the axial trapping frequency
is $\omega_z/2\pi\simeq 700\,\mbox{kHz}$. Then we get the recoil frequency 
$E_r/h\simeq 2.33\,\mbox{kHz}$ and $\eta=\sqrt{E_r/\hbar\omega_z}\simeq 0.06$.

\begin{table}[htb]
\begin{center}
\begin{tabular}{|c|cc|cc|}
\hline
& 
\multicolumn{2}{|c|}{$T_{{\cal B}}\ [\mu\mbox{s}]$} &
\multicolumn{2}{|c|}{$T\ [\mbox{ms}]$}\\[0.75mm] 
$\ \ N\ \ $ & 
$\ F=99\%\ $ & 
$\ F=75\%\ $ &  
$\ F=99\%\ $ & 
$\ F=75\%\ $\\[0.75mm]
\hline\hline
2  & 124 & 24.8 & 0.50 & 0.10\\[0.75mm]\hline
3  & 152 & 30.3 & 0.91 & 0.18\\[0.75mm]\hline
6  & 214 & 42.9 & 2.58 & 0.52\\[0.75mm]\hline
9  & 263 & 52.5 & 4.74 & 0.98\\[0.75mm]\hline 
10 & 277 & 55.4 & 5.55 & 1.12\\[0.75mm]\hline
\end{tabular}
\end{center}
\caption{{\footnotesize
$N$ is the total number of the ions confined in the trap and also the number
of the ions involved in the realization of the multi-qubit CNOT gate, 
${T}_{{\cal B}}$ is the duration of the operation corresponding to 
the $\pi$-pulse on the first red sideband given by Eq.~(\ref{sp1})
calculated for two values of the fidelity
($F=99\%, F=75\%$) and $T$ is the total minimal time
[Eq.~(\ref{sp2})] for the realization of the multi-qubit CNOT gate on $N$ ions 
[see Eq.~(\ref{qg21})] evaluated for two different fidelities.}}
\label{tab}
\end{table}

The multi-qubit CNOT on $Q$ ions (\ref{qg21}) differs
from the two-qubit
CNOT gate only in the number of laser pulses required for its realization.
The multi-qubit CNOT gate corresponds to two $\pi/2$-pulses on the carrier
[$\hat{{\cal A}}_{m_{q+1}}^{1/2}(0)$, $\hat{{\cal A}}_{m_{q+1}}^{1/2}(\pi)$], 
a single $2\pi$-pulse on the first red sideband
($\hat{{\cal B}}_{m_{q+1}}^{2,II}$) and $(2Q-2)$ $\pi$-pulses also on the
first red sideband 
($\hat{{\cal B}}_{m_1}^{1,I}$, $\hat{{\cal B}}_{m_2}^{1,II}$, $\dots$,
$\hat{{\cal B}}_{m_{q}}^{1,II}$). Thus, it requires all together $2Q+1$
laser pulses, where $Q=q+1$ refers to the number of the ions
involved in the gate ($q$ control ions, one target ion).
Then, in the spirit of the previous discussion 
the minimal total time for the realization of the multi-qubit CNOT gate
on $Q$ ions reads
\be
\label{sp2}
T=2(T_{{\cal A}}+QT_{{\cal B}})\,,
\ee
where we assume $T_{{\cal A}}=5\,\mu\mbox{s}$ and $T_{{\cal B}}$ is given by
Eq. (\ref{sp1}) for a total number $N$ ($Q\leq N$) of the ions confined 
in the trap. We give some estimations for the realization of
multi-qubit CNOT gates in TABLE \ref{tab}, where the number~$Q$ of the ions
involved in the gate and the total number~$N$ of the ions in the
trap are equal ($N=Q$). We stress this point because there is a difference if
we realize a two-qubit CNOT gate ($Q=2$) having just two ions in the trap
($N=2$), then we get for the total time of the gate 
\be
\label{sp3}
T_1=2(T_{{\cal A}}+2T_{{\cal B}}^{N=2})\simeq 0.5\,\mbox{ms}\,,
\ee 
or having a larger register of ten ions in the trap ($N=10$), what gives
\be
\label{sp4}
T_2=2(T_{{\cal A}}+2T_{{\cal B}}^{N=10})\simeq 1.1\,\mbox{ms}
\ee 
at the same fidelity $F=99\%$.

Any multi-qubit gate on a register of size $N$ can be decomposed into a network
of single-qubit rotations and two-qubit CNOT gates \cite{95-8}. However,
there might be a possibility to realize this multi-qubit directly, if 
a given physical system allows it. For instance, the multi-qubit CNOT gate on six
qubits can be decomposed into the network of 12 two-qubit CNOT gates
including three additional auxiliary qubits \cite{95-8}. 
In the case of cold trapped
ions it requires the total time for the realization of the whole network
($Q=2$, $N=9$)
\be
\label{sp5}
T_3=12\times 2(T_{{\cal A}}+2T_{{\cal B}}^{N=9})\simeq 12.7\,\mbox{ms}\,.
\ee
However, the direct implementation (\ref{qg21}) would take only ($Q=N=6$)
\be
\label{sp6}
T_4=2(T_{{\cal A}}+6T_{{\cal B}}^{N=6})\simeq 2.6\,\mbox{ms}\,,
\ee
which is about five-times less than the former case. 
We have again assumed almost
the perfect fidelity $F=99\%$ of the operation. We conclude that
there can be quantum systems that may support a direct implementation of
multi-qubit gates, rather than their decomposition into fundamental gates,
what may bring advantages at experimental realization as well as in
quantum state synthesis \cite{ms}.


\section{Discussion}


\subsection{Decoherence}

Throughout the paper we have discussed many aspects of cold trapped ions for
quantum computing but we have not dealt with the decoherence which appears
to be a main obstacle in achievements of experimental quantum computing. 
The reason was that our main goal has been to give a basic review and the
discussion on decoherence sources and effects would refer more to an
advanced study \cite{plen1, plen2, garg, hughes, dfv, res, dfqm}.
Nevertheless, for the sake of completeness we would like to
mention on this place some decoherence aspects met in the lab. We will
follow a detailed study of experimental issues in quantum manipulations with
trapped ions given by Wineland et al. \cite{98-5}. The decoherence will be
met in a more general usage of this term. Thus, by the {\it decoherence} 
we mean any effect that limits the fidelity (the match between
desired and achieved realization). Further, we will distinguish three
categories.
\begin{itemize}

\item
{\it Motional state decoherence} is the most troublesome source of 
the decoherence in ion trap experiments and refers to the relaxation of two
vibrational states $|n=0\r$ and $|n=1\r$ of a given motional mode used as
the quantum data bus. The ions are cooled to the ground motional state 
$|n=0\r$ and the excitation to the $|n=1\r$ state  is used for the transfer
of information on a distinct ion. However, this scenario is not ideal
for several reasons:
\begin{itemize}

\item[$\circ$] Instability of trap parameters. We are simply not able to
control all voltages as they undergo fluctuations and dephasing.

\item[$\circ$] We also have to count on (i) the micromotion, (ii) the
Coulomb repulsion between the ions making the motional modes 
(except the COM mode) anharmonic in reality and (iii) stray electrode fields
causing possible excitations of the ion motion.

\item[$\circ$] We have considered just a single motional mode in our
approach, but there are also other $3N-1$ modes present and the
cross-coupling between the modes appears. If spectator $3N-1$ modes are not
cooled to their ground motional states, the energy can be transferred to the
mode of interest. This happens because the trapping potential is anharmonic
in real and these higher anharmonic terms are responsible for the
cross-coupling.

\item[$\circ$] We should mention also inelastic and elastic collisions with the
background gas, even though experiments are carried out in an excellent    
environment ($p\simeq 10^{-8}\,\mbox{Pa}$).

\end{itemize}

\item
{\it Internal state decoherence} corresponds to the evolution when 
a pure state of the ion $|\psi\r=\a|g\r+\b|e\r$ transforms into a mixture
$\hat{\rho}=|\a|^2|g\r\l g|+|\b|^2|e\r\l e|$. The ions demonstrate internal
decoherence times of the order of seconds (Calcium) up to minutes and hours
(Beryllium). The type of the decoherence discussed here can be eliminated by
a proper choice of metastable excited states with long lifetimes.

\item
{\it Operational decoherence} refers to the precision of coherent laser-ion
manipulations. There are several aspects that we have to consider:
\begin{itemize}

\item[$\circ$]
When the ion is illuminated with a laser beam, one has to
control the pulse duration and the phase adjustment in order to avoid the
preparation of unwanted states. 

\item[$\circ$]
Due to the laser spatial intensity
profile, there is a probability (if the ions are spaced too closely) that
the state of a neighbouring ion will be affected. 

\item[$\circ$]
If we consider the standing-wave configuration we have to take care of 
the precise position of the ion in the node or the antinode of the standing 
wave what seems to be very troublesome.

\item[$\circ$]
Finally, off-resonant transitions are always present and we have to control
the laser power very carefully to avoid their excitations. 

\end{itemize}

\end{itemize}
However, there is a way to eliminate the effect of the decoherence. 
We can encode information into a~decoherence-free subspace whose states are
invariant under coupling to the environment \cite{dfs, knight1, knight2}.

\subsection{Ion trap systems}

We have been discussing cold trapped ions so far. {\it Cold} refers to the
fact that all motional modes have to be cooled to their ground motional
states because the dynamics assumes the precise control over the motional
state. However, there have appeared other proposals referring to {\it warm}
or {\it hot} trapped ions which assume an arbitrary motional state.
\begin{itemize}

\item Poyatos et al.\,\cite{poyatos} proposed a scheme for the realization of
two-qubit CNOT gates between two trapped ions using ideas from the atomic
interferometry. They split the wavepacket of the control ion into two
directions depending on its internal state with a laser pulse. 
Then they address one of the
wavepackets of the target ion changing conditionally its internal state and
finally they bring together the wavepackets of the ions using another laser
pulse. The ions communicate through the Coulomb repulsion and under ideal
conditions the scheme is independent on the motional state of the ions.

\item Milburn et al.\,\cite{milburn} described two schemes for manipulations
with warm trapped ions. Firstly, they use the adiabatic passage for 
the conditional phase shift, i.e. the phase of the ion is flipped if the
motional mode is in the superposition of odd number states and the ion is
excited. The COM mode in an arbitrary vibrational state is used for quantum
logic but all other motional modes are assumed to be cooled to their ground
motional states. Secondly, they apply the idea of the collective spin 
\cite{collspin} for faster gates. This idea has been also used for 
the introduction of multi-qubit gates for quantum computing \cite{coll}.

\item S{\o}rensen and M{\o}lmer \cite{hot1, hot2}
proposed a novel scheme based on the idea of
bichromatic light ($\omega_1, \omega_2$).
Realizing the two-qubit CNOT gate they illuminate two
ions with the bichromatic light coupling the states $|gg\r|n\r$ and
$|ee\r|n\r$. They choose detunings far enough from the resonance with 
the first red and blue sidebands such that the intermediate states 
$|eg\r|n\pm 1\r$ and $|ge\r|n\pm 1\r$ are not populated in the process.
The scheme is not sensitive on fluctuations of the number of phonons in the
relevant motional mode.
It is also possible to illuminate with the bichromatic light more ions and 
generate a multiparticle entangled state. Actually, these experiments were
already realized in NIST \cite{sackett}
and they generated the GHZ state with two ions
$|\psi_2\r=1/\sqrt{2}(|gg\r-i|ee\r)$ with a fidelity $F=83\%$ and also the
GHZ state with four ions $|\psi_4\r=1/\sqrt{2}(|gggg\r+i|eeee\r)$
with a fidelity $F=57\%$ using Beryllium ions and the Raman scheme.
Jonathan and Plenio proposed light shift induced quantum gates for trapped
ions insensitive on phonon number in motional modes (thermal motion) 
\cite{jon2}.

\item Finally, there has appeared a proposal of a scalable quantum computer
with the ions in an array of microtraps by Cirac and Zoller \cite{micro1}
detailed in Ref. \cite{micro2}. The ions are placed in a 2D array of
independent ion microtraps \cite{devoe}
and there is another ion (head) that moves above
this plane. If we position the head above a particular ion from the array
and switch on the laser in the perpendicular direction, we can realize a
two-qubit gate. This operation allows us to swap the state of the ion 
to the head which can be moved immediately above a distinct ion in the array 
and transfer the information onto it. The ions oscillating in the microtraps
are not assumed to be cooled to their ground motional states. However, 
their motion can couple to the environment. It becomes relevant during 
the time when the ion interacts with the head but not in the case
when the head moves.
 
\end{itemize}


\section{Conclusion}


In this paper we have tried to review achievements accomplished in the field
of cold trapped ions. In the first part we have discussed in detail  
the ion loading and trapping process, the collective vibrational motion 
of the ions. We have also given a detailed
derivation of the Hamiltonian governing the dynamics of the system including
the discussion of weak coupling and Lamb-Dicke regime. Further, 
we have reviewed laser cooling techniques and a detection process with 
experimental illustrations on Calcium ions. In the second part we have
discussed the implementation of quantum computing using cold trapped ions. 
In particular,
we have described how to realize single-qubit, two-qubit and multi-qubit quantum
logic gates. Finally, we have estimated the speed of quantum gates 
with cold trapped ions. The aim of this
paper is to give an introduction to this field with many references on
relevant papers and studies.


\section*{Acknowledgments}


We would like to thank 
Peter Knight, Danny Segal, Martin Plenio, Andrew Steane 
and Miloslav Du\v{s}ek 
for their helpful comments and suggestions. 
We are also  grateful to Rainer Blatt, Giovanna Morigi and David
Kielpinski for sending us original files of their figures. 
This work was supported by the European Union projects
QUBITS (IST-1999-13021) and QUEST (HPRN-CT-2000-00121).
We acknowledge the special support from 
the Slovak Academy of Sciences.
One of us (M.\v{S}.) is thankful for the support from the  ESF via the 
{\em Programme on Quantum information and quantum computation}.


\appendix
\section{Weak coupling regime}
\label{Off}


The expression (\ref{int9}) corresponds to the complete Hamiltonian 
in the sense that it includes also off-resonant transitions. For instance,
even though a sufficiently intense laser is tuned on the carrier,
the off-resonant transitions on the sidebands are also present and they cause
imprecisions and perturbations in the dynamics. This can be avoided by
setting the laser intensity $I\propto |\lambda|^2$,
i.e. the laser coupling constant [Eq.~(\ref{int8}) or (\ref{int8.1})], 
sufficiently small.
In what follows we
will determine conditions (characterizing the {\it weak coupling regime})
under which we can neglect off-resonant transitions.

\subsection{Off-resonant transitions}

\begin{figure}[htb]
\centerline{\epsfig{width=11cm,file=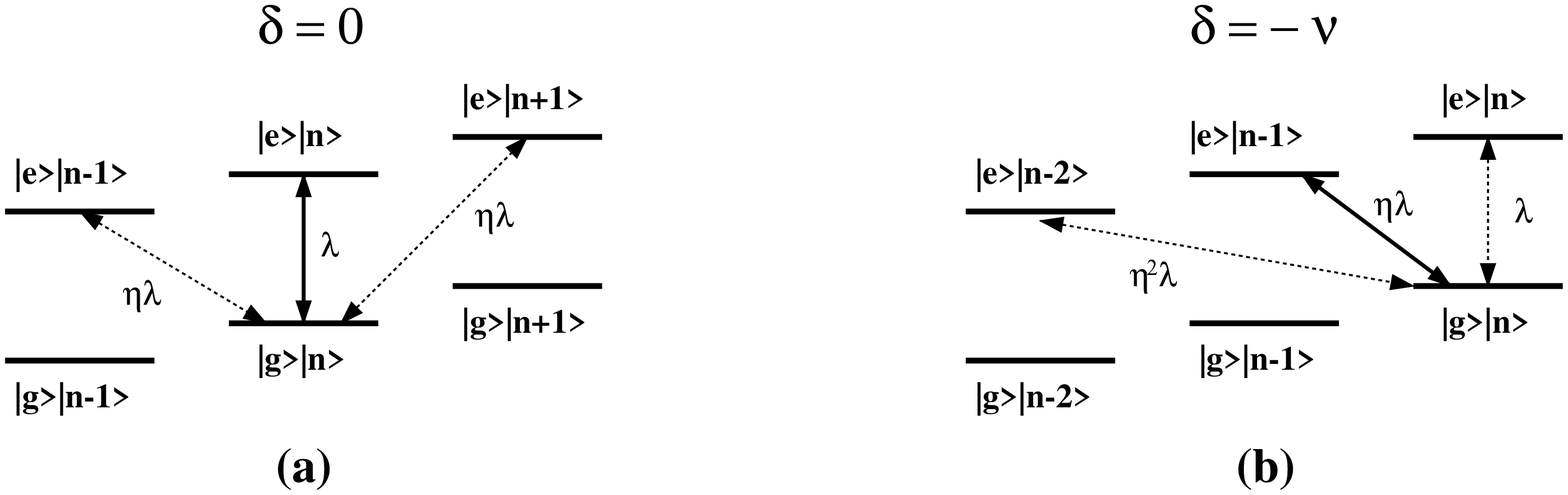}}
\caption{{\footnotesize
(a) The laser is tuned on the carrier ($\delta=0$) with the coupling
constant $\lambda$. The off-resonant transitions are present on the first
blue and red sideband with the coupling constant $\eta\lambda$. 
(b) The laser is tuned on the first red sideband ($\delta=-\nu$) with 
the coupling constant $\eta\lambda$. The off-resonant transitions are
present on the carrier with the coupling constant $\lambda$ and on the weak
second red sideband ($\eta^2\lambda$).}}
\label{off}
\end{figure}

The implementation of quantum gates on cold trapped ions requires laser
pulses on the carrier [FIG.\,\ref{3freq}(a)] and on the first red sideband
[FIG.\,\ref{3freq}(b)]. Therefore, we will discuss the dynamics on these two
spectral lines.
At first, let us assume that the laser is tuned on the carrier $(\delta=0)$.
The closest off-resonant transitions (detuned by the frequency~$\nu$) are on
the first blue and on the first red sideband [FIG.\,\ref{off}(a)].
We will consider only the respective terms in the Hamiltonian
(\ref{int9}) and drop down the index $j$. In the Lamb-Dicke limit we can write
\be
\label{off1}
\hat{{\cal H}}_1=
\frac{\hbar\lambda}{2}\hat{\sigma}_++
\frac{\hbar\lambda}{2}(i\eta)\hat{\sigma}_+\co\,e^{i\nu t}+
\frac{\hbar\lambda}{2}(i\eta)\hat{\sigma}_+\ao\,e^{-i\nu t}+
\mbox{H.c.}\,.
\ee
Further, we assume the ion to be initially in the state $|g\r|n\r$, we apply
the Hamiltonian (\ref{off1}) and calculate the probability of 
off-resonant transitions. The dynamics governed by a time-dependent
Hamiltonian $\hat{{\cal H}}(t)$ is described to the first order by the
unitary evolution operator
\be
\label{off3}
\hat{U}(t,t_0)\approx
\openone-
\frac{i}{\hbar}\int_{t_0}^t\hat{{\cal H}}(t^{\prime})dt^{\prime}\,.
\ee
Then the probability to find the ion (initially prepared in the state
$|g\r|n\r$) in the state $|e\r|n+1\r$ (i.e. undergoing the off-resonant
transition on the first blue sideband) is
\be
\label{off4}
P_B=|\,\l e|\l n+1|\,\hat{U}_1\,|g\r|n\r\,|^2=
\frac{|\lambda|^2\eta^2(n+1)}{\nu^2}
\sin^2\left[\frac{\nu(t-t_0)}{2}\right]\,,
\ee
where $\hat{U}_1$ is given by Eq. (\ref{off3}) for the Hamiltonian
(\ref{off1}). The probability to find the ion in the state $|e\r|n-1\r$
corresponding to the off-resonant transition on the first red sideband is
given as
\be
\label{off5}
P_R=|\,\l e|\l n-1|\,\hat{U}_1\,|g\r|n\r\,|^2=
\frac{|\lambda|^2\eta^2n}{\nu^2}
\sin^2\left[\frac{\nu(t-t_0)}{2}\right]\,.
\ee
If there is no population transferred via the off-resonant transitions to
the states $|e\r|n+1\r$ and $|e\r|n-1\r$, i.e. $P_B\ll 1$ and $P_R\ll 1$ at
any time $t$, we can neglect these off-resonant transitions. Then we get the
conditions of the weak coupling regime for the transition on the carrier
in the form
\be
\label{off6}
|\lambda|\eta\sqrt{n+1}\ll\nu
\ee
and
\be
\label{off7}
|\lambda|\eta\sqrt{n}\ll\nu\,.
\ee 
We can also avoid the off-resonant transitions by setting the laser beam
perpendicular to the $z$ axis ($\vartheta=\pi/2$). Then the Lamb-Dicke
parameter [see Eq. (\ref{int5})] is equal to zero 
($\eta\simeq\kappa\cos\vartheta z_0$) and the coupling on the off-resonant
transitions vanishes. 

Analogically, we can assume the laser to be tuned on the first red sideband
($\delta=-\nu$) and the closest off-resonant transitions are on the carrier
and on the second red sideband [FIG.\,\ref{off}(b)]. However, the strength
of the second red sideband is of the order of $\eta^2$ and we can omit it in
the Lamb-Dicke limit.
Then the respective Hamiltonian is given as
\be
\label{off8}
\hat{{\cal H}}_2=
\frac{\hbar\lambda}{2}(i\eta)\hat{\sigma}_+\ao+
\frac{\hbar\lambda}{2}\hat{\sigma}_+e^{i\nu t}+
\mbox{H.c.}\,.
\ee
The probability to find the ion (initially in the state $|g\r|n\r$) in the
state $|e\r|n\r$ (after the off-resonant transition on the carrier) can be
calculated as
\be
\label{off9}
P_C=|\,\l e|\l n|\,\hat{U}_2\,|g\r|n\r\,|^2=
\frac{|\lambda|^2}{\nu^2}\sin^2\left[\frac{\nu (t-t_0)}{2}\right]\,,
\ee
where $\hat{U}_2$ is given by Eq. (\ref{off3}) for the Hamiltonian
(\ref{off8}). 
We can neglect the off-resonant dynamics if $P_C\ll 1$ at any time $t$. 
Then for the weak coupling regime on the first red sideband applies
\be
\label{off11}
|\lambda|\ll\nu\,.
\ee
Even though the transition on the carrier is off-resonant, 
it has stronger coupling in the Lamb-Dicke limit than the first red sideband. 
Therefore, it is very important to follow in the experiment the constraint given 
by Eq. (\ref{off11}).

\subsection{Stark light shifts}

Besides the population of the off-resonant levels there is another 
source of imprecisions in the state manipulation. 
However, it is weaker and it doesn't
require any special constraints on physical parameters except those for the
weak coupling regime. When the laser drives a transition between two levels
there appears a frequency shift called {\it Stark light shift}
caused by the presence of other spectator levels. Therefore, in the experiment we
have to consider the detuning (\ref{int9.0}) rather in the form
\be
\label{off20}
\tilde{\delta}=\omega_L-\omega_0-\Delta\omega\,,
\ee
where
$\Delta\omega$ corresponds to the Stark light shift. The higher the laser
intensity, the more significant the light shift is. We can correct for this
effect by shifting the laser frequency 
($\omega_L\rightarrow\omega_L+\Delta\omega$) which tunes the transition back
to the resonance. Further, we will estimate the frequency shift
$\Delta\omega$.

Let us consider the Hamiltonian (\ref{int3}) and transform it to the
interaction picture (\ref{int6}) with $\hat{U}_0=\exp(-i\hat{H}_0t/\hbar)$, where 
$\hat{H}_0=(\hbar\omega_L/2)\hat{\sigma}_z$. Then we get
\be
\label{off21}
\hat{{\cal H}}=\hat{{\cal H}}_0+\hat{{\cal V}}= 
\left(-\frac{\hbar\delta}{2}\hat{\sigma}_z+\hbar\nu\co\ao\right)+
\left(
\frac{\hbar\lambda}{2}\hat{\sigma}_+\,e^{i\eta(\ao+\co)}+\mbox{H.c.}
\right)\,, 
\ee
where in the Lamb-Dicke limit the interaction term reduces to
\be
\label{off22}
\hat{{\cal V}}\approx
\frac{\hbar\lambda}{2}\hat{\sigma}_++
\frac{\hbar\lambda}{2}(i\eta)\hat{\sigma}_+\co+
\frac{\hbar\lambda}{2}(i\eta)\hat{\sigma}_+\ao+
\mbox{H.c.}
\ee
with the first term corresponding to the transition on the carrier, the
second term to the first blue sideband and the last one to the transition on
the first red sideband. 
In the second order of the time-independent perturbation theory (the first
order gives no contribution) we can write for the shift of the energy levels
\be
\label{off23}
\Delta E_{|g\r|n\r}&=&
\sum_{m\atop \delta\neq\nu (m-n)}
\frac{\l g|\l n|\hat{{\cal V}}|e\r|m\r\l e|\l m|\hat{{\cal V}}|g\r|n\r}
{\hbar\delta+\hbar\nu (n-m)}
\ee
and
\be
\label{off24}
\Delta E_{|e\r|n\r}&=&
\sum_{m\atop \delta\neq\nu (n-m)}
\frac{\l e|\l n|\hat{{\cal V}}|g\r|m\r\l g|\l m|\hat{{\cal V}}|e\r|n\r}
{-\hbar\delta+\hbar\nu (n-m)}\,.
\ee
For instance, for the transition on the first red sideband ($\delta=-\nu$)
we can calculate
\be
\label{off25}
\Delta E_{|g\r|n\r}\approx -\frac{\hbar |\lambda|^2}{4\nu}\,,\qquad
\Delta E_{|e\r|n-1\r}\approx\frac{\hbar |\lambda|^2}{4\nu}
\ee
and the corresponding light shift can be estimated as
\be
\label{off26}
\Delta\omega=
\bigg(\Delta E_{|e\r|n-1\r}-\Delta E_{|g\r|n\r}\bigg)/\hbar\approx
\frac{|\lambda|^2}{2\nu}\,.
\ee
If we choose $|\lambda|/2\pi=50\,\mbox{kHz}$ and 
$\nu/2\pi=700\,\mbox{kHz}$ so
that the condition (\ref{off11}) holds, then 
$\Delta\omega/2\pi\simeq 1.8\,\mbox{kHz}$.


\section{Lamb-Dicke regime}
\label{ldl}


In the relation for the coupling constant (\ref{int15}) 
we can expand the exponential function to the Taylor series 
about the value $\eta_j=0$ and use the expression (\ref{int16}) 
for the Laguerre polynomial. Then we get
\be
\label{ldl1}
\Omega_j^{n,k}(\eta)&=&
\lambda_j
\left(i\eta_j\right)^{|k|}
\sqrt{\frac{n!}{(n+|k|)!}}
\left[1-\frac{\eta_j^2}{2}+{\cal O}(\eta_j^4)\right]\\
&\times&
\left[\left({n+|k|\atop n}\right)
-\eta_j^2\left({n+|k|\atop n-1}\right)+{\cal O}(\eta_j^4)\right]
\nonumber\,,
\ee
where ${\cal O}(\eta_j^4)$ denotes the terms proportional to the fourth and
higher powers of $\eta_j$.
In the {\it Lamb-Dicke regime}
we consider the dependence of $\Omega_j^{n,k}$ on the parameter $\eta_j$
only to its lowest order, i.e. in Eq. (\ref{ldl1})
we neglect all terms with any higher
power than $\eta_j^{|k|}$ and we get the coupling constant 
$\Omega_j^{n,k}$ given by the expression (\ref{int20}). 
This approximation can be done only if the conditions
\be
\label{ldl2}
\frac{\eta_j^2}{2}\ll 1
\ee
and
\be
\label{ldl3}
\eta_j^2
\left({n+|k|\atop n-1}\right)\ll
\left({n+|k|\atop n}\right)\qquad\Rightarrow\qquad
\eta_j^2\frac{n}{|k|+1}\ll 1
\ee
are satisfied. 
We will refer to the condition ($|k|=0$)
\be
\label{ldl4}
\eta_j\sqrt{\l n \r+\frac{1}{2}}\ll 1
\ee
as the {\it Lamb-Dicke limit}, where $\l n \r$ is the average number of
phonons in the respective vibrational mode \cite{jonathan}.

\begin{itemize}

\item
The Lamb-Dicke limit corresponds physically to the situation where
the spatial extent of the vibrational motion of the ion $z_0$ 
is much smaller than the wavelength $\Lambda$ of the laser,
where $\eta_j\simeq\kappa z_0$ and 
$\kappa=2\pi/\Lambda$ [see def. in Eq.~(\ref{int5})].

\item
The Lamb-Dicke limit can be physically interpreted also from a different
point of view. We may rewrite the Lamb-Dicke parameter of the $j$ ion of $N$
ions in the COM mode to the form $\eta_j^2=E_r/\hbar\omega_z$, 
where $E_r=\hbar^2\kappa^2/2mN$ is the recoil energy. 
It can be shown \cite{tan2} that the trapped ion emits spontaneously photons 
of the average energy
$\hbar\omega_0-E_r$, where $\omega_0$ is for the atomic frequency. Taking
into account the Lamb-Dicke limit ($E_r\ll\hbar\omega_z$) we may say that
during the spontaneous emission the change in the vibrational state of the
ion is very unlikely. In other words, the trapped ion in the Lamb-Dicke
regime decays spontaneously mostly on the carrier 
$(\omega_0\approx\omega_0-E_r/\hbar)$.

\item
The Lamb-Dicke parameter for a single trapped ion equals to $\bar{\eta}$
and for an ion from the string of $N$ ions in the COM mode is given as
$\eta_j=\bar{\eta}/\sqrt{N}$. It means that we can reach the Lamb-Dicke
limit (\ref{ldl4})
for $N$ ions even if the limit is not fulfilled for single ions 
\cite{morigi}.

\end{itemize}



\thebibliography{99}


\bibitem{NIST} {\tt http://www.bldrdoc.gov/timefreq/ion} 

\bibitem{95-5} J.I. Cirac and P. Zoller, 
{\it Quantum computation with cold trapped ions}, 
Phys. Rev. Lett. {\bf 74}, 4091 (1995)

\bibitem{sam} Fortschritte der Physik {\bf 48}, Number 9 -- 11 (2000)

\bibitem{00-4} A. Ekert, P. Hayden, H. Inamori
{\it Basic concepts in quantum computation},
{\tt quant-ph/0011013} (2000)

\bibitem{divin} D.P. DiVincenzo,
{\it The physical implementation of quantum computation} 
Fortschritte der Physik {\bf 48}, 771 (2000) 

\bibitem{loss} Private communication with Daniel Loss.

\bibitem{95-8} 
A. Barenco, Ch. Bennett, R. Cleve, D.P. DiVincenzo, N. Margolus, P. Shor, T.
Sleator, J.A. Smolin and H. Weinfurter,
{\it Elementary gates for quantum computation},                      
Phys. Rev. {\bf A52}, 3457 (1995)

\bibitem{russia} V.I. Balykin, V.G. Minogin and V.S. Letokhov,
{\it Electromagnetic trapping of cold atoms},
Rep. Prog. Phys. {\bf 63}, 1429 (2000)

\bibitem{90-1} W. Paul, 
{\it Electromagnetic traps for charged and neutral particles}
Reviews of Modern Physics {\bf 62}, 531 (1990)

\bibitem{ghosh} P.K. Ghosh, 
{\it Ion traps} (Clarendon Press, Oxford 1995)

\bibitem{LesH} edited by J. Dalibard et al., 
{\it Fundamental systems in quantum optics} (Elsevier, Amsterdam 1992)

\bibitem{innsbruck} {\tt http://heart-c704.uibk.ac.at}

\bibitem{nagerl} H.Ch. N\"{a}gerl, 
{\it PhD thesis} (Innsbruck 1998)

\bibitem{roos} Ch.F. Roos, 
{\it PhD thesis} (Innsbruck 2000)

\bibitem{00-3} H.Ch. N\"{a}gerl, C. Roos, H. Rohde, D. Leibfried, J. Eschner, 
F. Schmidt-Kaler and R. Blatt,
{\it Addressing and cooling of single ions in Paul traps},
Fortschritte der Physik {\bf 48}, 623 (2000) 

\bibitem{blatt1} D. Leibfried, C. Roos, P. Barton, H. Rohde, S. Gulde, 
A. Mundt, G. Reymond, M. Lederbauer, F. Schmidt-Kaler, J. Eschner 
and R. Blatt,
{\it Experiments towards quantum information with trapped Calcium ions},
{\tt quant-ph/0009105} (2000)

\bibitem{blatt2} H. Rohde, S.T. Gulde, C.F. Roos, P.A. Barton, D. Leibfried, 
J. Eschner, F. Schmidt-Kaler and R. Blatt,
{\it Sympathetic ground state cooling and coherent manipulation with 
two-ion crystals},
{\tt quant-ph/0009031} (2000)

\bibitem{zig-zag} 
D.G. Enzer, M.M. Schauer, J.J. Gomez, M.S. Gulley, M.H. Holzscheiter, 
P.G. Kwiat, S.K. Lamoreaux, C.G. Peterson, V.D. Sandberg, D. Tupa, 
A.G. White, R.J. Hughes and D.F.V. James,
{\it Observation of power-law scaling for phase transitions in linear
trapped ion crystals},
Phys. Rev. Lett. {\bf 85}, 2466 (2000)

\bibitem{oxf} A. Steane,
{\it The Ion Trap Quantum Information Processor},
Appl. Phys. {\bf B64}, 623 (1997) 

\bibitem{98-7} D.F.V. James,
{\it Quantum dynamics of cold trapped ions with application to quantum
computation}, 
App. Phys. {\bf B66}, 181 (1998)

\bibitem{louisell}  W.H. Louisell, 
{\it Quantum statistical properties of radiation} 
(John Wiley \& Sons, New York 1973

\bibitem{98-5} D.J. Wineland, C. Monroe, W.M. Itano,
D. Leibfried, B.E. King BE and D.M. Meekhof,
{\it Experimental issues in coherent quantum-state manipulation 
of trapped atomic ions},
Journal of Research of the National Institute of Standards and Technology
{\bf 103}, 259 (1998), (see also {\tt quant-ph/9710025})

\bibitem{99-2} H.Ch. N\"{a}gerl, D. Leibfried, H. Rohde, G. Thalhammer, 
J. Eschner, F. Schmidt-Kaler and R. Blatt,
{\it Laser addressing of individual ions in a linear ion trap},
Phys. Rev. {\bf A60}, 145 (1999)

\bibitem{99-7} D. Kielpinski, B.E. King, C.J. Myatt, C.A. Sackett, 
Q.A. Turchette, W.M. Itano, C. Monroe, D.J. Wineland and W.H. Zurek,
{\it Quantum Logic Using Sympathetically Cooled Ions},
{\tt quant-ph/9909035} (1999)

\bibitem{00-11} F. Schmidt-Kaler, C. Roos C, H.Ch. N\"{a}gerl, H. Rohde,
S. Gulde, A. Mundt, M. Lederbauer, G. Thalhammer, T. Zeiger, P. Barton, 
L. Hornekaer, G. Reymond, D. Leibfried, J. Eschner and R. Blatt,
{\it Ground state cooling, quantum state engineering and study of 
decoherence of ions in Paul traps},
Journal of Modern Optics {\bf 47}, 2573 (2000)

\bibitem{exp} D.J. Wineland, C. Monroe, W.M. Itano, B.E. King, D. Leibfried, 
D.M. Meekhof, C. Myatt and C. Wood,
{\it Experimental primer on the trapped ion quantum computer},
Fortschritte der Physik {\bf 46}, 363 (1998) 

\bibitem{garching} {\tt http://www.mpq.mpg.de/laserphysics.html}

\bibitem{IBM} {\tt http://researchweb.watson.ibm.com/quantuminfo}

\bibitem{imperial} {\tt http://www.lsr.ph.ic.ac.uk/iontrap}

\bibitem{JPL} {\tt http://horology.jpl.nasa.gov}

\bibitem{LANL} {\tt http://p23.lanl.gov/Quantum}


\bibitem{oxford} {\tt http://www.qubit.org/research/IonTrap}

\bibitem{aarhus} {\tt http://www.ifa.au.dk/iontrapgroup}

\bibitem{hamburg}
{\tt http://www.physnet.uni-hamburg.de/ilp/english/research.html}

\bibitem{mainz} {\tt http://www.physik.uni-mainz.de/werth}

\bibitem{lanl} R.J. Hughes, D.F.V. James, J.J. Gomez, M.S. Gulley, 
M.H. Holzscheiter, P.G. Kwiat, S.K. Lamoreaux, C.G. Peterson, M.M. Sandberg,                 
V.D. Schauer, C.M. Simmons, C.E. Thorburn, D. Tupa, P.Z. Wang and A.G. White,
{\it The Los Alamos trapped ion quantum computer experiment},
Fortschritte der Physik {\bf 46}, 329 (1998)

\bibitem{oct} M. Roberts, P. Taylor, G.P. Barwood, P. Gill, 
H.A. Klein and W.R.C. Rowley,
{\it Observation of an Electric Octupole Transition in a Single Ion},  
Phys. Rev. Lett. {\bf 78}, 1876 (1997)

\bibitem{plen1} M.B. Plenio and P.L. Knight,
{\it Realistic lower bounds for the factorization time of large numbers 
on a quantum computer},
Phys. Rev. {\bf A53}, 2986 (1996)

\bibitem{plen2} M.B. Plenio and P.L. Knight,
{\it Decoherene limits to quantum computation using trapped ions},
Proc. R. Soc. Lond. {\bf A453}, 2017 (1997)

\bibitem{raman} J. Steinbach, J. Twamley and P.L. Knight,
{\it Engineering two-mode interactions in ion traps}, 
Phys. Rev. {\bf A56}, 4815 (1997)

\bibitem{97-2} S.A. Gardiner, J.I. Cirac and P. Zoller,
{\it Nonclassical states and measurement of general motional 
observables of a trapped ion}, 
Phys. Rev. {\bf A55}, 1683 (1997)

\bibitem{98-8} S. Chu, 
{\it The manipulation of neutral particles},
Reviews of Modern Physics {\bf 70}, 685 (1998)

\bibitem{98-9} C.N. Cohen-Tannoudji, 
{\it Manipulating atoms with photons},
Reviews of Modern Physics {\bf 70}, 707 (1998)

\bibitem{98-10} W.D. Phillips, 
{\it Laser cooling and trapping of neutral atoms},
Reviews of Modern Physics {\bf 70}, 721 (1998)

\bibitem{metcalf} H.J. Metcalf and P. van der Straten,
{\it Laser cooling and trapping} (Springer-Verlag, New York 1999)

\bibitem{86-1} S. Stenholm, 
{\it The semiclassical theory of laser cooling},
Reviews of Modern Physics {\bf 58}, 699 (1986)

\bibitem{zeil} D. Bouwmeester, A. Ekert and A. Zeilinger, 
{\it The Physics of Quantum Information} (Springer-Verlag, Berlin 2000)

\bibitem{morigi0} G. Morigi, J.I. Cirac, K. Ellinger and P. Zoller,
{\it Laser cooling of trapped atoms to the ground state: A dark state in
position space}, 
Phys. Rev. {\bf A57}, 2909 (1998)

\bibitem{89} F. Diedrich, J.C. Bergquist, W.M. Itano and D.J. Wineland,
{\it Laser cooling to the zero-point energy of motion}, 
Phys. Rev. Lett. {\bf 62}, 403 (1989)

\bibitem{king} B.E. King, C.S. Wood, C.J. Myatt, Q.A. Turchette, 
D. Leibfried, W.M. Itano, C. Monroe and D.J. Wineland,
{\it Cooling the Collective Motion of Trapped Ions to Initialize a Quantum
Register},
Phys. Rev. Lett. {\bf 81}, 1525 (1998)

\bibitem{95-11} C. Monroe, D.M. Meekhof, B.E. King, W.M. Itano 
and D.J. Wineland,
{\it Demonstration of a fundamental quantum logic gate},  
Phys. Rev. Lett. {\bf 75}, 4714 (1995)

\bibitem{jonathan} D. Jonathan, 
{\it PhD thesis} (Imperial College, London 2000)

\bibitem{tan2} C. Cohen-Tannoudji, J. Dupont-Roc and G. Grynberg,
{\it Atom-photon interactions: Basic processes and applications} 
(John Wiley \& Sons, New York 1992)

\bibitem{morigi} G. Morigi, J. Eschner, J.I. Cirac and P. Zoller,
{\it Laser cooling of two trapped ions: Sideband cooling beyond 
the Lamb-Dicke limit}, 
Phys. Rev. {\bf A59}, 3797 (1999)

\bibitem{comm1} Eq. (14) in Ref. \cite{morigi}

\bibitem{98-16} B.E. King, C.S. Wood, C.J. Myatt, Q.A. Turchette, D. Leibfried, 
W.M. Itano, C. Monroe and D.J. Wineland,
{\it Cooling the Collective Motion of Trapped Ions to Initialize a Quantum
Register},  
Phys. Rev. Lett. {\bf 81}, 1525 (1998)

\bibitem{symp2} G. Morigi and H. Walther, 
{\it Low temperature dynamics and laser-cooling of two-species Coulomb
chains for quantum logic},
{\tt quant-ph/0005082} (2000)    

\bibitem{dfs} D. Kielpinski, A. Ben-Kish, J. Britton, V. Meyer, M.A. Rowe,
C.A. Sackett, W.M. Itano, C. Monroe and D.J. Wineland,
{\it Recent Results in Trapped-Ion Quantum Computing},
{\tt quant-ph/0102086} (2001)

\bibitem{symp1} D. Kielpinski, B.E. King, C.J. Myatt, C.A. Sackett, 
Q.A. Turchette, W.M. Itano, C. Monroe, D.J. Wineland and W.H. Zurek,
{\it Sympathetic cooling of trapped ions for quantum logic},
Phys. Rev. {\bf A61}, 032310 (2000)

\bibitem{In} E. Peik, J. Abel, Th. Becker, J. von Zanthier and H. Walther, 
{\it Sideband cooling of ions in radio-frequency traps},
Phys. Rev. {\bf A60}, 439 (1999)

\bibitem{EIT1} G. Morigi, J. Eschner and Ch. H. Keitel,
{\it Ground State Laser Cooling Using Electromagnetically Induced
Transparency},
Phys. Rev. Lett. {\bf 85}, 4458 (2000)

\bibitem{EIT2} C.F. Roos, D. Leibfried, A. Mundt, F. Schmidt-Kaler, 
J. Eschner and R. Blatt,
{\it Experimental Demonstration of Ground State Laser Cooling
with Electromagnetically Induced Transparency},
Phys. Rev. Lett. {\bf 85}, 5547 (2000)

\bibitem{EIT3} F. Schmidt-Kaler, J. Eschner, G. Morigi, C.F. Roos, 
D. Leibfried, A. Mundt and R. Blatt,
{\it Laser cooling with electromagnetically induced transparency:
Application to trapped samples of ions or neutral atoms},
{\tt quant-ph/0107087} (2001)

\bibitem{comm2} Eq. (4) in Ref. \cite{EIT3}

\bibitem{danny} Private communication with Danny Segal.

\bibitem{comm3} Eq. (126) in Ref. \cite{98-5}

\bibitem{86-2} W. Nagourney, J. Sandberg and H. Dehmelt,
{\it Shelved optical electron amplifier: Observation of quantum jumps},
Phys. Rev. Lett. {\bf 56}, 2797 (1986)

\bibitem{chuang} A.M. Childs and I.L. Chuang, 
{\it Universal quantum computation with two-level trapped ions},
{\tt quant-ph/0008065} (2000)    

\bibitem{monroe} C. Monroe, D. Leibfried, B.E. King, D.M. Meekhof, 
W.M. Itano and  D.J. Wineland,
{\it Simplified quantum logic with trapped ions},
Phys. Rev. {\bf A55}, R2489 (1997)

\bibitem{stn} Private communication with Andrew Steane.

\bibitem{jon} D. Jonathan, M.B. Plenio and P.L. Knight,
{\it Fast quantum gates for cold trapped ions}, 
Phys. Rev. {\bf A62}, 42307 (2000)

\bibitem{speed} A. Steane, Ch.F. Roos, D. Stevens, A. Mundt, D. Leibfried, 
F. Schmidt-Kaler and R. Blatt,
{\it Speed of ion-trap quantum-information processors},
Phys. Rev. {\bf A62}, 042305 (2000) 

\bibitem{gen} D.M. Meekhof, C. Monroe, B.E. King, W.M. Itano 
and D.J. Wineland,
{\it Generation of Nonclassical Motional States of a Trapped Atom},
Phys. Rev. Lett. {\bf 76}, 1796 (1996) 

\bibitem{conc} W.K. Wootters,
{\it Entanglement of Formation of an Arbitrary State of Two Qubits},
Phys. Rev. Lett. {\bf 80}, 2245 (1998) 

\bibitem{ms} M. \v{S}a\v{s}ura and V. Bu\v{z}ek, 
{\it Multiparticle entanglement with quantum logic networks: 
Application to cold trapped ions},
Phys. Rev. {\bf A64}, 012305 (2001)

\bibitem{garg} A. Garg,
{\it Decoherence in Ion Trap Quantum Computers},
Phys. Rev. Lett. {\bf 77}, 964 (1996)

\bibitem{hughes} R.J. Hughes, D.F.V. James, E.H. Knill, R. Laflamme 
and A.G. Petschek,
{\it Decoherence Bounds on Quantum Computation with Trapped Ions},
Phys. Rev. Lett. {\bf 77}, 3240 (1996) 

\bibitem{dfv} D.F.V. James,
{\it Theory of Heating of the Quantum Ground State of Trapped Ions},
Phys. Rev. Lett. {\bf 81}, 317 (1998)

\bibitem{res} Q.A. Turchette, C.J. Myatt, B.E. King, C.A. Sackett, 
D. Kielpinski, W.M. Itano, C. Monroe and D.J. Wineland,
{\it Decoherence and decay of motional quantum states of a trapped atom
coupled to engineered reservoirs},
Phys. Rev. {\bf A62}, 053807 (2000)

\bibitem{dfqm} D. Kielpinski, V. Meyer, M.A. Rowe, C.A. Sackett, W.M. Itano,
C. Monroe and D.J. Wineland,
{\it A decoherence-free quantum memory using trapped ions},
Science {\bf 291}, 1013 (2001)

\bibitem{knight1} M.B. Plenio, S.F. Huelga, A. Beige and P.L. Knight,
{\it Cavity-loss-induced generation of entangled atoms},
Phys. Rev. {\bf A59}, 2468 (1999)

\bibitem{knight2} A. Beige, S. Bose, D. Braun, S.F. Huelga, P.L. Knight, 
M.B. Plenio, V. Vedral,
{\it Entangling atoms and ions in dissipative environments},
Journal of Modern Optics {\bf 47}, 2583 (2000)

\bibitem{poyatos} J.F. Poyatos, J.I. Cirac and P. Zoller,
{\it Quantum Gates with "Hot" Trapped Ions}, 
Phys. Rev. Lett. {\bf 81}, 1322 (1998)

\bibitem{milburn} G.J. Milburn, S. Schneider and D.F.V. James,
{\it Ion trap quantum computing with warm ions},  
Fortschritte der Physik {\bf 48}, 801 (2000)

\bibitem{collspin} G.J. Milburn,
{\it Simulating nonlinear spin models in an ion trap},
{\tt quant-ph/9908037} (1999)

\bibitem{coll} X. Wang, A. S{\o}rensen and Klaus M{\o}lmer, 
{\it Multibit Gates for Quantum Computing},
Phys. Rev. Lett. {\bf 86}, 3907 (2001)

\bibitem{hot1} K. M{\o}lmer and A. S{\o}rensen,
{\it Multiparticle Entanglement of Hot Trapped Ions},
Phys. Rev. Lett. {\bf 82}, 1835 (1998)

\bibitem{hot2} A. S{\o}rensen and A. S{\o}rensen, 
{\it Quantum Computation with Ions in Thermal Motion},
Phys. Rev. Lett. {\bf 82}, 1971 (1998)

\bibitem{jon2} D. Jonathan and M.B. Plenio,
{\it Light-shift-induced quantum gates for ions in thermal motion},
{\tt quant-ph/0103140} (2001)

\bibitem{sackett} C.A. Sackett, D. Kielpinski, B.E. King, C. Langer, V. Meyer,
C.J. Myatt, M. Rowe, Q.A. Turchette, W.M. Itano, D.J. Wineland and I.C. Monroe,
{\it Experimental entanglement of four particles},
Nature {\bf 404}, 256 (2000)

\bibitem{micro1} J.I. Cirac and P. Zoller, 
{\it A scalable quantum computer with ions in an array of microtraps},
Nature {\bf 404}, 579 (2000)

\bibitem{micro2} T. Calarco, H.J. Briegel, D. Jaksch, J.I. Cirac and P. Zoller,
{\it Quantum computing with trapped particles in microscopic potentials}, 
Fortschritte der Physik {\bf 48}, 945 (2000), 
(see also {\tt quant-ph/0010105}) 

\bibitem{devoe} R.G. DeVoe,
{\it Elliptical ion traps and trap arrays for quantum computation},
Phys. Rev. {\bf A58}, 910 (1998)
\end{document}